\pdfoutput=1
\documentclass[%
twoside,%
openright,%
titlepage,%
headinclude,%
footinclude=true,%
cleardoublepage=empty,%
abstractoff, 
BCOR=5mm,%
fontsize=11pt,%
english,%
manychapters,%
listsseparated,%
]{scrreprt}

\PassOptionsToPackage{%
				 eulerchapternumbers,%
				 listings,%
				 floatperchapter,
				 linedheaders,%
				 subfig,%
				 beramono,%
				 dottedtoc,
				 manychapters,%
				  }{mio-classicthesis}										

\usepackage{ifthen}
\newboolean{enable-backrefs} 
\setboolean{enable-backrefs}{false} 

\newcommand{\myTitle}{%
Some Observable Effects of Modified Gravity in Cosmology and Astrophysics}

\newcommand{\myName}{Lorenzo Reverberi\xspace}

\newcommand{\myFaculty}{Faculty of Science\xspace}

\newcommand{\myUni}{University of Ferrara\xspace}

\newcommand{\myVersion}{-- L. Reverberi\xspace}

\providecommand{\mLyX}{L\kern-.1667em\lower.25em\hbox{Y}\kern-.125emX\@}

\PassOptionsToPackage{utf8}{inputenc}	
 \usepackage{inputenc}				

\PassOptionsToPackage{english}{babel}
 \usepackage{babel}

\PassOptionsToPackage{square,numbers,%
sort&compress,
}{natbib}
\usepackage{natbib}

\PassOptionsToPackage{fleqn}{amsmath}	
 \usepackage{amsmath}

	\usepackage{fontenc}     
\usepackage{textcomp} 
\usepackage{scrhack} 
\usepackage{xspace} 
\usepackage{mparhack} 
\usepackage{fixltx2e} 

\usepackage{tabularx} 
\PassOptionsToPackage{font={small},%
                      labelfont={sc},%
                      labelformat=simple,
                      margin=20pt,%
                      }{caption}
\usepackage{caption}
\DeclareCaptionLabelSeparator{mysep}{$\,-\,$}
\newboolean{@MySwitchCaption}
\setboolean{@MySwitchCaption}{true}
\ifthenelse{\boolean{@MySwitchCaption}}%
{
\captionsetup{labelsep=mysep,indention=-1.26cm}
}%
{
\captionsetup{labelsep=newline,format=plain}
}
\usepackage{subfig}
\usepackage{textcomp}

\usepackage{listings} 
	\usepackage{graphicx} 

\newcommand{\backrefnotcitedstring}{\relax}
\newcommand{\backrefcitedsinglestring}[1]{(Cited on page~#1.)}
\newcommand{\backrefcitedmultistring}[1]{(Cited on pages~#1.)}
\ifthenelse{\boolean{enable-backrefs}}%
{%
		\PassOptionsToPackage{hyperpageref}{backref}
		\usepackage{backref} 
		   \renewcommand*{\backref}[1]{}  
		   \renewcommand*{\backrefalt}[4]{
		      \ifcase #1 %
		         \backrefnotcitedstring%
		      \or%
		         \backrefcitedsinglestring{#2}%
		      \else%
		         \backrefcitedmultistring{#2}%
		      \fi}%
}{\relax}

\makeatletter
\@ifpackageloaded{babel}%
    {%
       \addto\extrasamerican{%
				}%
       \addto\extrasngerman{%
				}%
    }{\relax}
\makeatother

\let\myoldchap=\chapter
\usepackage{bmpsize}

\usepackage{mio-classicthesis} 

\usepackage{type1ec,%
array,%
dcolumn,%
amsfonts,%
amssymb,%
amsthm,%
color,%
calc,%
eso-pic,%
eepic,%
type1cm,%
url,%
empheq,%
multirow,%
wrapfig,%
float,%
scrpage2,%
tocbibind,   
}%

\usepackage[version=3]{mhchem}

\usepackage{hyperref}

\hypersetup{%
    colorlinks=true, linktocpage=true, pdfstartpage=3, pdfstartview=FitV,%
    breaklinks=true, pdfpagemode=UseNone, pageanchor=true, pdfpagemode=UseOutlines,%
    plainpages=false, bookmarksnumbered, bookmarksopen=true, bookmarksopenlevel=1,%
    hypertexnames=true, pdfhighlight=/O,
    urlcolor=myhypercolor, linkcolor=myhypercolor, citecolor=myhypercolor, 
    pdftitle={\myTitle},%
    pdfauthor={\textcopyright\ \myName, \myUni, \myFaculty},%
    pdfsubject={},%
    pdfkeywords={},%
    pdfcreator={pdfLaTeX},%
    pdfproducer={LaTeX with hyperref and mio-classicthesis}%
}

\renewcommand*{\chapter}{\secdef{\Chap}{\ChapS}}
\renewcommand\ChapS[1]{\myoldchap*{#1}}%
\renewcommand\Chap[2][]{%
    \myoldchap[\texorpdfstring{\spacedlowsmallcaps{#1}}{#1}]{#2}%
         }

\ifthenelse{\boolean{@a5paper}}
{%

\linespread{.95} 
\areaset[current]{280pt}{520pt} %
\setlength{\marginparwidth}{7em}%
\setlength{\marginparsep}{2em}%
}
{%
\linespread{1.05} 
\areaset[current]{312pt}{761pt} 
\setlength{\marginparwidth}{7em}%
\setlength{\marginparsep}{2em}%
}

\usepackage[left=11.2em,right=14.8em,top=5em,bottom=10em]{geometry}

\allowdisplaybreaks

\def\it{\textit}
\def\bf{\textbf}
\def\mb{\mathbf}

\def\mc{\mathcal}

\def\Lemaitre{Lema\^{\i}tre}

\def\idx{\int d^4 x}
\def\mn{{\mu\nu}}
\def\-g{\sqrt{-g}\,}
\def\mpl{m_{Pl}}
\def\so{\quad\Rightarrow\quad}
\def\L{\mathcal L}

\def\to{\rightarrow}
\def\D{\nabla} 

\def\dal{\square}
\def\blank{\thispagestyle{empty}
\
\newpage
}

\newcommand\rif[1]{(\ref{#1})}
\newcommand{\be}{\begin{equation}}
\newcommand{\ee}{\end{equation}}
\newcommand{\bea}{\begin{eqnarray}}
\newcommand{\eea}{\end{eqnarray}}

\renewcommand{\rho}{\varrho}
\renewcommand{\tilde}{\widetilde}

\def\R{R}
\def\T{T}
\def\sun{\odot}
\def\rhoc{10^{29}\text{ g cm}^{-3}}
\def\mywidthsingle{.7\textwidth}
\def\mywidthdouble{.52\textwidth}

\newcommand\citazione[2]{
\vspace*{1.5em}
\begin{flushright}
#1
\end{flushright}
\begin{flushright}
\lowsmallcapsspacing{#2}\\
\end{flushright}
}

\newcommand\AlCentroPagina[3]{
\AddToShipoutPicture*{\AtPageCenter{
\makebox(0,0){\includegraphics[width=#1\paperwidth,angle=#2]{#3}}}}}

\protected\def\titlertwo{$R^2\text{ }$}

\protected\def\titlefr{$f(R)\text{ }$}
\protected\def\titlefR{$f(R)\text{ }$}
\protected\def\titleLambda{$\Lambda$}

\graphicspath{{Images/}}

\newcommand{\titolino}[1]{%
\markboth{\spacedlowsmallcaps{#1}}{\spacedlowsmallcaps{#1}}
\markright{\spacedlowsmallcaps{#1}}
}

\ifthenelse{\boolean{@a5paper}}%
{
 }%
{
}%

\newboolean{@MySwitchTitolo}
\setboolean{@MySwitchTitolo}{true}  

\newboolean{@unifeshit}
\setboolean{@unifeshit}{true} 

\begin{document}

\pagestyle{empty}
\pagenumbering{roman}

\AlCentroPagina{1}{0}{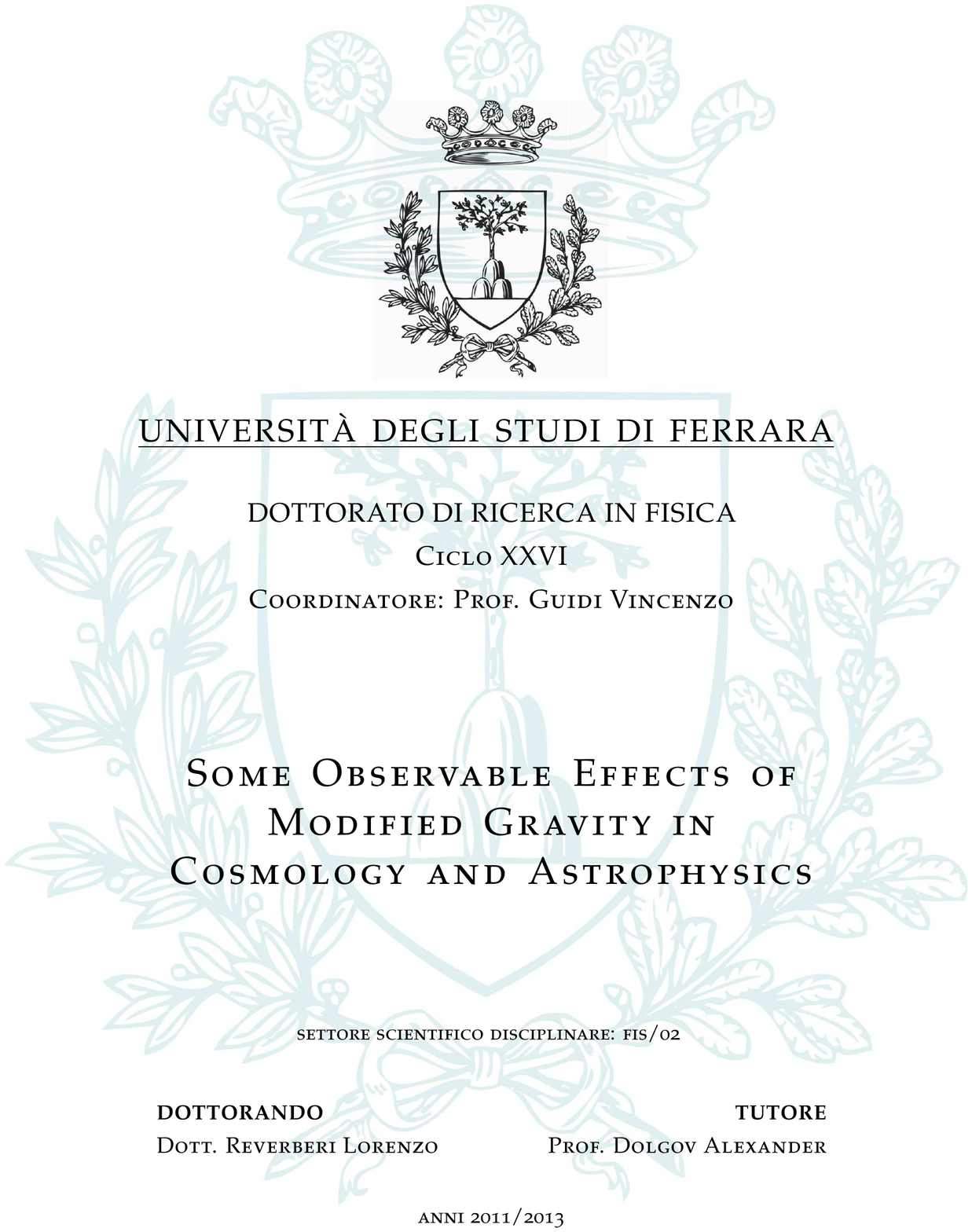}
\thispagestyle{empty}

\newpage

\blank

\cleardoublepage
\pagestyle{scrheadings}


\newboolean{@LoFLoT}
\setboolean{@LoFLoT}{true}     

\setcounter{tocdepth}{2} 
\setcounter{secnumdepth}{3} 
\manualmark
\markboth{\spacedlowsmallcaps{\contentsname}}{\spacedlowsmallcaps{\contentsname}}
\tableofcontents 
\automark[section]{chapter}
\renewcommand{\chaptermark}[1]{\markboth{\spacedallcaps{#1}}{\spacedlowsmallcaps{#1}}}  
\renewcommand{\sectionmark}[1]{\markright{\thesection\enspace\enspace\spacedlowsmallcaps{#1}}} 
    \vspace*{8ex}

\ifthenelse{\boolean{@LoFLoT}}%
{
\begingroup 
    \let\clearpage\relax
    \let\cleardoublepage\relax
    \let\cleardoublepage\relax
%
\manualmark
\markboth{\spacedlowsmallcaps{\contentsname}}{\spacedlowsmallcaps{\contentsname}}
    \listoffigures
    \vspace*{8ex}
\cleardoublepage


\manualmark
\markboth{\spacedlowsmallcaps{\contentsname}}{\spacedlowsmallcaps{\contentsname}}

    \listoftables
        
    \vspace*{8ex}
                        
\endgroup
}%
{\relax}

\cleardoublepage

\chapter*{Preface}
\addcontentsline{toc}{chapter}{Preface}
\titolino{Preface}

\section*{General Relativity and Its Limits}

The General Theory of Relativity (GR) is possibly the most beautiful physical theory ever developed. After being born almost one hundred years ago, it has been extensively studied theoretically and tested observationally, and most of its prediction have been verified to an astounding precision. The universe expansion, the formation of light nuclei and structures, the presence and features of the cosmic microwave radiation, the motion of planets about the Sun, black holes, etc., are only a few of the innumerable consequences of this remarkable theory. Probably no other theory has had the same success.

This is not even all. It is fair to say that general relativity has changed the world even for non-specialists. Albert Einstein has indeed been one of the most famous and influential characters for most of last century, and still is. Suppose you are travelling to a city for vacation: whichever the transport means, GPS devices are likely to be used, and such devices could not operate were it not for the laws of GR. What is more, once we have arrived to our destination, it is hard not to find a souvenir shop selling at least a fridge magnet with one of Einstein's famous quotes, or a t-shirt with the famous image of him sticking his tongue out printed on the chest.

Nevertheless, the emergence of the ``dark universe'' picture during the last decades of observations and precision cosmology has questioned the validity of general relativity at all scales and energies. Dark matter and dark energy appear to account for about 95\% of the total content of the universe. Einstein's dream of a theory unifying quantum and gravitational phenomena drives a great many efforts in the scientific community, but we are probably still very far from such an accomplishments.

These and other questions stimulated an impressive amount of work on extensions to the standard theory. Indeed, modified theories of gravity were born soon after GR, and some of them were even influenced by Einstein's work. More recently, a great variety of models have been proposed to explain the unexpected features of our universe, and the scientific activity in this sector is as lively as ever. Among these theories, I have focused on $f(R)$ gravity, a large class of models in which the Einstein-Hilbert action is replaced by a non-linear function of the Ricci scalar $R$. Testing these theories with cosmology, astrophysics and astroparticle physics is a fundamental path to approach solutions of many problems of modern gravitational physics.

\section*{The Organization of this Thesis}

This thesis is the result of roughly three years of work at the University of Ferrara under the supervision of Prof. A.D. Dolgov and in collaboration with E.V. Arbuzova at Dubna University. This work resulted in a few published papers~\cite{Arbuzova:2011fu, Arbuzova:2012su, Reverberi:2012ew, Reverberi:2013goa, Arbuzova:2013ina, Arbuzova:2014aa} and was presented at several international conferences. 

In Chapter~\ref{ch:vacuum_energy_puzzle} I present the current picture of the universe and briefly review the cosmological constant problem and some of the theories proposed to solve it. The following Chapters essentially contain the published papers with some modifications and additions: in Chapter~\ref{ch:cosm_evol_R2_grav} I study the radiation-dominated epoch in $R^2$ gravity, gravitational particle production and its implications for the early and contemporary universe; in Chapter~\ref{ch:curv_sing_grav_contr} I investigate the formation of curvature singularities in contracting astronomical systems; in Chapter~\ref{ch:grav_part_prod_struct_form}, I discuss some mechanisms to prevent such singularities and calculate the related cosmic-ray production; finally, in Chapter~\ref{ch:spher_symm_grav_repuls}, I deal with spherically symmetric solutions and explore the possibility of gravitational repulsion.

\hyphenation{Sa-bi-no}

\ifthenelse{\boolean{@ringraziamenti}}
{
\section*{Acknowledgements}

This work would not have been possible without the help of many people. First and foremost, I wish to thank my supervisor Prof. Alexander D. Dolgov for his guidance during my PhD, for his inspiring ideas and for teaching me plenty about being a researcher, and Elena V. Arbuzova for her hard work and a fruitful collaboration. I would also like to thank Dr.~Denis Comelli and Dr.~Luigi Pilo for stimulating discussions regarding (but not limited to) gravitation, and Prof.~Roland Triay for his support and for organising the most delightful conferences in Cargèse.

In addition, a thanks to Prof.~Salvatore Capozziello and Prof.~Sabino Matarrese for kindly accepting to refereeing my work and for their nice words.

A special thanks to my family. I am forever indebted to my parents Luisa and Roberto, who never stopped supporting me and believing in me, and to my sister Serena, for making me very proud and for the great fun together. Your love has been truly invaluable to me.

A huge thanks to Giuseppe Aquilino, Paolo Cardarelli, Valentina Rolando, Valentina Sarti Mantovani and all the friends at the Physics department for giving me such a good time at the university and outside of it. Sometimes you were my only motivation to come to work, so thank you very much!

A very big thanks to Agnese Balestra, Francesco Bottoni, Stefano Mannella, Nicola Manto, and all the many friends who brought me joy during these years. Thank you so much, my life would not be the same without you guys.

Finally, very special thanks to Chiara, to whom this work is dedicated. It is simply impossible to express with words how grateful I am for these years together. I love you.\\

Grazie a tutti.\\\\

\begin{flushright}
Ferrara, 10 March 2014\\$\,$\\
Lorenzo Reverberi
\end{flushright}
}%
{\relax}

\chapter*{Notations and Conventions}
\addcontentsline{toc}{chapter}{Notations and Conventions} 
\titolino{Notations and Conventions}

\begin{table}[h]

\begin{tabular}{>{\scshape} l >{$} l <{$}}

Covariant Coordinates		& x_\mu  \\
Contravariant Coordinates	& x^\mu  \\
Metric 				& g_\mn  \\
Inverse Metric 			& g^\mn  \\
Partial Derivative 		& \partial_\mu f = f_{,\mu} = \partial f/\partial x^\mu\\
				& \partial^\mu f = f^{,\mu} =  \partial f/\partial x_\mu \\
Covariant Derivative		& \D_\mu = f_{;\mu} \\
				& \D^\mu f = f^{;\mu} \\
D'Alambertian			& \dal = g^\mn\D_\mu\D_\nu\\
Affine Connection		& \Gamma^\alpha_\mn\\
Christoffel Symbols		& \{^\alpha_\mn\}\\
Riemann Tensor			& \R^\alpha_{\,\,\mu\beta\nu}\\
Ricci Tensor			& \R_\mn \\
Ricci (Curvature) Scalar	& \R \\
Einstein Tensor			& G_\mn = R_\mn - \frac{1}{2}\,g_\mn\,R\\
Energy-Momentum Tensor		& \mc T_\mn = 8\pi G_N\,T_\mn = \frac{8\pi}{\mpl^2}\,T_\mn

\end{tabular}

\end{table}

We use the ``time-like'' conventions of \cite{Landau_Lifshitz:2} or $\left(-,+,+\right)$ in the sign convention of~\cite{Misner-Thorne-Wheeler:1973}, assuming the Einstein's summation convention unless otherwise stated. The Minkowski metric is
\[
\eta_\mn = \text{diag}\left(1,-1,-1,-1\right)\,,
\]
and the relations between metric, affine connection, Ricci tensor and scalar, and Riemann tensor are:
\[
\begin{aligned}
&\Gamma^\alpha_\mn = \frac{1}{2}g^{\alpha\beta}\left(g_{\mu\beta,\nu} + g_{\beta\nu,\mu} - g_{\mn,\beta}\right)\,,\\
&R^\alpha_{\,\,\mu\beta\nu} = \Gamma^\alpha_{\mn,\beta} - \Gamma^\alpha_{\mu\beta,\nu} + \Gamma^\lambda_{\mn}\Gamma^\alpha_{\lambda\beta} - \Gamma^\lambda_{\mu\beta}\Gamma^\alpha_{\lambda\nu}\,,\\
& R_\mn = R^\alpha_{\,\,\mu\alpha\beta}\,,\\
& R = g^\mn R_\mn\,.
\end{aligned}
\]
We use natural units
\[
c = \hbar = k_B = 1\,,
\]
in which we define the Plank mass in terms of Newton's constant as
\[
\mpl \equiv G_N^{-1/2}\,.
\]
Hence, $R$ is negative for vacuum- and matter-dominated Universes, and Einstein's equations read
\[
G_\mn \equiv R_\mn - \frac{1}{2}\,R\,g_\mn = (+)\frac{8\pi}{\mpl^2}\,T_\mn.
\]
A dot denotes derivative with respect to coordinate time $t$, and a prime denotes derivative with respect to the radial coordinate $r$, except if otherwise stated. Greek indices run from 0 to 3, Latin indices run from 1 to 3.

\cleardoublepage
\pagestyle{scrheadings}
\pagenumbering{arabic}

\chapter{The Vacuum Energy Puzzle}\label{ch:vacuum_energy_puzzle}

\section{The \titleLambda CDM Universe}

\subsection{The FLRW Metric}

Our universe is extremely homogeneous and isotropic. It may be hard to tell since we live on a clump of dense matter, rotating about a star with many kilometres of essentially ``nothing'' separating us, and since looking at the night sky we do not observe a uniform, faint glow but rather a spectacular show of point-like, luminous stars hanging on a dark blue ceiling. On very small scales such as these, the universe is in fact very far from being homogeneous and isotropic, and thankfully so - I would not be writing this thesis otherwise. 

However, on scales of the order of the Megaparsec (Mpc), it looks the same everywhere and in all directions~\cite{Wu:1998ad}. This is even more evident when we look at the Cosmic Microwave Background radiation (CMB), which appears isotropic at the order of $10^{-5}$~\cite{Ade:2013iscmb}. 

We describe our universe on large enough scales using a very simple metric, the Friedmann-\Lemaitre-Robertson-Walker metric~\cite{Friedmann:1922,Friedmann:1924bb,Lemaitre:1927zz,Robertson:1935zz,Robertson:1936zz,Robertson:1936zza,Walker:1937}. Although originally found by Friedmann as a solution of the Einstein field equations, it can be derived on the basis of homogeneity and isotropy alone (see e.g. the classic~\cite{Weinberg:Gravitation_and_Cosmology}). It gives the line-element
\be\label{eq:FLRW_line_element}
ds^2 = g_\mn\,dx^\mu dx^\nu = dt^2 - a(t)^2\left(\frac{dr^2}{1-kr^2} + r^2d\Omega\right)\,,
\ee
where $a(t)$ is the cosmic \it{scale factor} and the possible values $k=+1,0,-1$ correspond to a spherical, Euclidean and hyperbolic space, respectively. The quantity $k/a^2$ is the spatial curvature\footnote{Also called \it{Gauss curvature}.} of the universe. 

This is in fact the most general homogeneous and isotropic metric (up to an arbitrary coordinate transformation). The assumption of homogeneity and isotropy implies that the spatial metric be \it{locally} a 3-sphere $\mathbb S^3$, a 3-hyperboloid $\mathbb H^3$ or euclidean space $\mathbb R^3$, while the space can have different \it{global} properties. For instance, euclidean space may be $\mathbb R^3$ and infinite or $\mathbb T^3$ (3-torus) and finite\footnote{This is accomplished by identifying the opposite sides of the fundamental 3-volume element.}. However, this has been strongly constrained by the non-observation of the expected fluctuation patterns in the CMB~\cite{Starobinsky:1993yx,Stevens:1993zz}.

The FLRW metric describes an \it{expanding} (or contracting) spacetime of ``radius'' $a(t)$, where all points are dragged apart from each other like marks on an inflating balloon. In fact, the meaning of the scale factor $a$ is clarified by calculating the proper distance between an object located at comoving coordinate $r$ and an observer in $r=0$, at a given cosmological time $t$:
\be
d(t,r)= a(t)\int_0^r dr'\,\sqrt{g_{rr}}
= a(t)\times
\begin{cases}
\sin^{-1} r &(k=+1)\\
r &(k=0)\\
\sinh^{-1} r &(k=-1)
\end{cases}
\ee
Each comoving object has a constant $r$-coordinate, so the distance between comoving points simply goes like the scale factor $a$.

\subsection{Redshift and Universe Expansion}

When speaking of an expanding universe, we cheated a little bit. Nothing in the geometric properties of the FLRW allows us to determine if $a$ is increasing, decreasing or constant with time. To this purpose, we need to look at the sky, that is at light signals coming from distant sources.

Consider a photon emitted at a given time $t_{em}$ in the past at comoving coordinate $r_{em}$, and let us assume for simplicity that the observer is in $r_{obs}=0$. The time at which the photon will reach the observer, keeping in mind that massless particles follow null geodesics [such that $ds^2 = 0$, cf.~\rif{eq:FLRW_line_element}], is given by
\be
\int_{t_{em}}^{t_{obs}} \frac{dt}{a(t)} = \int_0^{r_{em}}\frac{dr}{\sqrt{1-kr^2}}\,.
\ee
Differentiating this relation, one finds that the time interval between two emitted light signals (or indeed two crests of the light wave) at emission and observation are related by
\be
\frac{\delta t_{em}}{a_{em}} = \frac{\delta t_{obs}}{a_{obs}}\,,
\ee
where $a_{em}\equiv a(t_{em})$ and analogously for $t_{obs}$. If $a$ increases (decreases) with time, it determines an increase (decrease) of the light wavelength by a factor
\be\label{eq:z_lambda}
\frac{\lambda_{obs}}{\lambda_{em}} = \frac{a_{obs}}{a_{em}} \equiv 1 + z\,,
\ee
which defines the \it{redshift} $z$, a fundamental quantity in cosmology and cosmography. ``Today'' corresponds to $z=0$ and $z$ increases up to $z\to\infty$ moving towards the past singularity $a=0$ through the cosmological history.

The observation of a redshift in the light received from distant galaxies is an indication of an expanding universe. For nearby sources we have
\be
a(t) = a_{em}[1 + H_{em}(t-t_{em}) + \cdots]\,,
\ee
where
\be
H = \frac{\dot a}{a}
\ee
is the \it{Hubble parameter}. Similarly, \rif{eq:z_lambda} gives
\be
z = H_{em}(t-t_{em}) + \cdots \,.
\ee
Notice that for nearby sources the quantity $(t-t_{em})$ is essentially the distance $d$ between the source and the observer (in units with $c=1$), and $H_{em}\simeq H_{obs} \equiv H_0$ so\footnote{From now on, a subscript ``${}_0$'' will denote the present value of a given quantity.} we find the famous \it{Hubble's law}:
\be
z \simeq H_0\,d\,.
\ee
Naturally, one missing ingredient is the actual measurement of the distance $d$, but luckily, there are a number of methods and sources allowing us to do so. A more complete discussion on the astronomically relevant distance indicators is way beyond the scope of this thesis; we refer the interested reader to~\cite{Feast:2004rx} and references therein.

The Hubble ``constant'' $H_0$, that is the present value of the Hubble parameter, has the value~\cite{Ade:2013zuv,Ade:2013ktc}
\be
H_0 = (67.80\pm 0.77)\text{ km s}^{-1}\text{Mpc}^{-1}\,.
\ee
In 1929, Edwin Hubble made his famous discovery~\cite{Hubble:1929ig} and announced the world that our universe is expanding, since $H_0>0$. The hypothesis of an expanding universe had already been raised by Friedmann~\cite{Friedmann:1922,Friedmann:1924bb} and de Sitter~\cite{deSitter:1916zz,deSitter:1916zza,deSitter:1917zz}, whereas Einstein himself had neglected this possibility and instead worked a static-universe model adding a \it{cosmological constant} to the picture. Infamously, he dubbed this attempt the ``biggest blunder'' of his life~\cite{Gamow:my_world_line}. Little did he know that the cosmological constant was far from disappearing from physics; we will come back to this very soon enough.

We can also define the \it{deceleration parameter} $q$:
\be\label{eq:deceleration_parameter}
q \equiv -\frac{\ddot a}{aH^2} = -\left(1+\frac{\dot H}{H^2}\right)\,,
\ee
which is positive (negative) if the expansion is decelerating (accelerating).

\subsection{Energy and Matter}

After dealing with the geometric properties of the universe, it is time to look at its content. The properties of homogeneity and isotropy of geometry must reflect on the properties of the \it{energy-momentum tensor} describing matter and its behaviour. The simplest realization of such a tensor is that of a perfect fluid
\be\label{eq:T_mn}
T^\mn = (\rho + P)u^\mu u^\nu - Pg_\mn\,.
\ee
Note that in the comoving frame the 4-velocity is simply $u^\mu = (1,0,0,0)$. The quantities $\rho$ and $P$ are the \it{energy density} and \it{pressure} of the fluid. The relation between the two, in principle arbitrary, is in most cases of interest parametrised by a single constant, the \it{equation of state} $w$:
\be
P = w\,\rho\,.
\ee
The covariant conservation of the $0-\nu$ component of the energy-momentum tensor
\be\label{eq:Energy_Conservation}
\D_\mu T^{\mu 0} = 0\,,
\ee
leads in the FLRW metric to:
\be\label{eq:Energy_Conservation_FLRW}
\dot\rho + 3H(\rho + P) = \dot\rho + 3H(1+w)\rho = 0\,,
\ee
which is easily solved by
\be
\rho\sim a^{-3(1+w)}\,.
\ee
See table~\ref{tab:eq_state} for the most commonly considered cases. As we see, for non-relativistic matter ($w=0$) the dilution of density due to proper volume increase leads to the intuitive scaling $\rho\sim a^{-3}$, while an additional factor $a^{-1}$ enters for radiation ($w=1/3$) due to the redshifting of momentum during expansion. The more exotic vacuum energy ($w=-1$) has instead a constant energy density.

\begin{table}[t]
\centering
\begin{tabular}{l l l}
				& $w$		& Evolution\\
\hline
Non-relativistic matter		& $0$		& $\rho\sim a^{-3}$\\
Relativistic matter (radiation)	& $1/3$		& $\rho\sim a^{-4}$\\
Vacuum energy			& $-1$		& $\rho\sim$ const.\\
\hline
\end{tabular}
\caption{The most common equations of state and their corresponding cosmological evolution.}
\label{tab:eq_state}
\end{table}

The ``concordance'' model of contemporary cosmology is the $\Lambda$CDM model, according to which our universe contains essentially three components of matter/energy~\cite{Ade:2013zuv}:
\begin{enumerate}
\item cold, non-relativistic matter, in the form of ordinary baryonic matter and (mostly) dark matter;
\item $\Lambda$, or vacuum energy;
\item relativistic matter (radiation), mostly in the form of photons (CMB).
\end{enumerate}
Despite the impressive agreement of this picture with most observational evidence, the $\Lambda$CDM model fails in explaining the \it{nature} of both dark matter and dark energy, which collectively account for roughly 95\% of the overall energy budget.

\subsection{Cosmological Dynamics}

The universe, and all matter within it, has to obey the Einstein field equations:
\be\label{eq:Einstein_eqs}
R_\mn - \frac{1}{2}\,R\,g_\mn - \Lambda\,g_\mn = \frac{8\pi}{\mpl^2}\,T_\mn\,.
\ee
As usual, $R_\mn$ is the Ricci tensor and $R=g^\mn R_\mn$ is the Ricci- or curvature scalar. Comparing this expression with~\rif{eq:T_mn}, it is clear that the term containing $\Lambda$ can be thought of as describing a perfect fluid with
\be
\rho_\Lambda = \frac{\mpl^2\Lambda}{8\pi}\,,\qquad P_\Lambda = -\rho_\Lambda\,.
\ee
This justifies the statement that the equation of state of vacuum energy is $w=-1$ (table~\ref{tab:eq_state}).

Einstein's equations~\rif{eq:Einstein_eqs} can be derived by an action principle, provided that one identifies the gravitational lagrangian with the spacetime curvature $R$, that is
\be\label{eq:Einstein_Hilbert_action}
A_{grav} = -\frac{\mpl^2}{16\pi}\int d^4x\,\sqrt{-g}(R + 2\Lambda)\,.
\ee
In both~\rif{eq:Einstein_eqs} and~\rif{eq:Einstein_Hilbert_action}, we have included a $\Lambda$ term for generality and because it is the main subject of this chapter and in fact of this thesis altogether. This term was absent in the original formulation of the theory, but even then there was no \it{a priori} reason not to include it. Indeed, the left-hand side of~\rif{eq:Einstein_eqs} is now the most general local, symmetric, divergenceless two-index tensor that can be constructed solely from the metric and its first and second derivatives.

In a FLRW spacetime, the set of equations~\rif{eq:Einstein_eqs} can be rewritten as the two independent \it{Friedmann equations}
\begin{subequations}\label{eq:Friedmann_eqs}
\begin{align}
& H^2 = \frac{8\pi\rho}{3\mpl^2} + \frac{\Lambda}{3} - \frac{k}{a^2}\,, \label{eq:Friedmann_H_2}\\
& \frac{\ddot a}{a} = -\frac{4\pi}{3\mpl^2}(\rho+3P) + \frac{\Lambda}{3}\,. \label{eq:Friedmann_ddot_a}
\end{align}
\end{subequations}
It is useful to define the \it{critical energy density} $\rho_c$:
\be
\rho_c\equiv \frac{3\mpl^2H^2}{8\pi}\,,
\ee
whose present value is of the order of
\be
\rho_{c,0} \sim 10^{-29}\text{ g cm}^{-3} \sim 10^{-11}\text{ eV}^4\,.
\ee
The Universe is perfectly Euclidean (spatially flat, i.e. $k=0$) if $\rho = \rho_c$, see eq.~\rif{eq:Friedmann_H_2}. Notice that this is indeed the case for our universe, at least at the percent level\footnote{Apparently, the elementary school teacher who taught us that the sum of the angles of a triangle is always 180° was not completely wrong after all!}~\cite{Ade:2013zuv}. 

This spatial flatness is most likely the result of an \it{inflationary} period of accelerated expansion in the very early universe. Of course, a thorough discussion on inflation is beyond the scope of this work, but we refer the interested reader to the beautiful and comprehensive~\cite{Martin:2013tda}.

The contribution of the various forms of matter can be described through the dimensionless parameter $\Omega_s$, with $s=$ radiation, non-relativistic matter, vacuum energy, defined as the ratio of the energy density of the specific species of matter in units of the critical energy density:
\be
\Omega_s \equiv \frac{\rho_s}{\rho_c}\,.
\ee
The statement that the Universe be (almost) Euclidean then becomes
\be\label{eq:Omega_flatness}
\Omega_r + \Omega_m + \Omega_\Lambda \simeq 1\,.
\ee
At a given redshift $z$, the expansion rate $H$ can be expressed as:
\be
H(z)^2 = H_0^2\left[\Omega_{m,0}(1+z^3) + \Omega_{r,0}(1+z)^4 + \Omega_{\Lambda,0} + \Omega_{k,0}(1+z)^2\right]\,.
\ee
In the case of our universe, where~\rif{eq:Omega_flatness} holds, the last term can essentially be neglected. Evidently, the importance of a given matter type can change greatly during the cosmological evolution due to the different dependence on $z$. Our universe, after the initial inflationary period mentioned above, underwent a period of radiation-domination until redshifts of about $z\sim 3600$, then a period of matter-domination until $z\sim 0.4$, and today it appears to be dominated by vacuum energy (see below).

We can also rewrite the deceleration parameter $q$~\rif{eq:deceleration_parameter} as:
\be
\begin{aligned}
q  &= \frac{1}{2}(1+3w_{tot})\left(1+\frac{k}{a^2H^2}\right)\,,\\
& = \frac{\Omega_m}{2} - \Omega_\Lambda\,.
\end{aligned}
\ee
This clearly shows that a Universe whose collective equation of state is $w>-1/3$ decelerates, so any indication of accelerated expansion means that the dominant component of the cosmic fluid has a large, negative equation of state. Equivalently, if the energy density of vacuum exceeds twice that of non-relativistic matter, again the universe expansion would be accelerated. Interestingly, this is precisely what a number of observations are currently telling us.

\subsection{Evidence for a Cosmological Constant}

The presence of a ``dark'' component in the Universe has been known at least since the 1930's, when F. Zwicky measured a luminous matter deficit in the Coma cluster~\cite{Zwicky:1933gu}. Similarly, later observations of the flattening of rotation curves in spiral galaxies~\cite{Salucci:1996bf,Salucci:1997} were incompatible with the observed luminous matter distribution and instead suggested the presence of a dark matter halo with density profile $\sim r^{-2}$. Analogous results can be independently derived from the motion of galaxies, the X-ray temperature of galactic gas, and weak lensing measurements. All of these effects are explained by dark matter halos.

Dark energy, on the other hand, has been around since the 1980's. Inflation predicts a very Euclidean Universe, and it was already clear that there was a severe shortage of non-relativistic matter in the total energy budget; for instance, estimates of the age of globular clusters of 12-14 billion years is incompatible with a matter-dominated Universe~\cite{Turner:1984nf}. 

More recently, the birth of ``precision cosmology'' has led to additional compelling evidence for a dominant negative equation of state:

\begin{itemize}
\item The luminosity distance of Supernovae Ia is consistent with a cosmological constant, and incompatible with a flat, matter-dominated Universe or an open Universe\footnote{For this discovery, S. Perlmutter, A. Riess and B. Schmidt were awarded the Nobel Prize in Physics in 2011.}~\cite{Riess:1998cb,Perlmutter:1997zf,Perlmutter:1998np,Riess:2004nr}, and later results constrained the equation of state of dark energy~\cite{Guy:2010bc,Conley:2011ku,Sullivan:2011kv} to:
\be
w = -1.068^{+0.080}_{-0.082}\,.
\ee

\item Measurements of the CMB~\cite{Ade:2013zuv} combined with WMAP polarization data~\cite{Hinshaw:2012aka} and Baryon Acoustic Oscillations (BAO) data~\cite{Percival:2009xn,Blake:2011en,Beutler:2011hx,Anderson:2012sa,Padmanabhan:2012hf} favour a model with
\be
\begin{aligned}
\Omega_\Lambda &= 0.686 \pm 0.020\\
w &= -1.013^{+0.013}_{-0.010}
\end{aligned}
\ee

\item The cross-correlation between the Integrated Sachs-Wolfe (ISW) effect and large-scale structure favours
\be
w = 1.01^{+0.30}_{-0.40}
\ee
at about $4\sigma$~\cite{Giannantonio:2008zi,Ho:2008bz}.

\item the distribution of galaxy clusters as a function of redshift disfavours a flat, matter-dominated Universe. In particular, the presence of massive clusters at relatively high redshift points towards an accelerated expansion starting around $z\simeq 2$~\cite{Allen:2011zs}.

\end{itemize}

\subsection{Shortcomings of $\Lambda$CDM}

\subsubsection*{The Smallness Problem}
The most serious problem of the $\Lambda$CDM model is probably the \it{smallness problem}: a naive argument sets the value of the cosmological constant as zero-point fluctuation of fundamental quantum fields of mass $m$:
\be
\langle\rho\rangle \simeq \int_0^\infty \frac{d^3k}{2(2\pi)^3}\,\sqrt{\mb k^2+m^2} = \infty^4\,.
\ee
If one believes that QFT be true up to an ultraviolet cutoff scale $k_{UV}$, then the previous integral gives
\be
\langle\rho\rangle \sim k_{UV}^4\,.
\ee
Choosing $k_{UV}\sim \mpl$ yields the extraordinary value often quoted in the literature:
\be
\langle\rho\rangle\simeq 10^{76}\text{ GeV}^4\simeq 10^{123}\rho_{obs}
\ee
This has led some authors to dub this, probably rightfully so, the ``worst theoretical prediction in the history of physics''~\cite{Hobson_Efstathiou_Lasenby:2006}.

However, we must stress that imposing a cutoff scale on the three-momentum alone breaks Lorentz invariance and as a result this is not to be regarded as a rigorous calculation. In fact, it was shown that the quartic divergence for the energy density and pressure of a scalar field does not describe vacuum energy but rather \it{homogeneous background radiation}~\cite{DeWitt:1975ys,Akhmedov:2002ts,Martin:2012bt}, because one finds that $\langle p \rangle = \langle\rho\rangle/3$. Indeed, using dimensional regularization one finds
\be
\langle\rho\rangle\simeq \frac{m^4}{64\pi^2}\,\ln\frac{m^2}{k_{UV}^2}\,,
\ee
which goes roughly as the fourth power of the mass of the fundamental particle, not of the cutoff scale. Although not quite 120 orders of magnitude, the discrepancy with the observed value is nevertheless found to be still outstanding~\cite{Martin:2012bt}.

In addition to this, phase-transitions also give a finite but possibly large contribution to the zero-point energy. This is sometimes referred to as the ``classical'' cosmological constant problem, to distinguish it from the purely quantum effect discussed above. For instance, consider the case of the spontaneous breaking of electroweak symmetry, starting with the Higgs lagrangian:
\be
\mc L_{Higgs} \sim g^\mn D_\mu\Phi^\dagger D_\nu\Phi - V(\Phi,\Phi^\dagger)
\ee
where $D_\mu$ denotes covariant\footnote{Under transformations belonging to the gauge group SU(2)$_L\times$U(1)$_Y$.} derivative and the Higgs potential reads
\be\label{eq:Higgs_potential}
V(\Phi,\Phi^\dagger) = V_0 + \frac{m}{2}\,\Phi^\dagger\Phi + \frac{\lambda}{4}(\Phi^\dagger\Phi)^2\,.
\ee
Before the phase transition, $m^2>0$ so that the minimum is in $\Phi = 0$, while after the phase transition $m<0$ and the new vacuum expectation value of the Higgs field $\Phi$ is 
\be
v = \sqrt{-\frac{m^2}{\lambda}}\,.
\ee
The corresponding shift in energy is
\be
\Delta V = -\frac{m^4}{4\lambda}\,,
\ee
which is negative. The constant term $V_0$ in~\rif{eq:Higgs_potential} may be chosen so that the Higgs potential vanishes \it{after} the phase transition (i.e. today), but in principle it is completely arbitrary, and of course it cannot be chosen to vanish both before and after the transition. The contributions of the electro-weak and QCD phase transitions are of the order of~\cite{Martin:2012bt}
\be
\begin{aligned}
& \rho^{EW}_{vac} \sim - 10^{55}\,\rho_{obs}\,,\\
& \rho^{QCD}_{vac} \sim 10^{45}\,\rho_{obs}\,.
\end{aligned}
\ee

Moreover, a no-go theorem by Weinberg~\cite{Weinberg:1988cp} states that the vacuum energy cannot be cancelled without fine-tuning in any effective 4-dimensional theory satisfying the following conditions~\cite{Dvali:2002fz}:
\begin{enumerate}
\item General Covariance is preserved
\item standard gravity is mediated by a massless graviton
\item the theory contains a finite number of fields below the cutoff scale
\item the theory is ghost-free
\item the fields are assumed spacetime-independent at late times
\end{enumerate}

\subsubsection*{The Coincidence Problem}
On top of the smallness problem, one needs to answer another serious question: why is the present value of the energy density associated with the cosmological constant so close to the present value of the energy density of matter? That is, why 
\be
\left.\Omega_\Lambda\right|_{today} \sim \left. \Omega_m \right|_{today}\,?
\ee
This joins another \it{coincidence problem}, i.e. between ordinary baryonic matter and dark matter (DM)
\be
\Omega_{baryons} \sim \Omega_{DM}\,,
\ee
which is actually rather serious in its own right. In fact, baryons are produced non-thermally whereas DM is usually thought to be created in thermal equilibrium\footnote{This does not apply to \it{all} models of DM, though. For instance, \it{axions} are produced non-thermally.}, since weak interaction cross-sections naturally lead to the correct DM abundance (``WIMP Miracle'')\footnote{Actually, thermal freeze-out giving the right relic energy density is not peculiar to the electroweak scale, see e.g. the ``WIMPless miracle'' of~\cite{Feng:2008ya}, which however remains a rather ``natural'' place for the miracle to occur.}. Solutions to this problem have been proposed, but usually involve additional particles to those forming dark matter, and there is no widely accepted mechanism.

As for the former coincidence problem, one can invoke the \it{anthropic argument}~\cite{Weinberg:1987dv} as a possible solution, which states that a biological observer is most likely to observe the Universe when~\cite{Vilenkin:1994ua}
\be
\Omega_\Lambda \lesssim 10\,\Omega_m\,.
\ee
However, this approach has been widely criticised and even considered non-scientific, since in many of its formulations it is neither verifiable nor falsifiable. The discussion is still very much open on the subject.

What is sure is that the $\Lambda$CDM model agrees remarkably with observations and seems to correctly describe the entire history of the universe after inflation. Its failure in accounting for the value of such parameters, particularly that of the cosmological constant, has however lead to a great many attempts to make such cosmological term \it{dynamical}, e.g. as a result of additional/modified fields, interactions, or gravitation. 

We can divide these proposed models into two classes, depending on which side of the Einstein equations~\rif{eq:Einstein_eqs} they modify: those which modify the right-hand side are usually called ``dark energy models'', and alter the field content of the theory or their gravitational interactions; those which modify the gravitational part of the theory are called, with obvious choice of terminology, ``modified gravity models''.

\section{Dark Energy}

In this section, we will briefly discuss some of the most popular dark energy models, while modified gravity models and in particular $f(R)$ gravity will be considered in more detail in the following section.

\subsection{Quintessence}

Perhaps the leading class of dark energy models is \it{quintessence}~\cite{Fujii:1982ms,Ratra:1987rm,Caldwell:1997ii,Wetterich:1987fm}. These models introduce a minimally-coupled scalar field $\phi$, similarly to scalar-field inflationary models accounting for the early-universe epoch of accelerated expansion. 

The action of the theory is
\be\label{eq:quintessence_action}
\begin{aligned}
A&= A_{GR} + A_M + A_{Quin} \\
A_{Quin} &= \int d^4x\,\sqrt{-g}\,\left[\frac{1}{2}g^\mn\partial_\mu\phi\,\partial_\nu\phi - V(\phi)\right]\,,
\end{aligned}
\ee
where $A_{GR}$ is the usual Einstein-Hilbert action~\rif{eq:Einstein_Hilbert_action} with no bare cosmological term and $A_M$ is the action of ordinary matter fields. Notice the use of the subscript ``${}_\text{M}$'' to indicate a general mixture of radiation and non-relativistic matter, as opposed to ``${}_\text{m}$'' to indicate non-relativistic matter alone. 

So-called ``extended'' quintessence models~\cite{Perrotta:1999am,Baccigalupi:1999dd,Chiba:2001xx} also include a non-minimal coupling between gravity and the field $\phi$, much like scalar-tensor theories (see later. section~\ref{sec:BD-ST}).

The energy-momentum tensor of $\phi$, appearing on the r.h.s. of~\rif{eq:Einstein_eqs}, is
\be
T_\mn^{(\phi)} = \partial_\mu\phi\,\partial_\nu\phi - \frac{1}{2}g_\mn \left[g^{\alpha\beta} \partial_\alpha\phi \, \partial_\beta\phi - V(\phi)\right]\,.
\ee
Taking a flat FLRW universe, the corresponding energy-density and pressure are:
\begin{subequations}\label{eq:quintessence_rho_P}
\begin{align}
\rho_\phi & = \frac{1}{2}\,\dot\phi^2 + V(\phi)\,,\\
P_\phi & = \frac{1}{2}\,\dot\phi^2 - V(\phi)\,.
\end{align}
\end{subequations}
The cosmological evolution is found substituting the expressions into~\rif{eq:Friedmann_eqs}:
\begin{subequations}\label{eq:quintessence_Friedmann_eqs}
\begin{align}
& H^2 = \frac{8\pi}{3\mpl^2}\left[\frac{1}{2}\,\dot\phi^2 + V(\phi) + \rho_M\right]\,, \label{eq:quintessence_H_2}\\
& \frac{\ddot a}{a} = -\frac{4\pi}{3\mpl^2}\left[2\dot\phi^2 - 2V(\phi) + \rho_M + 3P_M\right]\,,\label{eq:quintessence_ddot_a}
\end{align}
\end{subequations}
while varying of~\rif{eq:quintessence_action} with respect to $\phi$ yields
\be\label{eq:quintessence_phi_eq}
\ddot\phi + 3H\dot\phi + V'(\phi) = 0\,,
\ee
where a prime denotes derivative with respect to $\phi$. This equation can also be derived from~\rif{eq:Energy_Conservation_FLRW}.

The equation of state of $\phi$ is:
\be
w = \frac{P}{\rho} = \frac{\dot\phi^2 - 2V(\phi)}{\dot\phi^2 + 2V(\phi)}\,,
\ee
which gives the constraints
\be\label{eq:quintessence_w}
-1\leq w \leq 1\,.
\ee
The vacuum energy equation of state $w=-1$ is recovered when the field is static, in which case its energy density is simply the value of the potential in the equilibrium point, but in the general case we have more interesting features. The condition~\rif{eq:quintessence_w} and the energy conservation equation~\rif{eq:Energy_Conservation_FLRW} also imply
\be
1 \leq \frac{\rho_\phi(z)}{\rho_{\phi,0}} \leq (1+z)^6\,.
\ee
Of course, the shape of the potential $V(\phi)$ is crucial in determining the evolution of $\phi$ and its possible implications for dark energy. Equation~\rif{eq:quintessence_ddot_a} shows that if
\be\label{eq:quintessence_accel_condition}
\dot\phi^2 < V(\phi)\,,
\ee
then the effect of the field $\phi$ will be to accelerate the expansion. One can easily see the analogy with~\rif{eq:deceleration_parameter}: acceleration starts when $w=-1/3$, which is precisely the equation of state~\rif{eq:quintessence_w} when~\rif{eq:quintessence_accel_condition} holds. This relates completely to the \it{slow-roll} condition typical of the inflaton field. In fact, the potential must be sufficiently flat in order to have effective acceleration: if the potential is so steep that $\dot\phi^2 \gg 2V(\phi)$ everywhere during the cosmic evolution, then $\rho_\phi\sim a^{-6}$ decreases much faster than the background density, hence the late-time effect would be completely negligible.

Unfortunately, it is quite difficult to accommodate a light scalar field such as that needed for quintessence in particle physics, with $m_\phi\sim H_0\sim 10^{-33}$ eV~\cite{Carroll:1998zi,Kolda:1998wq}, but there have been attempts in this direction, especially in the framework of supersymmetric theories.

Quintessence models can be divided into two large families~\cite{Caldwell:2005tm}: ``freezing models'' and ``thawing models''.

\subsubsection*{Freezing Models}
In freezing models, the field slows down due to the shallowness of the potential at late times. Some prototypical models are:
\begin{itemize}
\item $ V(\phi) = M^{4+\alpha}\phi^{-\alpha}$
The first potential has no local minimum so the field rolls down indefinitely, gradually slowing down~\cite{Ratra:1987rm,Peebles:1987ek}. There exists a so-called \it{tracker} solution with an almost constant equation of state
\be
w \simeq -\frac{2}{2 + \alpha}
\ee
during the matter era~\cite{Zlatev:1998tr}. Independently of the initial conditions, solutions approach the tracker solution before relaxing to $w\to -1$ at later times. Analogous potentials arise e.g. in the fermion condensate model as dynamical supersymmetry breaking~\cite{Binetruy:1998rz}.

\item $ V(\phi) = M^{4+\alpha}\phi^{-\alpha}\,\exp(\phi^2/\phi_0^2) $

This potential can arise in the framework of supergravity~\cite{Brax:1999gp}. The addition of the exponential term produces a true minimum for the potential, hence the field eventually freezes to $w=-1$.

\item $ V(\phi) = M_1^4\,\exp(-\lambda_1\phi/\mpl) + M_2^4\,\exp(-\lambda_2 \phi/\mpl)$.

This potential is a candidate belonging to a sub-class of freezing models other than tracking models~\cite{Steinhardt:1999nw,Baccigalupi:2000je,Baccigalupi:2001aa}, namely ``scaling'' models~\cite{Copeland:1997et}. During most of the matter era the field equation of state scales as the equation of state of the background fluid. Suitable values of the dimensionless constants are $\lambda_1\gg 1$ and $\lambda_2\lesssim 1$~\cite{Barreiro:1999zs}.

\end{itemize}

\subsubsection*{Thawing Models}

In thawing models the field is initially frozen by Hubble friction [the $H\dot\phi$ term in~\rif{eq:quintessence_phi_eq}], and starts evolving at late times when the Hubble parameter $H$ drops below the field mass $m_\phi$. Accordingly, the equation of state of the field, initially equal to $w=-1$, starts \it{increasing} at late times to values $w>-1$. 

Models belonging to this class are
\begin{itemize}
\item $V(\phi) = V_0 + M^{4-\alpha}\,\phi^\alpha$.

This potential resembles that of chaotic inflation for $\alpha=2,4$ and $V_0=0$~\cite{Linde:1983gd}, but of course with a very different mass scale $M$.

\item $V(\phi) = M^4[1 \pm \cos(\phi/\phi_0)]$.

This potential arises as potential of the Pseudo-Nambu-Goldstone boson~\cite{Frieman:1995pm} or of the axion~\cite{Nomura:2000yk,Choi:1999xn}.
\end{itemize}

\subsection{Phantom Fields}

Current observations are compatible with $w<-1$, so it is rather natural to devise theories reproducing such behaviour. The first attempt is to give a ``wrong-sign'' kinetic term to a scalar field, now called \it{phantom}~\cite{Caldwell:1999ew}:
\be
\begin{aligned}
A &= A_{GR} + A_M + A_{Ph} \\
A_{Ph} &= \int d^4x\,\sqrt{-g}\,\left[-\frac{1}{2}\,g^\mn\,\partial_\mu\phi\,\partial_\nu\phi - V(\phi)\right]\,.
\end{aligned}
\ee
The equation of state is
\be
w_\phi = \frac{P}{\rho} = \frac{\dot\phi^2 + 2V(\phi)}{\dot\phi^2 - 2V(\phi)}\,,
\ee
so values $w_\phi<-1$ are allowed for $\dot\phi^2 < 2 V(\phi)$. These models may be motivated by $S$-brane construction in string theory~\cite{Chen:2002yq}, but are unfortunately plagued by terrible ultraviolet instabilities against the production of ghosts and positive energy fields~\cite{Carroll:2003st}. In short, this is due to the fact that the energy of the field is not bounded from below. Even assuming phantom models as low-energy effective theories, thus valid up to an energy cutoff $E_{max}$, observations strongly constrain such limit energy at the level of the MeV~\cite{Cline:2003gs}.

\subsection{$k$-Essence}\label{sec:k-essence}

So-called $k$-essence models can be thought of as a generalization of quintessence models in which non-canonical kinetic terms for the new scalar field $\phi$ are allowed. The general action is
\be\label{eq:k-essence_action}
\begin{aligned}
A &= A_{GR} + A_M + A_{k-ess} \\
A_{k-ess} &= \int d^4x\,\sqrt{-g}\,P(\phi,X)\,,
\end{aligned}
\ee
where
\be\label{eq:X_kinetic_energy}
X \equiv \frac{1}{2}\,g^\mn\,\partial_\mu\phi\,\partial_\nu\phi
\ee
is the usual expression for the kinetic energy of the field $\phi$. The function $P$ is an arbitrary function of the field ($\sim$ potential) and of its kinetic energy. What is interesting in these models is that the latter, not the ``potential'' term, is responsible for the universe acceleration.

These models were initially proposed in the context of inflation~\cite{ArmendarizPicon:1999rj}, while their application to the late-times dark energy was first carried out in~\cite{Chiba:1999ka} and later extended in~\cite{ArmendarizPicon:2000dh,ArmendarizPicon:2000ah}, where the name ``$k$-essence'' first appeared.

The energy-momentum tensor of $k$-essence is
\be
T_\mn^{k-ess} = P_{,X}\,\partial_\mu\phi\,\partial_\nu\phi - P\,g_\mn\,.
\ee
where as usual the subscript ``${}_{,P}$'' denotes partial derivatives with respect to $X$. The choice of the letter $P$ for the arbitrary lagrangian of $\phi$ is then clarified looking at the expressions for energy density and pressure:
\begin{subequations}
\begin{align}
\rho_\phi &= 2X P_{,X} - P\,, \\
P_\phi &= P\,,
\end{align}
\end{subequations}
so the equation of state is
\be
w_\phi = \frac{P_\phi}{\rho_\phi} = \frac{P}{2XP_{,X} - P}\,.
\ee
As long as $|2XP_{,X}|\ll |P|$, the equation of state will be close to $-1$, hence mimicking a cosmological term.

As in the case of quintessence, $k$-essence models can also display a tracking behaviour, see e.g.~\cite{Chiba:2002mw}.

Non-standard kinetic terms appear in several particle-physics frameworks:
\begin{itemize}
\item Low-energy effective string theory~\cite{Gasperini:2007vw}
\be\label{eq:string_low_action}
P(\phi,X) = L(\phi)X + K(\phi)X^2 + \cdots\,.
\ee

\item Ghost condensate model~\cite{ArkaniHamed:2003uy} and dilatonic ghost condensate model~\cite{Piazza:2004df}
\begin{subequations}
\begin{align}
P_{GC}(\phi,X) &= -X + \frac{X^2}{M^4}\,,\\
P_{DGC}(\phi,X) &= -X + \frac{X^2}{M^4}\,e^{\lambda\phi/\mpl}\,.
\end{align}
\end{subequations}

\item Tachyon fields\cite{Garousi:2000tr,Sen:2002nu}
\be
P(\phi,X) = -V(\phi)\sqrt{-\det(g_\mn + \partial_\mu\phi\,\partial_\nu\phi)}\,,
\ee
where typical potentials are:
\begin{subequations}
\begin{align}
V(\phi) &\sim [\cosh(\lambda\phi/\mpl)]^{-1}\,, && \text{\cite{Kutasov:2003er}} \\
V(\phi) &\sim \phi^{-\alpha}\,,\quad (\alpha\leq 2) && \text{\cite{Padmanabhan:2002cp,Copeland:2004hq}}\\
V(\phi) &\sim \exp(\lambda\phi^2/\mpl^2)\,. && \text{\cite{Garousi:2004uf}}
\end{align}
\end{subequations}

\item Dirac-Born-Infeld (DBI) theories~\cite{Silverstein:2003hf,Alishahiha:2004eh,Martin:2008xw,Guo:2008sz}
\be
P(\phi,X) = -f(\phi)^{-1}\sqrt{1-2f(\phi)X} + f(\phi)^{-1} - V(\phi)\,.
\ee

\end{itemize}

We will not work out the full cosmological dynamics of these theories; details can be found in the cited papers and e.g. in~\cite{Amendola_Tsujikawa:Dark_Energy,Wang:Dark_Energy} and references therein.

\subsection{Chaplygin Gas}

The Chaplygin equation of state
\be
P = -\frac{A}{\rho}\,,
\ee
which can be generalised to 
\be\label{eq:chaplygin_general}
P = -\frac{A}{\rho^\alpha}\,,
\ee
was first introduced in 1904 in connection with aerodynamics~\cite{Chaplygin:1904}. It is of interest for particle physics as it can be motivated by supersymmetry~\cite{Hoppe:1993gz,Jackiw:2000cc} and string theory~\cite{Bordemann:1993ep}, while its application for dark energy was investigated in~\cite{Kamenshchik:2001cp}.

In FLRW background, \rif{eq:chaplygin_general} and~\rif{eq:Energy_Conservation_FLRW} yield
\be
\rho = \left[A + \frac{B}{a^{3(1+\alpha)}}\right]^\frac{1}{1+\alpha}\,,
\ee
with $B$ a constant. The equation of state of the generalised Chaplygin gas is
\be
w = \frac{P}{\rho} = -\left[1 + \frac{B\,a^{-3(1+\alpha)}}{A}\right]^{-1}\,,
\ee
which gives the following interesting behaviour:
\begin{itemize}
\item at early times, when $a$ is small, the Chaplygin gas behaves as cold matter: $w\simeq 0$, $\rho\sim a^{-3}$;
\item at later times, for
\be
a\gtrsim \left(\frac{B}{A}\right)^\frac{1}{3(1+\alpha)}\,,
\ee 
the Chaplygin gas behaves as a cosmological constants: $w\simeq -1$, $\rho\sim$ const.
\end{itemize}
It seems that this theory might realise an interesting unification between dark energy and dark matter, but unfortunately there are severe problems with the matter power spectrum in the absence of cold dark matter, hence ruling out Chaplygin gas as a DM candidate~\cite{Sandvik:2002jz}, see also~\cite{Perrotta:2004ye}.\\

\section{Modified Gravity}\label{ch:Modified_Theories_of_Gravity}

After briefly reviewing some of the leading ideas for dark energy models, let us take a deeper look at modified gravity theories.

\subsection{Brief History of Modified Gravity}

Modified theories of gravitation were proposed soon after the formulation of General Relativity (GR)~\cite{Weyl:1919fi, Eddington:1922}, indeed immediately after the first striking \it{confirmations} of Einstein's theory, namely the discovery of gravitational lensing~\cite{Dyson:1920}. It was scientific curiosity, rather than evidence, that led these seminal works; today, almost a century later, the interest in such theories is constantly growing, and is strongly motivated by astrophysical and cosmological data which appear to be in contrast with the predictions of GR.

Of course, a \it{good} modified gravitational theory should be constructed in such a way that it maintains some of the qualities of standard GR, namely that
\begin{enumerate}
 \item space-time is a differentiable four-manifold with a metric and a connection;
 \item it satisfies the Weak Energy Principle (WEP);
 \item its field equations can be derived from an action principle.
\end{enumerate}
The first statement restricts ourselves to ``metric'' theories of gravity; associating a metric with the space-time curvature directly leads to the equivalence of inertial and gravitational mass\footnote{Probed with very high accuracy in E\"otv\"os type experiments.} as test-particles (with negligible self-gravity) follow space-time geodesics.
Moreover, Newtonian theory must be recovered at small distances, in order to satisfy constraints coming from measurements in the Solar System and in the Earth-Moon system and from the most precise gravitational tests (Hughes-Drever experiments and WEP tests)~\cite{Will:1981cz,Will:1992ds}.\\
However, modified theories of gravity do present new features, and in particular may \it{not} satisfy the following conditions:
\begin{enumerate}
 \item[1a.] the Strong Equivalence Principle (SEP);
 \item[2a.] the space-time constancy of Newton's constant;
 \item[3a.] the linearity of the action in second derivatives of the metric.
\end{enumerate}
The SEP states that ``massive self-gravitating objects should follow geodesics of the space-time''. While this is true for test-particles, this is no longer assured for extended bodies with non-negligible self-gravity. These objects distort the background space-time, and in general may not follow geodesics\footnote{This violation of the SEP is sometimes referred to as the Nordtvedt effect.} in modified theories.\\
Of all possible modifications to GR, the most popular are scalar-tensor and $f(R)$ theories. Let us now consider these modified theories in more detail.

\subsection{Scalar-Tensor Theories}\label{sec:BD-ST}

Within the framework of scalar-tensor gravities, the coupling to the scalar curvature $R$ is no longer a constant but a dynamical field, or a function of such field. However, the only gravitational invariant present in the action, besides $\-g$, is $R$, so the field equations remain of second order in derivatives of the metric, unlike in $f(R)$ theories (see below).\\
The first scalar-tensor theory of gravitation was probably proposed by Jordan~\cite{Jordan:1949zz}, followed by many others~\cite{Fierz:1956zz,Brans:1961sx,Bergmann:1968ve, Nordtvedt:1970uv,Wagoner:1970vr}, who recognized the importance of the scalar field arising in the Kaluza-Klein compactification of a fifth dimension~\cite{Kaluza:1921tu,Klein:1926tv}. From a theoretical point of view, scalar-tensor theories are quite natural alternatives to standard GR. The attractive character of gravitation is preserved\footnote{Unlike in vector gravity which leads to an attractive matter-antimatter interaction but to \it{repulsive} matter-matter and antimatter-antimatter interactions. Please compare this to QED, in which the exchange of a photon (which is a vector) leads to repulsion between particles with the same charge.}, as well as the manifest covariance of the theory, and the WEP is also satisfied. Furthermore, such theories arise naturally in the low-energy limit of (super-)string theories~\cite{Green:1987a,Green:1987b} and from brane-world theories~\cite{Randall:1999ee,Randall:1999vf}.\\
The canonical Brans-Dicke action~\cite{Brans:1961sx} takes the form\footnote{Please note that the metric-signature here is opposed to that of the original paper.}
\be
A^\text{BD} = \frac{\mpl^2}{16\pi}\idx\-g\left(-\varphi R+\frac{\omega_0}{\varphi}\,g^\mn\partial_\mu\varphi\,\partial_\nu\varphi\right)\,,
\ee
where $\varphi$ is a scalar field, $\omega_0$ is a constant, often referred to as the ``Brans-Dicke parameter''. Ordinary matter and gravity couple as usual via the metric $g_\mn$. The choice $\omega(\varphi)=0$ is known in the literature as ``massive dilaton gravity'' or ``O'Hanlon gravity'', and was developed in order to generate a Yukawa term in the Newtonian limit~\cite{O'Hanlon:1972ya}. 

It is clear how the Brans-Dicke theory may lead to a space/time variation of Newton's constant: while the ``bare'' coupling $G$ is indeed a true constant, the ``effective'' coupling
\be
G_\text{eff}=G/\varphi\label{eq:geff}
\ee
may not be constant at all; as a matter of fact, Brans-Dicke gravity (as well as all other modified gravities) has been constructed precisely with the aim of allowing a dynamically variable $G_\text{eff}$. Any solution having $\varphi=\text{ const}$ is indeed trivial, and cannot be distinguished from standard GR.\\
The sole free parameter of the theory is $\omega_0$, which makes the theory remarkably easy to test; using the standard post-Newtonian expansion~\cite{Will:1981cz} and Solar System measurements, one can constrain $\omega_0$, finding~\cite{Bertotti:2003rm}
\be
|\omega_0|\gtrsim 40\,000\,.
\ee
This result is not very appealing, because usually one expects dimensionless parameters to be of order unity, therefore the original Brans-Dicke theory is not considered as a viable theory; nevertheless, the model is easily generalized by taking into account quite general kinetic (cfr. $k$-essence models of section~\ref{sec:k-essence}) and potential terms for the scalar field. The most general action hence reads
\be
A^\text{gBD}=\idx\,\frac{\-g}{16\pi G}\left[-\varphi R+\frac{\omega(\varphi)}{\varphi}(\partial\varphi)^2-U(\varphi)\right]\,.
\label{eq:gBD}
\ee
With obvious notation, $(\partial\varphi)^2 = g^\mn\partial_\mu\varphi\,\partial_\nu\varphi$. The action \rif{eq:gBD} can be rewritten, with a suitable redefinition of the scalar field (see Section \ref{sec:equivalence}), as the prototype scalar-tensor action:
\be
A^\text{ST} = \idx \-g\left[-\frac{\mpl^2\,f(\phi)}{16\pi} R + \frac{1}{2}\,(\partial\phi)^2-V(\phi)\right]\,.\label{eq:STact}
\ee
In this case, the scalar field has the canonical kinetic term and has the usual dimensions of [mass]; the most attractive scalar-tensor theories are those in which $f(\phi)$ tends to unity as $t\to t_0$ (today), so that GR is recovered. Possible constraints on the theory come from the residual energy density of the scalar field, which may account for the accelerated expansion.

Please note that in neither case does the matter action contain explicitly the scalar field, which means that matter couples only to metric, which is the condition for having a ``metric'' theory (see above). Furthermore, unlike in theories without self-interaction for the scalar field, the solution $\varphi=\varphi_0=$ const (or $\phi=\phi_0=$ const) may not be trivial if $U(\varphi_0)\neq 0$ [$V(\phi_0)\neq 0$], in which case the potential term would effectively act as a cosmological term.
\subsubsection*{Field Equations}
Varying the action \rif{eq:gBD} with respect to the metric and to the scalar field yields the field equations (see Appendix \ref{App:field_eqs_mod_grav} for the explicit derivation)
\begin{subequations}
 \begin{gather}
\begin{aligned}
  \varphi\,G_\mn = \,\,&8\pi G\,T_\mn^{(m)}+\frac{\omega}{\varphi}\,\partial_\mu\varphi\,\partial_\nu\varphi-\frac{1}{2}\,g_\mn\,\frac{\omega}{\varphi}(\partial\varphi)^2 + \\
& + \frac{1}{2}\,g_\mn\,U + (\D_\mu\,\D_\nu-g_\mn\dal )\varphi\,,\label{numerouno}
\end{aligned}\\\notag\\
\dal \varphi = \frac{1}{2}\,\left(\frac{1}{\varphi}-\frac{\omega'}{\omega}\right)(\partial\varphi)^2-\frac{\varphi}{2\omega}(R+U')\,.\label{numerodue}
 \end{gather}
\end{subequations}
Above, the prime ($'$) means derivative with respect to $\varphi$. Note that in the equation of motion for $\varphi$ matter is absent: as remarked earlier, $\varphi$ acts on standard matter only through geometry, hence the theory is metric and the WEP is obeyed.\\
It must also be stressed that, in modified gravitational theories, the inverse of the coefficient of $R$ in the lagrangian ($\sim G_\text{eff}$) and the quantity appearing in Newton's law, measurable in a Cavendish experiment, are not necessarily the same. The latter is~\cite{Nordtvedt:1968qr,Nordtvedt:1968qs}
\be
G_\text{eff}^\text{C} = \frac{G}{\varphi}\,\frac{2+2\omega\varphi}{3+2\omega\varphi}\,.
\ee
The field equations for the general scalar-tensor theory \rif{eq:STact} are
\begin{subequations}
\label{eq:st-fieldeq}
\begin{gather}
\begin{aligned}
 f(\phi)\,G_\mn &=8\pi G\left[ T_\mn^{(m)} + \partial_\mu\phi\,\partial_\nu\phi-g_\mn\left(\frac{1}{2}\,(\partial\phi)^2-V(\phi)\right)\right]+\\
&\qquad+\left(\D_\mu\D_\nu-g_\mn\dal \right)f(\phi)\,,\\
\end{aligned}\\
\dal \phi + V'(\phi)+\frac{f'(\phi)R}{16\pi G}=0\,.
\end{gather}
\end{subequations}

\subsection{$f(R)$ Theories}
Classical General Relativity is not renormalisable, and hence cannot be conventionally quantized. It has been shown, though, that one-loop renormalisation demands the presence of higher-order curvature terms~\cite{Utiyama:1962sn}, and that higher-order theories are indeed renormalisable (although not unitary)~\cite{Stelle:1976gc}. Initially, higher-order terms were considered to be relevant only in very strong gravity regimes. The pioneering works~\cite{Gurovich:1979xg,Starobinsky:1979ty,Starobinsky:1980te}, see also~\cite{Birrell:1982ix}, found that one-loop quantum corrections to the vacuum expectation value of $T_\mn$ generate terms with higher-order curvature invariants, such as $R^2$, $R_\mn R^\mn$, etc., with typical couplings of the order of $\mpl$ by the appropriate (negative) power. Therefore, such quantum corrections are effective only when curvature approaches the Planck scale, that is at very early times (if ever).

 This perspective has now changed. Recent results also show that when quantum corrections or string theory are taken into account, higher order curvature invariants may appear naturally in the effective low energy lagrangian~\cite{Birrell:1982ix,Buchbinder:1992rb,Vilkovisky:1992pb}. As widely discussed before, the vacuum energy problem boosted the interest in these theories, and ultimately scientific curiosity may well be enough for us to study them!

In $f(R)$ theories, the scalar curvature term is replaced by a function of $R$ itself:
\be
A^{f(R)} = -\frac{\mpl^2}{16\pi}\idx\,\-g\,f(R)\label{eq:f(R)act}\,,
\ee
and often one separates the ``GR'' and ``modified'' parts of the theory:
\be
f(R) = R + F(R)\,.
\ee
Higher order actions may include a huge variety of curvature invariants, like contractions of the Ricci and Riemann tensors $R_\mn R^\mn$, $R_{\alpha\beta\mn}R^{\alpha\beta\mn}$, $R_{\alpha\beta\mu\nu}R^{\alpha\beta}R^\mn$\dots, combinations of those, such as the Gauss-Bonnet invariant \cite{Nojiri:2005jg}
\be
\mc G = R^2 - 4R_\mn R^\mn + R_{\alpha\beta\mn}R^{\alpha\beta\mn}\,,
\ee
(although it seems difficult to find $f(\mc G)$ models that are consistent with observational data \cite{Li:2007jm,DeFelice:2009rw}), as well as terms containing derivatives such as $R\,\square R$, $R\,\square R^2$, \dots (see e.g.~\cite{Sotiriou:2007yd}). These possibilities will not be considered here.

In most cases, one takes functions of the form
\be
f(R) = \cdots + \frac{\alpha_2}{R^2}+\frac{\alpha_1}{R}-2\Lambda + R + \frac{R^2}{\beta_2}+\frac{R^3}{\beta_3}+\cdots\,,
\ee
where $\alpha_i$ and $\beta_i$ are constants of the correct dimensions. Please note the presence of the usual linear term, of course with coefficient equal to unity, since a different coefficient would simply result in a redefinition of the Planck mass, and of the constant term (= cosmological constant).

\subsubsection*{Field Equations}
The presence of non-linear terms yields field equations of order higher than two in derivatives of the metric, which read\footnote{Formally, in deriving the field equations from an action principle, there appears a Gibbons-York-Hawking-like surface term which cannot, in general, be cancelled out subtracting some surface term \it{before} performing the variation. Usually one simply neglects this term implicitly assuming that a suitable fixing has been chosen, but of course a rigorous discussion is possible (although not unique in the literature). In this regard, see e.g.~\cite{Guarnizo:2010xr}.} (see Appendix~\ref{App:field_eqs_mod_grav} for details):
\be
f_{,R}(R)\,R_\mn-\frac{1}{2}\,f(R)\,g_\mn+\left(g_\mn \dal - \D_\mu\D_\nu\right)f_{,R}(R) = 8\pi G\,T_\mn\,.\label{eq:f(R)feq} 
\ee
The last term in the left-hand side, which contains higher-order derivatives of the metric, disappears when $f_{,R}(R)$ is a constant, namely when $f(R)=R$ and standard GR is recovered. Taking the trace of \rif{eq:f(R)feq} yields
\be\label{eq:f_R_trace}
f_{,R}R)\,R-2f(R)+3\dal  f_{,R}(R) = 8\pi G\,T\,,
\ee
from which we can see that $R$ is related to the trace of the energy-momentum tensor $T$ differentially, and not algebraically as in GR. This an indication of the fact that $f(R)$ theories will admit more solutions than Einstein's theory: for instance, Birkhoff's theorem, which states that the unique vacuum spherically symmetric solution is the Schwarzschild solution, no longer holds in $f(R)$ gravity~\cite{Faraoni:2010rt}.

A brief remark is necessary: these equations are obtained in ``metric'' $f(R)$ theory, in which the gravitational lagrangian is a function of the metric alone; as a matter of fact, there are at least two more approaches to $f(R)$ gravities:
\begin{itemize}
 \item the \it{Palatini} approach\footnote{Despite the name, it was Einstein, not Palatini, to introduce it~\cite{Ferraris:1982}.}, in which the metric $g_\mn$ and the connection $\Gamma^\alpha_\mn$ are regarded as independent fields, and the matter lagrangian is assumed to depend only from the metric, while the Ricci scalar $\mc R=g^\mn\,\mc R_\mn$ is obtained contracting the Ricci tensor of the non-metric connection with the metric. In GR, both metric and Palatini approaches produce the same field equations, but for non-linear Lagrangians these may as well be different\footnote{The requirement that the metric and Palatini variations yield the same field equations selects the Lovelock gravity~\cite{Exirifard:2007da}, of which GR is a special case.}.
\item The \it{metric-affine} approach, in which the matter lagrangian also depends on the independent connection, which may be non-symmetric, leading to a torsion associated with matter. Metric-affine $f(R)$ gravity has not been thoroughly studied yet, particularly its cosmological implications, but this issue goes beyond the scope of this work.
\end{itemize}

\subsubsection{Accelerated Solutions and Cosmological Viability}

A large number of cosmological solutions in $f(R)$ gravity is known, and their stability has been extensively studied. The existence of a stable de Sitter solution corresponds to the existence of algebraic roots of the ``static'' trace equation~\rif{eq:f_R_trace} in vacuum~\cite{Barrow:1983rx}:
\be\label{eq:f(R)_de_Sitter}
f_{,R}(R_{dS})R_{dS} - 2f(R_{dS}) = 0\,.
\ee
If there exists such solution, then the de Sitter regime with
\be
R_{dS} = 4\Lambda
\ee
can be realised. Additional issues about the stability of the de Sitter regime in $f(R)$ gravity was studied in~\cite{Faraoni:2007yn,Quiros:2009xx}. Notice, for instance, that for the quadratic $R+R^2$ theory proposed in~\cite{Starobinsky:1980te} and mentioned earlier the condition~\rif{eq:f(R)_de_Sitter} is satisfied automatically.

The early works on $f(R)$ models of dark energy~\cite{Capozziello:2002rd, Capozziello:2003gx, Capozziello:2003tk, Carroll:2003wy, Nojiri:2003ft} relied on models with negative powers of $R$ (i.e. $f \sim R - \alpha/R^n$) to account for the present accelerated expansion; as $R$ drops down with the universe expansion, such terms become increasingly important and eventually dominate over the GR term. However, these models were soon shown to suffer from severe matter instabilities~\cite{Dolgov:2003px, Faraoni:2006sy}, due to the fact that the scalaron mass squared is negative (the scalaron is a tachyon, see later).

Recently, several works investigated additional conditions for the stability of cosmological FLRW solutions, for particular models~\cite{Carloni:2004kp, Clifton:2005aj, Abdelwahab:2007jp} and also in the general case~\cite{Amendola:2006we, Amendola:2007nt, Sawicki:2007tf, de_Souza:2007fq, Tsujikawa:2007xu, Li:2007xn, Bean:2010xw}.

One needs that:\\\\
\begin{tabular}{c p{7.8cm}}
$\cfrac{\partial^2 f(R)}{\partial R^2}<0$ & so that the scalaron is not a tachyon ($m^2>0$) and the dynamical behaviour is stable in the high-curvature regime;\\
$0<\cfrac{\partial f(R)}{\partial R}\leq1$ & so that the effective Newton's constant $G/f(R)$ is positive, hence gravitons are not ghosts (lower bound), and that GR is recovered in the early universe (upper bound).
\end{tabular}\vspace*{.5em}\\
The amount of work on the subject of $f(R)$ gravity is huge and keeps increasing by the day. For details and additional references, we refer the reader to the nice reviews~\cite{DeFelice:2010aj,Clifton:2011jh,Nojiri:2010wj,Bamba:2012cp} and the book~\cite{Capozziello_Faraoni:Beyond_Einstein_Gravity}.

\subsection{Equivalence of Theories}\label{sec:equivalence}
The equivalence between Brans-Dicke and $f(R)$ theories is well established, both in the metric~\cite{Lukash:1980,Chibisov:1982nx,Kodama:1985bj} and in the Palatini approach~\cite{Ellis:1989ju,Bruni:1991kb}. One says that two theories are (dynamically) equivalent when, under a suitable redefinition of the matter and gravitational fields, the field equations for the two theories can be made to coincide. The very same statement can be made at the level of the action. Therefore, equivalent theories give the same results in describing the evolution of the system to which they are applied. This may be extremely helpful, since results known for one theory can immediately be ``translated'' in analogous results for the other theory. For instance, one can use bounds on post-Newtonian parameters from Brans-Dicke theory to constrain $f(R)$ theories~\cite{Thorne:1970wv,Will:1971zzb,Will:1972zz,Nordtvedt:1972zz,Capone:2009xk}.

\subsubsection*{Brans-Dicke $\leftrightarrow$ Scalar-Tensor Theories}
We have already alluded (sec. \ref{sec:BD-ST}) to the equivalence between generalized Brans-Dicke theory and scalar-tensor theory. The generalized BD lagrangian
\be
\frac{\L^\text{gBD}}{\-g} = \frac{1}{16\pi G}\left(-\varphi R+\frac{\omega(\varphi)}{\varphi}g^\mn\partial_\mu\varphi\,\partial_\nu\varphi - U(\varphi)\right)\label{eq:genBD}
\ee
can be cast in the form
\be
\frac{\L'}{\-g}=-\frac{F(\sigma)}{16\pi G}R + K(\sigma)\,g^\mn\partial_\mu\sigma\,\partial_\nu\sigma-W(\sigma)\label{eq:STnon-stdkin}
\ee
under the substitutions
\begin{subequations}
 \begin{gather}
  \varphi = F(\sigma)\,,\\
 K(\sigma) = \frac{\omega(\varphi)}{16\pi G}\frac{(F'(\sigma))^2}{F(\sigma)}\,,\\
 W(\sigma) = \frac{U(\varphi)}{16\pi G}\,,
 \end{gather}
\end{subequations}
where $F'\equiv \partial F/\partial\sigma$. The more familiar form
\be
\frac{\L^\text{ST}}{\-g} = -\frac{f(\phi)}{16\pi G}R+\frac{1}{2}\,g^\mn\partial_\mu\phi\,\partial_\nu\phi-V(\phi)\label{eq:STgact}
\ee
is recovered with $\phi$ defined by
\be
\left(\frac{\partial\phi}{\partial\sigma}\right)^2=2 K(\sigma)\,.
\ee
This also shows the equivalence between scalar-tensor theories with the canonical kinetic term for the scalar field and scalar-tensor theories which have a non-standard kinetic term.\\
Passing from \rif{eq:STgact} to \rif{eq:genBD} is obtained defining
\begin{subequations}
 \begin{gather}
  \varphi = f(\phi)\,,\\
  \frac{\omega(\varphi)}{16\pi G\varphi} = \frac{1}{2(f'(\phi))^2}\,,\\
 U(\varphi) = 16\pi G\, V(\phi)\,.
 \end{gather}
\end{subequations}
Of course, there may be ``pathological'' situations which require more caution. The O'Hanlon theories, for instance, can be described by a theory of the form \rif{eq:STnon-stdkin} with $K(\sigma)=0$, but is incompatible with any theory of the form \rif{eq:STgact}.
\subsubsection*{Brans-Dicke $\leftrightarrow$ $f(R)$ Theories}
We can also prove the equivalence between Brans-Dicke theories (more precisely, O'Hanlon theories) and $f(R)$ theories. Taking the lagrangian \rif{eq:genBD} with $\omega = 0$ and performing the substitutions
\begin{subequations}
 \begin{gather}
  \varphi = f'(\phi)\\
 U(\varphi) = f(\phi) - \phi f'(\phi)
 \end{gather}
\end{subequations}
yields
\be
\frac{\L}{\-g} = -\frac{1}{16\pi G}\left(f(\phi)+(R-\phi)f'(\phi)\right)\,.
\ee
Setting $\phi = R$, we finally obtain the desired result
\be
\frac{\L}{\-g} = -\frac{f(R)}{16\pi G}\,,
\ee
which corresponds precisely to the action \rif{eq:f(R)act}.

\chapter{Cosmological Evolution in \titlertwo Gravity}\label{ch:cosm_evol_R2_grav}

\citazione{\footnotesize E.V. Arbuzova, A.D. Dolgov, {\color{myhypercolor}\bf{L. Reverberi}}, \it{JCAP} \bf{1202}, 049 (2012).}{}

\section{Introduction}\label{sec:R2_introduction}

As stated in the Introduction, one of the first models of extended gravity proposed was the Starobinsky model~\cite{Starobinsky:1980te}:
\be\label{eq:R^2_action}
A_{grav} = -\frac{\mpl^2}{16\pi}\int d^4x\,\-g\,\left(R - \frac{R^2}{6m^2} \right)\,,
\ee
where $m$ is the only additional parameter and has dimensions of [energy]. In this Chapter, we will examine some cosmological and particle-physics features of this simple theory, which we will use later in the Thesis for more complicated $f(R)$ models.

Cosmological models with an action quadratic in the curvature tensors were pioneered in~\cite{Gurovich:1979xg,Starobinsky:1979ty}. Such higher-order terms appear as a result of radiative corrections to the usual Einstein-Hilbert action after taking the expectation value of the energy-momentum tensor of matter in a curved background. In such models the universe may have experienced an exponential (inflationary) expansion without invoking phase transitions in the very early universe. This model has a graceful exit to matter-dominated stage which is induced by the new scalar degree of freedom, the \textit{scalaron} (curvature scalar), which becomes a dynamical field in $R^2$-theory.

The reheating process, due to gravitational particle production from scalaron (curvature scalar) oscillations, leads to a transition to a radiation-dominated FLRW universe. These features of the model are thoroughly discussed, for instance,
in refs.~\cite{Vilenkin:1985md,Zeldovich:1977,Mijic:1986iv,Suen:1987gu}. Cosmological dynamics of fourth-order gravity were investigated in several works, see e.g.\cite{Carloni:2009nc,Abdelwahab:2011dk} and references therein.

A somewhat similar study was performed in~\cite{Davidson:2004hh} where a version of massive Brans-Dicke (BD) theory without kinetic term (i.e. with BD parameter $\omega =0$) was considered. The Hubble parameter and curvature demonstrate oscillating behaviour which resembles the one found in our paper (and earlier in many others), but quantitative features are very much different.

Beside $R^2$-terms, terms containing the Ricci tensor squared $ R_{\mu\nu} R^{\mu\nu}$ are induced by radiative corrections as well, and with similar magnitude. In principle, there also appear terms quadratic in the Riemann tensor, $
\sim R^{\alpha\beta\mu\nu}R_{\alpha\beta\mu\nu}$, but the Gauss-Bonnet invariant
\be
\mc G = R^2 - 4R^\mn R_\mn + R^{\alpha\beta\mu\nu}R_{\alpha\beta\mu\nu}
\ee
is topological (total divergence) in 4 dimensions, therefore its variation does not contribute to the field equations~\cite{Back:1922aa,Lanczos:1938sf}. This implies that the most general quadratic theory can be expressed as the two-parameter family of theories
\be
f(R) = A\,R^2 + B\,R^\mn R_\mn\,.
\ee
However, the natural magnitude of such radiatively induced terms is quite small. The characteristic mass parameter, in fact, is of the order of the Planck mass in both cases, which makes this situation non-interesting for applications discussed below. On the other hand, $R^2$ cosmology (without  $ R_{\mu\nu} R^{\mu\nu} $) has been considered in the literature with much larger magnitude of $R^2$ than the natural value from radiative corrections. The assumption of large $R^2$ terms is made \textit{ad hoc} to formulate a model which could, for instance, cure singularities. We follow the spirit of those works. However, it could be worthwhile to study the consequences of more complicate models with both $R^2$ and $ R_{\mu\nu} R^{\mu\nu} $ terms. It may be a subject for future investigation.

In section~\ref{sec:eqs_of_motion}, we present the field equations for the model~\rif{eq:R^2_action} in a homogeneous, isotropic Friedmann Universe. After briefly considering inflation in $R^2$ gravity in Sec.~\ref{sec:inflation}, we will now move on to the radiation-dominated (RD) epoch, which represents the larger part of the history of the Universe (in terms of redshift, not time). Relativistic matter was dominant until redshifts of the order of $10^4$, which corresponds to the moment of the matter-radiation equality. From the observations of the abundances of light elements we know for sure that at the time of big bang nucleosynthesis (BBN) the Universe was dominated by relativistic matter with very good precision.

If earlier in the course of the Universe expansion and cooling down there were first order phase transitions in the primeval plasma, relatively short periods of vacuum-like matter dominance may have taken place. We also expect some (small) corrections to the RD regime due to the existence of particles in the plasma with masses comparable to the Universe temperature, and because of the trace anomaly in the energy-momentum tensor of matter which leads to $T_\mu^\mu \neq 0$ even for massless particles. The Universe might have even been in a practically pure MD regime after the post-inflationary RD stage. This regime could have been created by primordial black holes which evaporated early enough to bring the Universe back to the normal RD epoch~\cite{Dolgov:2000ht}. The last possibility is especially interesting in the case of $R^2$-inflation since the initial oscillations of $R$ would be damped due to particle production at the end of inflation, and the GR solution could be restored.

All these deviations from the pure RD regime would in turn induce deviations from the GR solution ($R=0$) and
give rise to oscillations of R even if they were initially absent.

In this chapter, we study the cosmological evolution in the $R^2$-model assuming rather general initial conditions for $R$ and $H$ and dominance of relativistic matter. In particular, we will not restrict ourselves to the case in which inflation was indeed driven by the $R^2$ term, but keep the inflationary scenario open to other possibilities.

\section{Friedmann Universe in \titlertwo Gravity}\label{sec:eqs_of_motion}

The modified Einstein equations for the model~\rif{eq:R^2_action} read
{\small
\be
R_{\mn} - \frac{1}{2}g_{\mn} R -
 \frac{1}{3m^2}\left(R_{\mn}-\frac{1}{4}R g_{\mn}+g_{\mn}\dal - \D_\mu \D_\nu\right)R
 =\frac{8\pi}{\mpl^2}T_\mn\,, \label{field_eqs}
\ee
}
where $\dal\equiv g^\mn \D_\mu\D_\nu$ is the covariant D'Alembert operator. We assume the
Friedmann-Robertson-Walker metric with line-element given by
\be
ds^2 = dt^2 - a^2(t)\left[\frac{dr^2}{1-kr^2}+r^2d\vartheta^2+r^2\sin^2\vartheta\,d\varphi^2\right]\,.
\label{FRW}
\ee
In what follows we will neglect the three-dimensional space curvature\footnote{This is a very good approximation, at least during the RD epoch.}, hence setting $k=0$. The curvature scalar $R$ is expressed through the Hubble parameter $H = \dot a/a$ as
\be
R=-6\dot H-12H^2\,.
\label{R-of-H}
\ee
Therefore, the time-time component of Eq.~(\ref{field_eqs}) reads
\be
\ddot H+3H\dot H - \frac{\dot H^2}{2H}+\frac{m^2 H}{2} = \frac{4\pi m^2}{3\mpl^2 H}\rho\,,
\label{eq:timetime}
\ee
where over-dots denote derivative with respect to physical time $t$.

Taking the trace of Eq.~(\ref{field_eqs}) yields
\be
\ddot R + 3H\dot R+m^2\left(R+\frac{8\pi}{\mpl^2}T^\mu_\mu\right)=0\,.
\label{eq:trace}
\ee
This equation is an oscillator equation for a homogeneous scalar field (the ``scalaron'') of mass $m$, with a source term proportional to the trace of the energy-momentum tensor of matter. General Relativity is recovered when $m\to\infty$. In this limit we expect to obtain the usual algebraic relation between the curvature scalar and the trace of the energy-momentum tensor of matter:
\be
\mpl^2 R_{GR} = - 8\pi T_\mu^\mu\,.
\label{G-limit}
\ee
However, unlike the usual GR, in higher-order theories curvature and matter are related to each other differentially, not simply algebraically. Therefore, the theory may approach GR as $m\rightarrow \infty$ in a non-trivial way or even not approach it at all.

For a perfect fluid with relativistic equation of state $P~=~\rho/3$, the trace of the energy-momentum tensor of matter $T^\mu_\mu$ vanishes and $R$ satisfies the homogeneous equation. The GR solution $R=0$ solves such equation, but if one assumes that either $R$ or $\dot R$ not vanish initially, the general solution for $R$ will be an oscillating function with a decreasing amplitude. The decrease of the amplitude is induced by the cosmological expansion (the second term in Eq.~\ref{eq:trace}) and by particle production by the oscillating gravitational field $R(t)$. The latter is not included in this equation and will be taken into account below in sec.~\ref{sec:particle_production}.

It can be easily shown that the left-hand side of Eq.~(\ref{field_eqs}) is covariantly conserved,
which in turn implies the covariant conservation
of the energy-momentum tensor of matter. The latter allows to write the evolution equation for the matter content, assuming it
to be a perfect fluid with energy density $\rho$ and pressure $P$:
\be
\dot\rho+3H(\rho+P)=0\,.
\label{eq:energy_density_evolution}
\ee
As is well known, only two of equations (\ref{eq:timetime}), (\ref{eq:trace}), and (\ref{eq:energy_density_evolution}) are independent.

From Eq.~(\ref{eq:energy_density_evolution}) it follows that relativistic matter, having equation of state $P=\rho/3$, satisfies
\be\label{eq:rho_evol}
\dot \rho_\text R + 4H\rho_\text R=0\,.
\ee

In what follows  we will use either the set of Eqs. (\ref{eq:timetime}) and (\ref{eq:rho_evol}) or
the set (\ref{R-of-H}) and (\ref{eq:trace}) as the basic equations. They are of course equivalent
but their numerical treatment may be somewhat different.

\subsection{Inflation}\label{sec:inflation}

First, let us all briefly review the inflationary scenario in $R^2$ gravity. Taking \rif{eq:timetime} and \rif{R-of-H}, we are left with the single evolution equation for the Hubble parameter $H$:
\be\label{eq:H_evol_R2}
\ddot H + 3H\dot H - \frac{\dot H^2}{2H} + \frac{m^2}{2}\,H = 0\,.
\ee
Of course, the relative importance of the various terms in the previous equation determine very different regimes; this will become particularly evident when we study the RD epoch later in the chapter. 

In order for inflation to take place, we need a sort of \it{slow-roll} conditions for $H$, namely:
\begin{subequations}\label{eq:infl_slow_roll}
\begin{align}
&\ddot H \ll H \dot H\,, \label{eq:slow_roll_ddot_H}\\
&\dot H \ll H^2\,. \label{eq:slow_roll_dot_H}
\end{align}
\end{subequations}
Under these assumptions, Eq.~\rif{eq:H_evol_R2} reduces to
\be
3H \dot H + \frac{m^2}{2}H\approx 0\,,
\ee
which is solved, apart from the trivial solution $H=0$, by
\be
H(t) = H_0 - \frac{m^2}{6}t \so a_1(t) = a_0\,\exp\left(H_0t - \frac{m^2t^2}{12}\right)\,.
\ee
This solution describes a quasi-de Sitter stage in which $H$ decreases linearly with time, with a relatively small scope. However, this phase is unstable. Indeed, combining this solution with the slow-roll condition~\rif{eq:slow_roll_dot_H}, we find
\be
H \gg m\,.
\ee
This ``large-field'' condition gives us the limit time $t_f$ until which the inflationary regime holds:
\be
H(t_f) \simeq m \so t_f \simeq \frac{6(H_0-m)}{m^2}\,.
\ee
For $t>t_f$, the inequality~\rif{eq:slow_roll_dot_H} is reversed and~\rif{eq:H_evol_R2} becomes
\be
\ddot H - \frac{\dot H^2}{2H} + \frac{m^2}{2}H\approx 0\,,
\ee
which has the solution:
\be
H(t) \sim \cos^2\left(\frac{mt}{2}\right)\,.
\ee
We can assume that including the effect of the $3H\dot H$ term in~\rif{eq:H_evol_R2} will result in a modulation of this solution, that is
\be
\cos^2\left(\frac{mt}{2}\right) \to h(t)\cos^2\left(\frac{mt}{2}\right)\,,
\ee
where $h(t)$ is slowly varying. In particular, neglecting $\ddot h$, $\dot{h}^2/h$ and $\dot{h}h$, we obtain the approximate solution
\be
h(t) \simeq \frac{2}{3(t-t_{*}) + 3\sin(mt)/m}\,,
\ee
where $t_{*}$ is some arbitrary integration constant. Hence, the Hubble parameter behaves practically as
\be
h(t)\sim \frac{2}{3t}\,,
\ee
which corresponds to a matter-dominated (MD) Universe. Indeed, the Universe is now completely dominated by the non-relativistic \it{scalaron}, having mass $m$. 

The total expansion during the quasi-de Sitter phase is:
\be
\frac{a(t_f)}{a_0} \simeq \exp\left(\frac{3H_0^2}{m^2}\right)\,.
\ee
In order to have a cosmologically relevant expansion, say 60 $e$-foldings, we only need
\be
\frac{H_0}{m}\simeq 4\,,
\ee
which is a quite reasonable value in this framework\footnote{Remember that $H_0>m$ is an essential condition for the very existence of the inflationary period.}.

\subsubsection{Reheating}

After the quasi-de Sitter stage is over, as we have seen, the Hubble parameter $H$ and consequently the scalar curvature $R$ display an oscillatory behaviour and Universe is in a scalaron-dominated, MD regime. Such oscillations will be damped as a result of gravitational production or relativistic SM-particles, and the Universe will eventually be filled with ordinary relativistic matter and expand as in GR. We will not resent a more accurate description of these mechanism, as gravitational particle production in this model will be extensively discussed in Sec.~\ref{sec:particle_production}.

For more details about inflation and reheating in $R^2$ gravity, see~\cite{Vilenkin:1985md,Mijic:1986iv} and the more general, but very comprehensive~\cite{Martin:2013tda}. A very interesting topic is also the equivalence of $R^2$-driven inflation and Higgs-inflation; for a detailed discussion, see e.g.~\cite{Bezrukov:2011gp}.

\subsection{Radiation-Dominated Epoch}
We now turn the attention to the RD epoch. After inflation, the reheating process filled the Universe with relativistic SM-particles. Some of the effects mentioned at the end of the introduction to this chapter (Sec.\ref{sec:R2_introduction}), however, may generate deviations from the perfectly GR expansion and induce non-trivial effects, which we will now investigate.

It is convenient to rewrite the equations in terms of the dimensionless quantities
\be
\begin{aligned}
& \tau\equiv H_0 t \,,\qquad && h \equiv \frac{H}{H_0} \,, \\
& r \equiv {R}{H_0^2} \,, && \omega \equiv \frac{m}{H_0}\,,
\end{aligned}
\qquad y \equiv \frac{8\pi\rho_m}{3\mpl^2H_0^2}\,,
\ee
where $H_0$ is the value of the Hubble parameter at some initial time $t_0$. Thus, equations~\rif{eq:timetime}, \rif{eq:trace} and~\rif{eq:rho_evol} can be recast as the two equivalent systems:
\begin{equation}\label{sys:hubble_evolution}
\begin{cases}
h'' + 3h h' - \cfrac{h'^2}{2h}+\cfrac{\omega^2}{2}\cfrac{h^2-y}{h}=0\,,\\
y' + 4hy = 0\,,
\end{cases}
\end{equation}
and
\begin{equation}\label{sys:curvature}
\begin{cases}
r''+3hr'+\omega^2r=0\,,\\
r+6h'+12h^2=0\,.
\end{cases}
\end{equation}
Here  prime indicates derivative with respect to dimensionless time $\tau$. If we impose the ``natural'' relativistic initial conditions:
\begin{equation}\label{eq:initial_conditions_1}
 \begin{aligned}
  &\tau_0 = 1/2\,,\\
&h_0=1\,,\\
&h'_0=-2\,,\\
&y_0=1\,,
 \end{aligned}
\end{equation}
we find that there exists the exact solution
\be\label{eq:exact_solution_RD}
h = \frac{1}{2\tau}\,,\qquad y = \frac{1}{4\tau^2}\,,\qquad r=0\,,
\ee
corresponding to the usual RD solution in GR:
\be
H = \frac{1}{2t}\,,\qquad \rho = \frac{3\mpl^2H^2}{8\pi}\,,\qquad R=0\,.
\ee
This solution, however, may deviate from GR because of deviations of the real expansion regime from the purely relativistic one. In principle, depending on the initial conditions, the solutions may oscillate around some value of $h\tau$, which may itself be different from $1/2$, and in fact this is what we will find later on.

Any non-negligible deviation from the GR solution may lead to observable effects, and hence to observational constraints on $m$, the only free parameter of the model. We will tackle these problems below both analytically and numerically.

\subsubsection{Approximate Analytical Solutions}\label{sec:analytical_estim_no_back_react}

First we assume that the deviations from GR are small and expand
\begin{subequations}
\begin{align}
& h = \frac{1}{2\tau} + h_1\,,\\
& y = \frac{1}{4\tau^2} + y_1\,,
\end{align}
\end{subequations}
assuming that $h_1/h \ll 1$ and $y_1/y \ll 1$, and linearize the system of equations.
It is convenient to introduce a new unknown function $z_1\equiv h_1'$, so we obtain three
first-order linear differential equations with time-dependent coefficients:
\begin{equation}\label{sys:linear_perturb}
\begin{cases}
  z_1' = -\cfrac{5}{2\tau}\,z_1+\left(\cfrac{1}{\tau^2}-\omega^2\right)h_1+\tau\omega^2 y_1\\
 h_1' = z_1\\
 y_1' = -\cfrac{1}{\tau^2}\,h_1-\cfrac{2}{\tau}\,y_1
 \end{cases}
\end{equation}
We can find an approximate analytical solution of this system in the limit of large times, or
$\omega\tau\gg 1$. In this limit we can treat the coefficients as approximately constant and find the eigenvalues
and eigenfunctions of the system of differential equations. This method essentially consists in separating "fast" and "slow" variables.

The characteristic polynomial of~\rif{sys:linear_perturb} is
\begin{equation}\label{polyn_1}
 P(\lambda)=\lambda^3+\frac{9\lambda^2}{2\tau}+ \lambda\left(\frac{4}{\tau^2}+\omega^2\right)
+ \frac{3\omega^2}{\tau} - \frac{2}{\tau^3}
\end{equation}
and the eigenvalues (for large $\omega \tau$) are approximately
\begin{equation}\label{lambda123}
 \lambda_1 \approx -\frac{3}{\tau}\,, \qquad\qquad \lambda_{2,3} \approx -\frac{3}{4\tau}\pm i\omega\,.
\end{equation}
The general solutions of the system (\ref{sys:linear_perturb}) is a linear combination of eigenvectors $V_j$:
\be
\left[h_1, z_1, y_1\right]= \sum C_j V_j \exp \left[ \int^\tau d\tau'\,\lambda_j (\tau') \right]\,,
\label{h1-z1-y1}
\ee
where
\be
\begin{aligned}
&V_1 = \left[1,\,-\frac{3}{\tau},\,\frac{1}{\tau}\right]\,,\\
&V_{2,3} = \left[1,\,-\frac{3}{4\tau}\pm i\omega,\,-\frac{1}{5\tau/4 \pm i\omega \tau^2}\right]
\end{aligned}
\ee
and
\be
\begin{aligned}
&\exp \left[ \int^\tau d\tau'\,\lambda_1 (\tau') \right] \sim \frac{1}{\tau^3}\\
&\exp \left[ \int^\tau d\tau'\,\lambda_{2,3} (\tau') \right] \sim \frac{e^{\pm i\omega \tau}}{\tau^{3/4}}\,.
\end{aligned}
\label{exp-lambda}
\ee
Since the solution must be real, one should consider only the real part.  Let us note that the eigenvectors $V_j$ are normalised to almost constant values, up to terms of order $1/\tau^2$. In principle the coefficients $C_j$ depend upon time but this dependence is quite weak, $C_j \sim C_{j0} + C_{j1}/\tau^2$, and asymptotically negligible. 

The correction to the GR solution corresponding to the first eigenvalue $\lambda_1$ quickly disappears, since
$h_1^{(1)} \sim 1/\tau^3$ and $y_1^{(1)} \sim 1/\tau^4$, and so it can be asymptotically neglected. The solutions for $h_1$ corresponding to $\lambda_{2,3}$, instead, oscillate and decrease more slowly than the GR solution, in fact
\begin{equation}
 h_1^{(2,3)} \sim \frac{\sin(\omega\tau+\varphi)}{\tau^{3/4}}\,, \label{eq:analytical_solution}
\end{equation}
while the solution for the energy density also oscillates but drops down faster than the GR one, $y \sim \tau^{-11/4}$. The complete asymptotic solution for $h$ has the form:
\begin{equation}\label{sol_h_linear}
h(\tau)\simeq \frac{1}{2\tau}+\frac{c_1\sin(\omega\tau+\varphi)}{\tau^{3/4}}\,.
\end{equation}
For sufficiently large $\tau$ the second term would start dominating and the linear approximation ceases to hold. Below we will obtain approximate analytical solutions even in the non-linear regime, in the high-frequency limit.

Before that, it would be anyhow instructive to find the solution for the equivalent set of equations (\ref{R-of-H}) and (\ref{eq:trace}) in the same approximation of small deviation from GR. We rewrite these equations in the form:
\begin{equation}\label{trace_dimensionless}
 \begin{cases}
 r=-6h'-12h^2\,,\\
 r''+3hr'+\omega^2r=0\,.
 \end{cases}
\end{equation}
Introducing the new function $q_1\equiv r'$, we obtain the system of three first-order linear
differential equations:
\begin{equation}\label{sys_rdot_r_h}
\begin{cases}
 q_1'=-\cfrac{3}{2\tau}\,q_1-\omega^2 r\\
r' = q_1\\
h_1'=-\cfrac{1}{6}\,r-\cfrac{2}{t}\,h_1
\end{cases}
\end{equation}
with characteristic polynomial
\begin{equation}\label{polyn_2}\nonumber
 P(\tilde \lambda)=\left(\tilde \lambda+\frac{2}{\tau}\right)
 \left(\tilde\lambda^2+\frac{3\tilde \lambda}{2\tau }+\omega^2\right)\,.
\end{equation}
Hence, the approximate eigenvalues for $\omega\tau\gg 1$ are
\begin{equation}\label{eigenval_2}
\tilde \lambda_1 = -\frac{2}{\tau}\,, \qquad\qquad \tilde\lambda_{2,3} \approx -\frac{3}{4\tau}\pm i\omega\,.
\end{equation}
As above, the solutions of the system (\ref{sys_rdot_r_h}) are linear combinations of eigenvectors $\tilde V_j$:
\be
[h_1, r, q_1] = \sum \tilde C_j \tilde V_j \exp \left[ \int^\tau d\tau'\,\tilde\lambda_j (\tau') \right]\,,
\label{h1-z1-y1_2}
\ee
where
\be
\begin{aligned}
&\tilde V_1 = \left[1,\, 0,\, 0\right]\,,\\
&\tilde V_{2,3} = \left[-\frac{\tau}{6(5/4\pm i\omega\tau)},\, 1,\, -\frac{3}{4\tau} \pm  i\omega \right]\,,
\end{aligned}
\ee
and
\be
\begin{aligned}
&\exp \left[ \int^\tau d\tau'\,\tilde\lambda_1 (\tau') \right] \sim \frac{1}{\tau^2}\,,\\
&\exp \left[ \int^\tau d\tau'\,\tilde\lambda_{2,3} (\tau') \right] \sim \frac{e^{\pm i\omega \tau}}{\tau^{3/4}}\,.
\end{aligned}
\label{exp-lambda_2}
\ee
So the oscillating solution for $h_1$ is the same as that of Eq.~(\ref{eq:analytical_solution}). However, the slowly
varying (non-oscillating) solution for $h_1$ decreases more slowly, namely as $1/\tau^2$ instead of $1/\tau^3$. Probably this difference is related to the freedom of the zeroth order GR solution with respect to the time shift:
\be
h_0 = \frac{1}{2(\tau +\delta)} =  \frac{1}{2\tau} - \frac{\delta}{2\tau^2 }\,.
\label{shift-h0}
\ee
Such freedom tells us that terms of order $1/\tau^2$ in the first-order corrections are in some sense arbitrary, so the solution $h_1 \sim 1/\tau^2$ is spurious and should be disregarded. Anyhow the non-oscillating solutions for $h_1$ quickly disappear asymptotically and can be neglected.

The solutions found above describe the oscillations of the Hubble parameter around the GR value $1/(2\tau)$. Moreover, the amplitude of such oscillations decreases more slowly than $1/\tau$, so at some stage the oscillations
will become large and the condition $h_1\ll h$ will no longer be satisfied. After this stage is reached, the
linear approximation is no longer valid and the method used above becomes inapplicable.

However, we can still find the asymptotic behaviour of the exact nonlinear equations (\ref{R-of-H}) and (\ref{eq:trace}) at large times looking for solutions in the form:
\begin{subequations}\label{expansion_H_R}
\begin{align}
&h(\tau) = A(\tau) + B_s(\tau) \sin \nu\tau + B_c (\tau) \cos \nu\tau\,, \label{H-of-t}\\
&r(\tau) = C(\tau) + D_s (\tau) \sin \nu \tau + D_c (\tau) \cos \nu \tau\,. \label{R-of-tCER2}
\end{align}
\end{subequations}
The coefficients $A,B,C$, and $D$ are slowly varying functions of time and $\nu$, assumed large\footnote{When compared to the variations of the slowly-varying coefficients, i.e. $\nu\gg \dot A/A\sim \dot B/B,~\dots$.}, may in principle be different from $\omega$. As we will see below, this is indeed the case due to radiative corrections (scalaron mass renormalisation). In the approximation taken here we find $\nu=\omega $. 

We obtain approximate equations for these functions equating the coefficients in front of the slowly varying terms, and in front of $\sin \omega \tau$ and $\cos \omega \tau$. These equations are approximate because we do not take into account higher-frequency terms which appear as a result of non-linearity, but the approximation happens to be quite accurate. Doing so, we find from equations (\ref{trace_dimensionless}):
\begin{subequations}\label{eqn-ABCD}
\begin{align}
&A' + 2 A^2 +B_s^2 + B_c^2 = -C/6\,,\label{eqn-ABCD_A}\\
&B_s' - B_c \nu + 4 A B_s = -D_s/6\,,\label{eqn-ABCD_Bs}\\
& B_c' + B_s \nu + 4 A B_c = -D_c/6\,,\label{eqn-ABCD_Bc}\\
&C'' + 3\left(A C' + \frac{1}{2} B_s  D_s'  + \frac{1}{2} B_c D_c'
 - \right.\nonumber\\
&\qquad\qquad\left. - \frac{1}{2}\nu  B_s  D_c + \frac{1}{2}\nu  B_c  D_s\right) + \omega^2 C = 0\,,\label{eqn-ABCD_C}\\
& D_s'' - 2 \nu D_c' -\nu^2 D_s  + 3A (D_s' - \nu D_c )
 +3 C' B_s + \omega^2 D_s = 0\,, \label{eqn-ABCD_Ds}\\
&D_c'' + 2 \nu D_s' -\nu^2 D_c  + 3A (D_c' + \nu D_s )
 +3 C' B_c + \omega^2 D_c = 0\,.\label{eqn-ABCD_Dc}
\end{align}
\end{subequations}
Assuming
\begin{equation}\label{expand_ABCD}
A = \frac{a}{\tau},\,\, B_s = \frac{b_s}{\tau},\,\,\  B_c = \frac{b_c}{\tau},\,\,\,
C=  \frac{c}{\tau^2},\,\, D_s = \frac{d_s}{\tau},\,\,\,  D_c = \frac{d_c}{\tau}\,,
\end{equation}
and keeping only the dominant terms (lowest powers of $1/\tau $) we obtain the solutions
\begin{subequations}\label{sol_ABCD}
\begin{align}
  &\nu^2=\omega^2\,,\label{sol_ABCD_omega}\\
 &d_s=6\omega b_c\,,\label{sol_ABCD_Ds}\\
&d_c=-6\omega b_s\,,\label{sol_ABCD_Dc}\\
&b_s^2+b_c^2=2a(2a-1)\,,\label{sol_ABCD_B}\\
&c=18a(1-2a)\label{sol_ABCD_C}\,.
 \end{align}
\end{subequations}
It is interesting that equation (\ref{sol_ABCD_B}) demands $a>1/2$, that is an expansion faster then normal, in presence of oscillations. We will see from the numerical solution that this is indeed the case.

To summarize, the situation is the following: we initially assume $h_1/h\ll 1$ and solve the linearised systems (\ref{sys:linear_perturb}) or (\ref{sys_rdot_r_h}), finding that the Hubble parameter oscillates around the value $1/(2\tau)$ with amplitude growing with time as $h_1/h\sim \tau^{1/4}$; eventually, such oscillations (and hence non-linear terms) become dominant and the linear approximation becomes invalid. However, we can proceed further using a sort of truncated Fourier expansion which allows us to take into account the non-linearity of the system in the limit $\omega\tau\gg 1$. As a result, we find that $h_1/h\to$ const. In other words, the amplitude of the oscillating part of $h$ asymptotically behaves as $1/\tau$, i.e. in the same way as the slowly-varying part of $h$.

To be completely sure about these analytical results we have to check if the exact numerical solution of the system (\ref{sys:hubble_evolution}) shows the same behavior. Still, the analytical estimates presented above are of great interest for the calculation of the evolution of $R$ and $H$ when particle production effects are taken into account.

\subsubsection{Numerical Solutions \label{ss-num-estim}}

We integrate the system of equations (\ref{sys:hubble_evolution}) starting at $\tau_0=1/2$, with the initial conditions
\be\label{eq:initial_conditions_2}
 \begin{aligned}
  &h_0=1+\delta h_0\\
  &h'_0=-2+\delta h'_0\\
 &y_0=1+\delta y_0\,,
 \end{aligned}
\ee
where $\delta h_0$, $\delta h'_0$ and $\delta y_0$ do not vanish simultaneously. As expected, numerical integration with the initial conditions given by Eq.~(\ref{eq:initial_conditions_1}) gives the usual GR solution $h=1/(2\tau)$ within numerical precision, so we are interested in the more general case in which the initial conditions deviate from the GR values.

\begin{figure}[b!]
{
\includegraphics[width=\mywidthdouble]{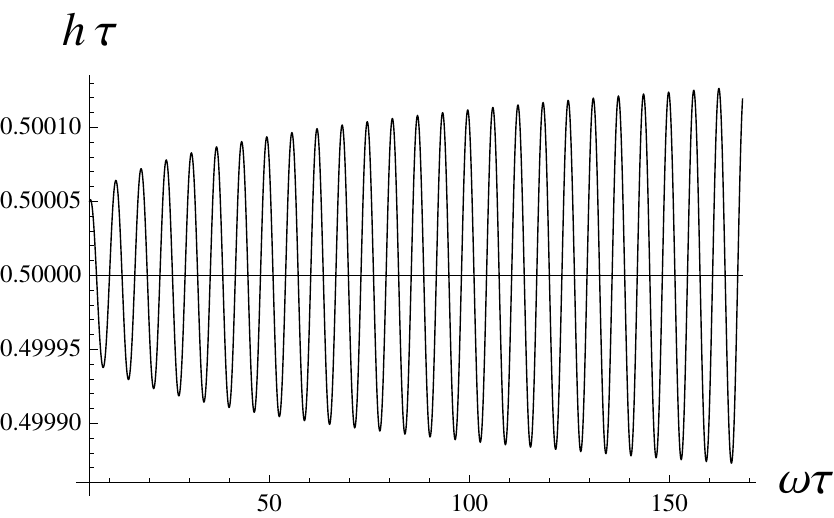}
\includegraphics[width=\mywidthdouble]{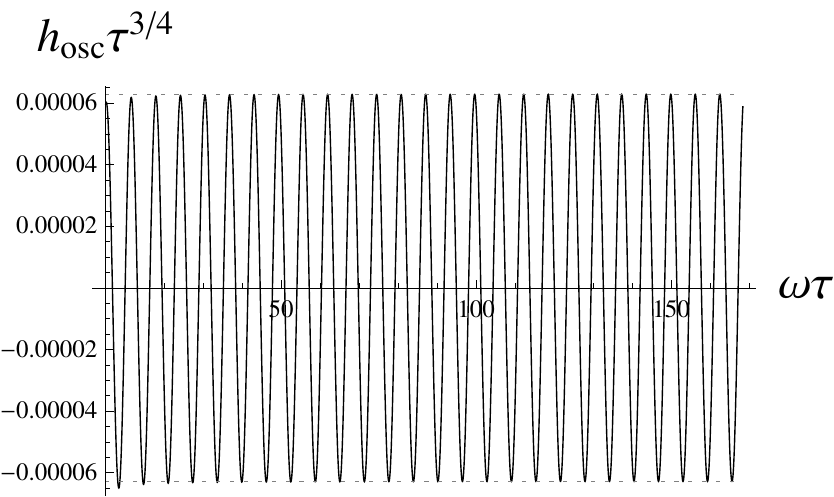}
}
\caption[Numerical solution of Eqs.~(\ref{sys:hubble_evolution}) with $\delta h_0=10^{-4}$, $\delta h'_0=0$, $y_0=1$,
and $\omega=10$.]{Numerical solution of Eqs.~(\ref{sys:hubble_evolution}) with $\delta h_0=10^{-4}$, $\delta h'_0=0$, $y_0=1$, and $\omega=10$. The best fit, for the functional form (\ref{h_fit_func}), is given by $\alpha\simeq 1$, $\beta\simeq 6.29\times 10^{-5}$.}
\label{fig:h0=1.0001}
\end{figure}
\begin{figure}[t]
{
\includegraphics[width=\mywidthdouble]{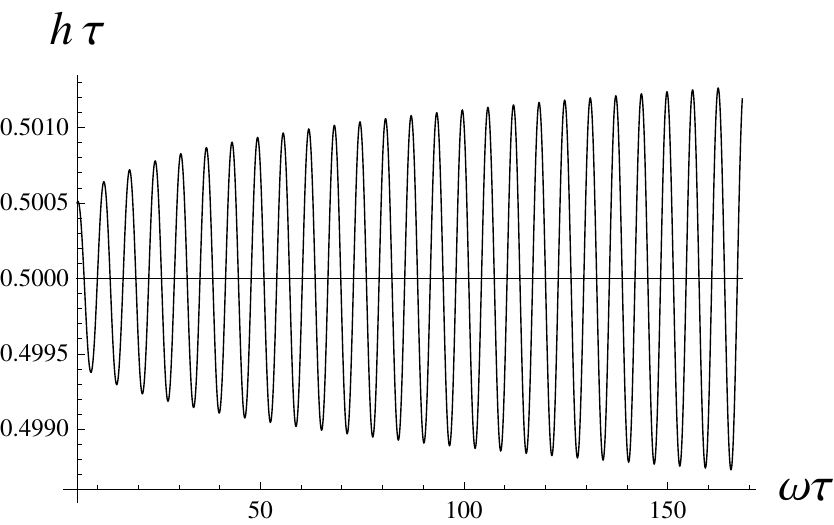}
\includegraphics[width=\mywidthdouble]{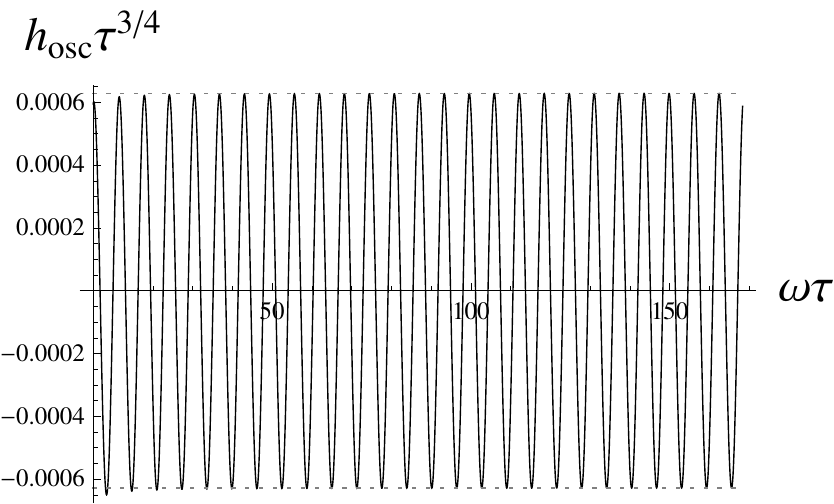}
}
\caption[Numerical solution of system (\ref{sys:hubble_evolution}) with $\delta h_0=10^{-3}$, $\delta h'_0=0$, $y_0=1$, $\omega=10$.]{Numerical solution of system (\ref{sys:hubble_evolution}) with $\delta h_0=10^{-3}$, $\delta h'_0=0$, $y_0=1$, $\omega=10$. The best fit is $\alpha\simeq 1$, $\beta\simeq 6.28\times10^{-4}$.}
\label{fig:h0=1.001}
\end{figure}
\begin{figure}[h!]
{
\includegraphics[width=\mywidthdouble]{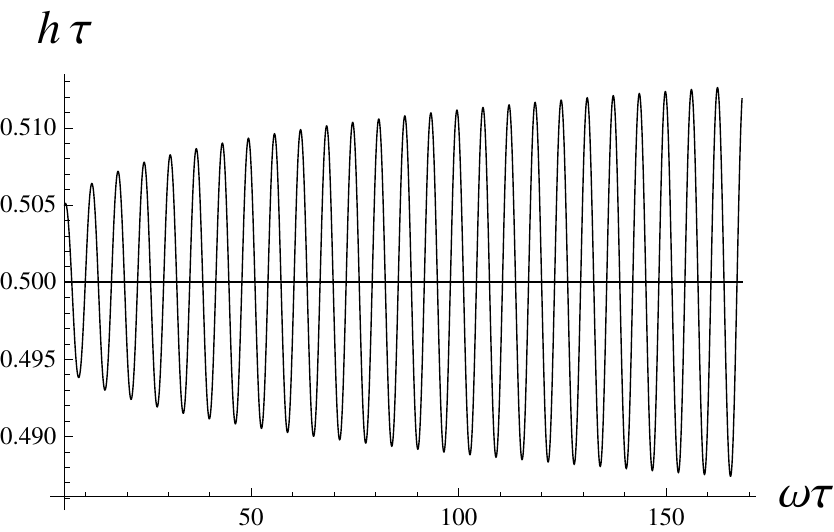}
\includegraphics[width=\mywidthdouble]{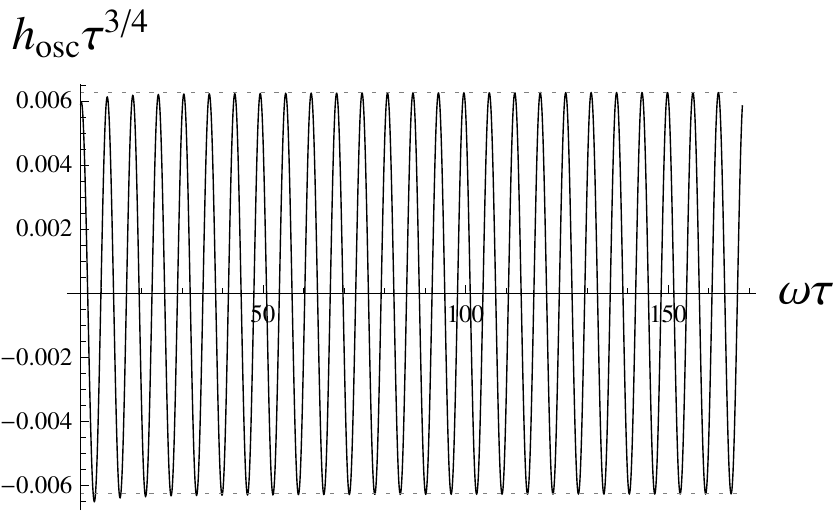}
}
\caption[Numerical solution of  system (\ref{sys:hubble_evolution}) with $\delta h_0=10^{-2}$, $\delta h'_0=0$, $y_0=1$, $\omega=10$.]{Numerical solution of  system (\ref{sys:hubble_evolution}) with $\delta h_0=10^{-2}$, $\delta h'_0=0$, $y_0=1$, $\omega=10$. The best fit is given by $\alpha\simeq 1$, $\beta\simeq 6.26\times 10^{-3}$.}
\label{fig:h0=1.01}
\end{figure}

As it was said earlier, the systems (\ref{sys:hubble_evolution}) and (\ref{sys:curvature}) are equivalent. However, for the numerical integration of these systems one has to specify initial values of different quantities. For the integration of system (\ref{sys:hubble_evolution}) one has to fix  $h_0$, $h'_0$, and $y_0$, while for the integration of (\ref{sys:curvature}) the values of $h_0$, $r_0$, and $r'_0$ must be specified. The expression of one set of initial values through the equivalent values of another set can be found using the equations under scrutiny. Indeed, once $h_0$, $h'_0$ and $y_0$ are chosen, $h''_0$ is uniquely determined through the first equation in (\ref{sys:hubble_evolution}), and consequently $r_0$ and $r'_0$ are specified as well, via (\ref{sys:curvature}). After all, both systems are equivalent to the same single third-order differential equation, whose Cauchy problem is determined by three initial conditions.

\begin{figure}[!t]
{
\includegraphics[width=\mywidthdouble]{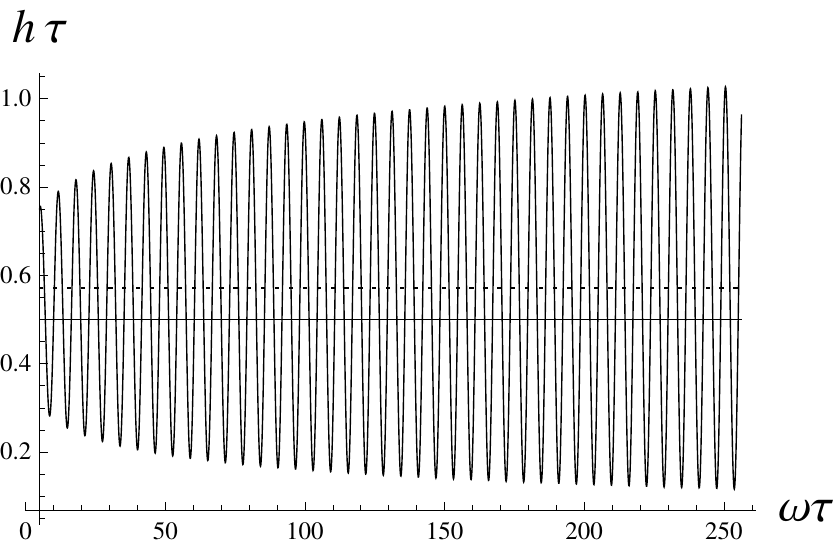}
\includegraphics[width=\mywidthdouble]{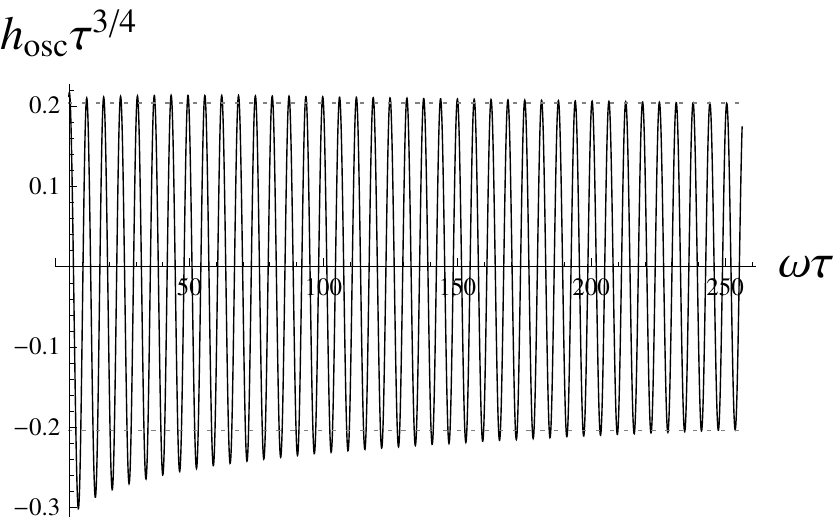}
}
\caption[Numerical solution of system (\ref{sys:hubble_evolution}) with $\delta h_0=0.5$, $\delta h'_0=0$, $y_0=1$, $\omega=10$.]{Numerical solution of system (\ref{sys:hubble_evolution}) with $\delta h_0=0.5$, $\delta h'_0=0$, $y_0=1$, $\omega=10$. The best fit at large times is $\alpha\simeq 1.14$ (dotted line in the left panel), $\beta\simeq 0.20$. Please also note, in the right panel, that the oscillating part of $h$ does not decrease \textit{exactly} as $\tau^{-3/4}$. The apparent up-down asymmetry in $h_{osc}$ is due to the fact that $\alpha$ is a function of time, and that we centered oscillations around its value at late times.}
\label{fig:h0=1.5}
\end{figure}

\begin{figure}[b]
\centering
\includegraphics[width=\mywidthsingle]{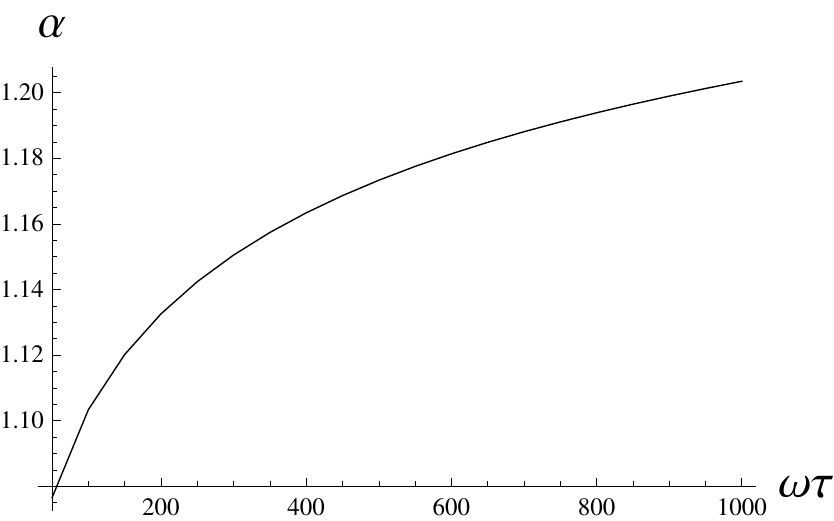}
\caption{Evolution of $\alpha$ with time. Initial conditions are those of figure \ref{fig:h0=1.5}.}
\label{fig:alpha_1.5}
\end{figure}

We have found that the numerical solutions of the system of equations~(\ref{sys:hubble_evolution}, \ref{eq:initial_conditions_2}) are in very good agreement with the previous analytical estimates in the linear regime, i.e. for initial conditions fulfilling the requirements  $\delta h_0/h_0\ll 1$, $\delta y_0/y_0\ll 1$, and $\omega\tau\gg 1$. In figures~\ref{fig:h0=1.0001}, \ref{fig:h0=1.001}, and \ref{fig:h0=1.01} we present the numerical results for the dimensionless Hubble parameter $h$ determined from the system of equations~(\ref{sys:hubble_evolution}), with $\omega=10$ and initial conditions
\begin{subequations}
\begin{align}
 \delta h_0 = 10^{-4}&\,, \quad \delta h'_0=0\,,\quad \delta y_0=0\qquad \text{(fig. \ref{fig:h0=1.0001})}\\
\delta h_0 = 10^{-3}&\,,\quad \delta h'_0=0\,,\quad \delta y_0=0\qquad \text{(fig. \ref{fig:h0=1.001})}\\
\delta h_0 = 10^{-2}&\,,\quad \delta h'_0=0\,,\quad \delta y_0=0\qquad \text{(fig. \ref{fig:h0=1.01})}
\end{align}
\end{subequations}
The function $h\tau$ is found to oscillate around the central value
$h=1/2$ with amplitude $h_1\sim\tau^{-3/4}$. As the deviation from the ideal GR behaviour increases, 
the average value of $h\tau$ also varies, and in general it
is no longer equal to 1/2. A very good functional form for fitting the solutions is
\begin{equation}\label{h_fit_func}
h(\tau)\simeq \frac{\alpha + \beta\,\tau^{1/4}\sin(\omega\tau+\varphi)}{2\tau}\equiv \frac{\alpha}{2\tau}+h_{osc}\,,
\end{equation}
where the dimensionless parameters $\alpha$ and $\beta$ are very slowly varying functions of time. This fit is shown, for instance,
in figures \ref{fig:h0=1.5} and \ref{fig:alpha_1.5}, where the numerical solution for $\delta h_0=0.5$ is presented.
A deviation from the analytical estimates of the linearised equations is to be expected in this situation, since the condition $\delta h_0/h_0 \ll 1$ is not fulfilled and non-linear terms in Eqs.~(\ref{sys:hubble_evolution}) are important.\\
For the moment, let us concentrate on the case of small $\delta h_0$, in which we can safely take $\alpha=1$. Qualitatively, one notices that parameter $\beta$, evaluated at the same value of $\omega\tau$, increases roughly linearly with the initial displacement $\delta h_0$. Moreover, when solving the system of equations (\ref{sys:hubble_evolution}) with $\delta h_0=0$ and varying $\delta h'_0$, we find again a (roughly) linear relation of the form $\beta\propto\delta h'_0$.

\begin{figure}[t]
{
\includegraphics[width=\mywidthdouble]{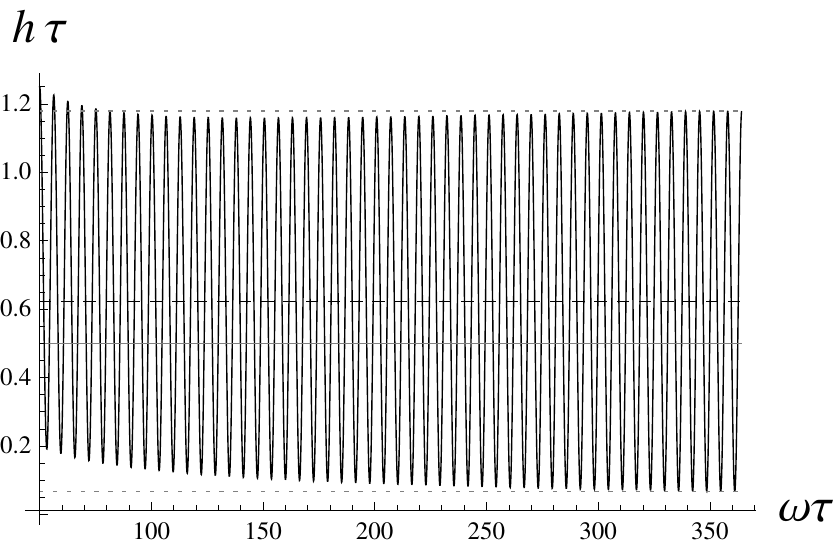}
\includegraphics[width=\mywidthdouble]{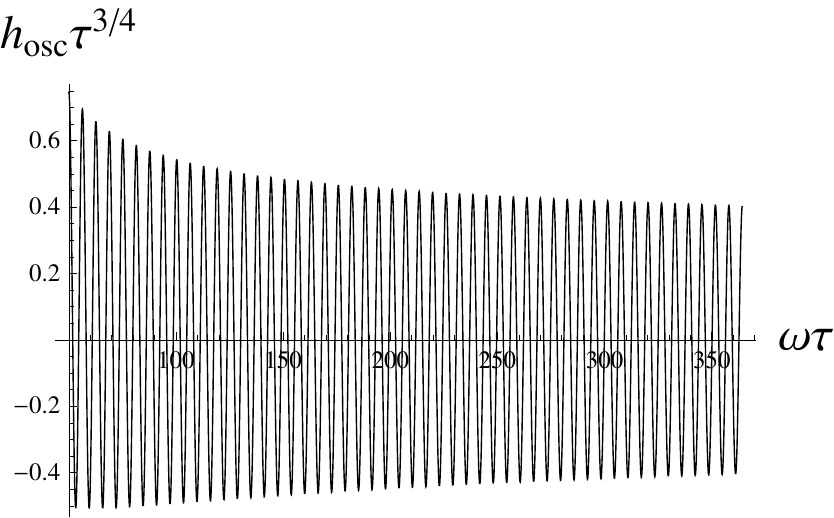}
}
\caption[Numerical solution for the dimensionless Hubble parameter $h$ with initial conditions (\ref{initial_cond_2.5}).]{Numerical solution for the dimensionless Hubble parameter $h$ with initial conditions (\ref{initial_cond_2.5}). Oscillations in the left panel are of almost constant amplitude, whereas in the right panel they are clearly decreasing.}
\label{fig:h0_2.5}
\end{figure}

\begin{figure}[h!]
{
\includegraphics[width=\mywidthdouble]{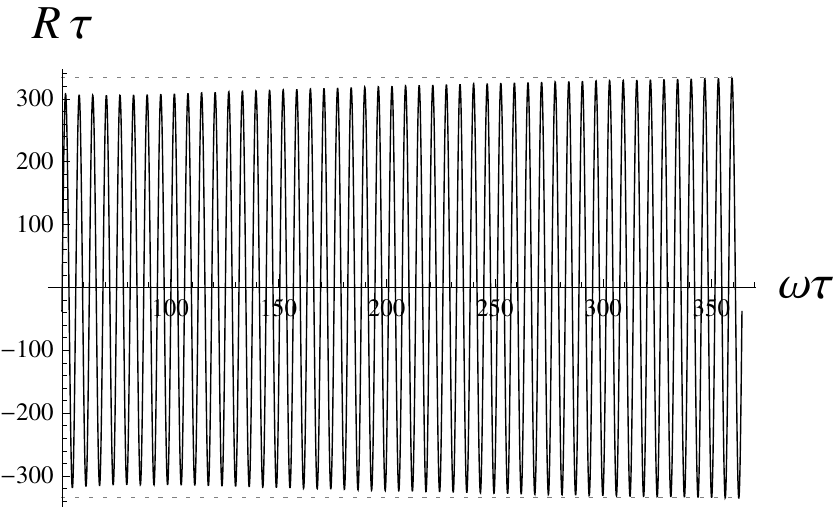}
\includegraphics[width=\mywidthdouble]{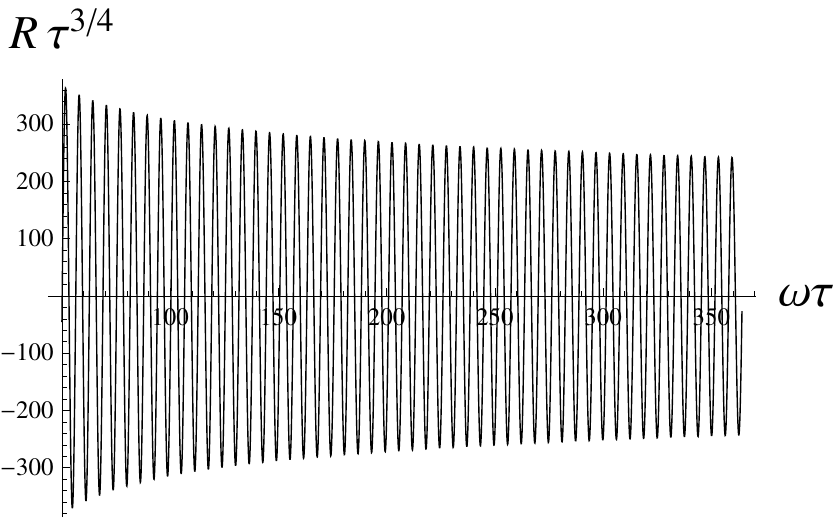}
}
\caption{Numerical solution for the dimensionless scalar curvature $r$ with initial conditions (\ref{initial_cond_2.5}).}
\label{fig:r0_2.5}
\end{figure}

In figures \ref{fig:h0_2.5} and \ref{fig:r0_2.5} we present the numerical results for the initial conditions
\begin{equation}\label{initial_cond_2.5}
  \delta h_0=1.5\,,\quad \delta h'_0=0\,,\quad\delta y_0=0\,,\quad\omega=100\,.
\end{equation}


Results are, at least qualitatively, in agreement with the analytical estimates made in the non-linear regime, for $\omega\tau\gg 1$. Evidently, the amplitudes of the oscillating terms of both $h$ and $r$ decrease faster than $\tau^{-3/4}$ (linear regime), and rather close to $\tau^{-1}$. Furthermore, the Hubble parameter does not oscillate around the GR value $h\tau=1/2$, but around a larger value, as expected from Eq.~(\ref{sol_ABCD_B}).
In fact, for these initial conditions, the best fit at the considered final integration time is given by $ \alpha\simeq 1.246$.

The Universe evolution in $R + R^2$ gravity after inflation, without relativistic matter, was considered in~\cite{Gurovich:1979xg,Starobinsky:1979ty}, where it was stated that the non-oscillating part of $h$ tends asymptotically to the GR matter-domination value $h\tau=2/3$, in contrast to our relativistic case. This scenario is in clear relation with the scalaron-dominated MD epoch at the end of an $R^2$-driven inflation (see the end of Sec.~\ref{sec:inflation} and reference therein).

Our result that $h_{osc}\sim t^{-3/4}$ (\ref{sol_h_linear}) in the linear regime agrees with what found in~\cite{Starobinsky:1979ty}, namely that $h_{osc}\sim a^{-3/2}$. This, however, is not in perfect agreement with our results in the non-linear regime, where we have $\alpha\simeq 1.246$ but $h_{osc}\sim 1/\tau$. This is probably due to the fact that we have not been able to reach the true asymptotic regime, but only a pre-asymptotic region.

\section{Particle Production and Back-Reaction} \label{sec:particle_production}

\subsection{Field Equations including Particle Production}

Particle production by an oscillating gravitational field in $R^2$ gravity was considered in~\cite{Starobinsky:1980te,Vilenkin:1985md}, where the particle production rate was estimated as $\Gamma \sim m^3/m_{Pl}^2$. In this section we present more rigorous calculations, which are essentially in agreement with~\cite{Vilenkin:1985md}. We will derive a closed equation of motion for the cosmological evolution of $R$, considering the back-reaction of particle production. To this end we consider a massless scalar field $\phi$ minimally-coupled to gravity. Its action can be written as:
\begin{equation}\label{phi_action}
 S_\phi=\frac{1}{2}\int d^4x\,\sqrt{-g}\,g^{\mu\nu}\partial_\mu\phi\,\partial_\nu\phi\,.
\end{equation}
In a spatially-flat FRW background (\ref{FRW}) it leads to the equation of motion:
\begin{equation}\label{eq_motion_phi}
 \dal\phi = \ddot \phi+3H\dot\phi-\frac{1}{a^2}\Delta\phi=0\,.
\end{equation}
The field $\phi$ enters the equation of motion for $R$ (\ref{eq:trace}) via the trace of its energy-momentum tensor:
\[
T^\mu_\mu (\phi)=-g^{\mu\nu}\partial_\mu\phi\,\partial_\nu\phi\equiv -(\partial\phi)^2\,.
\]
It is convenient to introduce the conformally rescaled field, $\chi\equiv a(t)\phi$, and the conformal time $\eta$,
such that $a\,d\eta=dt$. In terms of these quantities we can rewrite the equations of motion as:
\begin{subequations}
\begin{align}
& R'' + 2\cfrac{a'}{a}R' + m^2a^2R = \notag\\
& \quad = 8\pi\cfrac{m^2}{\mpl^2}\cfrac{1}{a^2}\left[\chi'^2-(\partial_i\chi)^2+\cfrac{a'^2}{a^2}\chi^2-\cfrac{a'}{a}(\chi\chi'+\chi'\chi)\right]\,, \label{R-diprime}\\
& R=-6a''/a^3\,, \label{R}\\
& \chi''-\Delta\chi+\cfrac{1}{6}\,a^2R\,\chi=0\,, \label{chi-diprime}
\end{align}
\end{subequations}
where $(\partial_i\chi)^2 \equiv \delta^{ij}\partial_i\chi\,\partial_j\chi$. The action (\ref{phi_action}) takes the form:
\begin{equation}\label{chi_action}
 S_\chi = \frac{1}{2}\int d\eta\,d^3x\,\left(\chi'^2-(\vec\nabla\chi)^2-\frac{a^2R}{6}\chi^2\right)\,.
\end{equation}
Here and above prime denotes derivative with respect to the conformal time, not to the radial coordinate.

Our aim is to derive a closed equation for $R$ taking the average value of the $\chi$-dependent quantum operators
in the r.h.s. of Eq.~(\ref{R-diprime}) over vacuum, in the presence of an external classical gravitational field $R$.
Our arguments essentially repeat those of~\cite{Dolgov:1998wz}, where the equations were derived in one-loop approximation.

We quantize the free field $\chi^{(0)}$ as usual:
\begin{equation}\label{quantiz_chi}
 \chi^{(0)}(x)=\int\frac{d^3k}{(2\pi)^3\,2E_k}\left[\hat a_k\,e^{-ik\cdot x}+\hat a^\dagger_k\,e^{ik\cdot x}\right]\,,
\end{equation}
where $x^\mu=(\eta,\mathbf x)$, $k^\mu=(E_k,\mathbf k)$, and $k_\mu k^\mu =0$. The creation-annihilation
operators satisfy the usual Bose commutation relations:
\begin{equation}\label{commutator}
 \left[\hat a_k,\hat a_k^\dagger\right]=(2\pi)^3\,2E_k\,\delta^{(3)}(\mathbf k-\mathbf k').
\end{equation}

Equation  (\ref{chi-diprime}) has the formal solution
\begin{equation}\label{formal_sol}
\begin{aligned}
 \chi(x)&=\chi^{(0)}(x)-\frac{1}{6}\int d^4y\,G(x,y)\,a^2(y)R(y)\chi(y)\\
&\equiv \chi^{(0)}(x)+\delta\chi(x)\,,
\end{aligned}
\end{equation}
where the massless Green's function is
\begin{equation}\label{green_func}
 G(x,y)=\frac{1}{4\pi|\mathbf x-\mathbf y|}\delta\left((x_0-y_0)-|\mathbf x-\mathbf y|\right)\equiv \frac{1}{4\pi r}\delta(\Delta\eta-r)\,.
\end{equation}
We assume that the particle production effects slightly perturb the free solution, so that $\delta\chi$ can be considered small and the Dyson-like series can be truncated at first order, yielding
\begin{equation}\label{approx_chi}
\begin{aligned}
 \chi(x) &\simeq \chi^{(0)}(x)-\frac{1}{6}\int d^4y\,G(x,y)\,a^2(y)R(y)\chi^{(0)}(y)\\
&\equiv\chi^{(0)}(x)+\chi^{(1)}(x)\,.
\end{aligned}
\end{equation}
We can now calculate the vacuum expectation values of the various terms in the right-hand side of equation (\ref{R-diprime}), keeping only first-order terms in $\chi^{(1)}$. All terms containing only $\chi^{(0)}$ and its derivatives do not affect particle production and can be re-absorbed by a renormalisation procedure into the parameters of the theory, so they are of little interest here. The other terms are calculated using formulas such as
\begin{subequations}
\begin{align}
& \partial_x\int d^4y\,G(x,y)\,a^2(y)R(y)\chi(y)=\nonumber\\
&\qquad\qquad = \int d^4y\,G(x,y)\left[a^2R\partial_y\chi+\partial_y(a^2R)\chi\right]\,,\\
& \int_0^\infty dk\,e^{i\alpha k}=\pi\delta(\alpha)+i\mc P\left(\frac{1}{\alpha}\right)\,.
\end{align}
\end{subequations}
Collecting all terms, we arrive to the expressions:
\begin{subequations}\label{trace_energy_momentum_tensor_chi}
 \begin{align}
  &\langle\chi^2\rangle\simeq -\frac{1}{48\pi^2}\int^\eta_{\eta_0}d\epsilon\,\frac{a^2(\epsilon)R(\epsilon)}{\eta-\epsilon}\,,\label{chi_squared}\\
 &\langle \chi'^2-(\partial_i\chi)^2\rangle\simeq-\frac{1}{96\pi^2}\int^\eta_{\eta_0}d\epsilon\,\frac{[a^2(\epsilon)R(\epsilon)]''}{\eta-\epsilon}\,,\label{chi_prime_squared}\\
&\langle\chi\chi'+\chi'\chi\rangle\simeq -\frac{1}{48\pi^2}\int^\eta_{\eta_0}d\epsilon\,\frac{[a^2(\epsilon)R(\epsilon)]'}{\eta-\epsilon}\,.\label{chi_chi_prime}
 \end{align}
\end{subequations}
Substituting these expressions into (\ref{R-diprime}), we obtain a closed integro-differential equation for $R$, for which we will find an approximate analytical solution. We also plan to find the exact numerical solution of this equation but this is a much more complicated problem. Still for our purpose the approximate analytical solution is accurate enough.

First of all, one has to note that despite having oscillating $H$ and $R$, the scale factor $a$ basically follows a power-law expansion, so it varies very little during many oscillation times $\omega^{-1}$. Thus, we expect that $d\eta/\eta\sim dt/t$ and that the dominant part in the integrals in (\ref{approx_chi}) be given by the derivatives of $R$, since $R'\sim \omega R+t^{-1}R\simeq\omega R$ and $\omega t\gg 1$. The dominant contribution to particle production is therefore given by Eq.~(\ref{chi_prime_squared}), which yields
\begin{equation}\label{R_with_back_reaction_approx}
\begin{aligned}
 \ddot R+3H\dot R+m^2R &\simeq -\frac{1}{12\pi}\frac{m^2}{\mpl^2}\frac{1}{a^4}\int_{\eta_0}^\eta d\epsilon\,\frac{[a^2(\epsilon)R(\epsilon)]''}{\eta-\epsilon}\\
&\simeq-\frac{1}{12\pi}\frac{m^2}{\mpl^2}\int_{t_0}^t du\,\frac{\ddot R(u)}{t-u}\,.
\end{aligned}
\end{equation}
The equation is naturally non-local in time since the impact of particle production depends upon all the history of the evolution of the system. The equation is linear in $R$, in contradiction with reference~\cite{Mijic:1986iv}, where the r.h.s. of the equation is quadratic in $R$. This latter result is physically doubtful because if the sign of $R$ changes, the effect of the particle production term would not be a damping of oscillations but rather their amplification.

\subsection{Effects of Back-Reaction}

We repeat the calculations of section \ref{sec:analytical_estim_no_back_react} using the expansion~(\ref{expansion_H_R}) and including the back-reaction effects in the form of equation~(\ref{R_with_back_reaction_approx}). The right-hand side of this equation can be written as:
\begin{align}
& g\int^t_{t_0}dt'\,\frac{\ddot R(t')}{t-t'} =g\int_0^{t-t_0}d\tau\,\frac{\ddot R(t-\tau)}{\tau}\notag\\
&= g\int_\epsilon^{t-t_0} d\tau\,\frac{\ddot R(t-\tau)}{\tau} + \int_0^\epsilon d\tau\,\frac{\ddot R(t-\tau)}{\tau}\notag\\
 &=g \int_\epsilon^{t-t_0}d\tau\,\frac{\ddot C}{\tau}+\notag\\
&+g \cos(m_1 t)\int_\epsilon^{t-t_0}d\tau\,\frac{1}{\tau}\left[F_c\cos(m_1\tau)-F_s\sin(m_1\tau)\right]+\notag\\
&+g\sin(m_1 t)\int_\epsilon^{t-t_0}d\tau\,\frac{1}{\tau}\left[F_c\sin(m_1\tau)+F_s\cos(m_1\tau)\right]\,, \label{eq:expansion_ddot_R_part_prod}
\end{align}
where
\begin{subequations}
\begin{align}
&g\equiv -\frac{1}{12\pi}\frac{m_1^2}{\mpl^2}\,,\\
 & F_c \equiv \ddot D_c+2 m_1\dot D_s-m_1^2D_c\,,\\
&F_s \equiv \ddot D_s-2m_1\dot D_c-m_1^2D_s\,,
\end{align}
\end{subequations}
and $\epsilon$ is an arbitrary, infinitesimal time. To avoid confusion we need also mention that $\tau$ here is merely an integration variable and has no relation with the dimensionless time $\tau$ of the previous sections.

We have  introduced here the new notation $m_1$ which is equal to $m$ plus radiative corrections
specified below, and which corresponds to $\nu$ in Eqs.~(\ref{H-of-t}, \ref{R-of-tCER2}). The difference between
$m$ and $m_1$ is not essential under the integral but it should be taken into account in the l.h.s. of Eq.~\rif{R_with_back_reaction_approx}, where
we should take $m_1$ instead of $m$.

Please note that the slowly-varying functions $C$, $D_s$ and $D_c$ inside the integrals are to be 
evaluated at $(t-\tau)$, and a dot denotes derivative with respect to (physical) time $t$, not $\tau$. Because of the $1/\tau$ factor,
the integral is logarithmically divergent, but this divergence can be
absorbed into the renormalisation of mass, $m$, and coupling, $g$. So we separate the integral into two parts: one where $\tau$ goes from 0 to some small parameter $\epsilon$, which determines the normalisation point at which
 the physical mass and coupling are fixed, and another, taken 
from $\epsilon$ to $(t-t_0)$, which gives corrections to
the physical qualities due to interactions. More details
can be found in~\cite{Dolgov:1998wz}.

Equating the coefficients multiplying the slow varying terms, $\sin m_1 t$, and $\cos m_1 t$ 
in the same way as it has been done in sec.~\ref{sec:analytical_estim_no_back_react}, see Eqs.~(\ref{eqn-ABCD}), we obtain 
the same first three equations~(\ref{eqn-ABCD_A}, \ref{eqn-ABCD_Bs}, \ref{eqn-ABCD_Bc}), where the effects of particle production do not directly appear, 
and the remaining three ones with the additional terms coming from Eq.~(\ref{R_with_back_reaction_approx}), see also the expansion (\ref{eq:expansion_ddot_R_part_prod}). 
The latter equations become integro-differential but they can be reduced to differential equations in the case of fast oscillations.
So the complete set of equations with the account of particle production has the form (for convenience we also include
the unchanged first three equations of (\ref{eqn-ABCD}):
\begin{subequations}\label{eqn-ABCD_with_back_react}
\begin{align}
&\dot A + 2 A^2 +B_s^2 + B_c^2 = -C/6\,,\\
&\dot B_s - B_c m_1 + 4 A B_s = -D_s/6\,,\\
& \dot B_c + B_s m_1 + 4 A B_c = -D_c/6\,,\\
&\ddot C + 3A \dot C + \frac{3}{2} B_s \dot D_s + \frac{3}{2} B_c \dot D_c
 - \frac{3}{2}m_1 B_s D_c + \notag\\
&\qquad + \frac{3}{2}m_1 B_c  D_s + m^2 C \simeq g\int_\epsilon^{t-t_0}d\tau\,\frac{\ddot C}{\tau}\,,\label{eqn-ABCD_with_back_react_C}\\
&\ddot D_s +(m^2 - m_1^2) D_s - 2 m_1 \dot D_c\,  +3A (\dot D_s - m _1 D_c) +\notag\\
& \qquad+3\dot C B_s \simeq g\int_\epsilon^{t-t_0}d\tau\,\frac{F_s\cos(m\tau)+F_c\sin(m\tau)}{\tau}\,,\label{eqn-ABCD_with_back_react_Ds}\\
&\ddot D_c +(m^2 - m_1^2) D_c
+ 2 m_1 \dot D_s\, + 3A (\dot D_c + m_1 D_s) +\notag\\
& \qquad+3\dot C B_c \simeq g\int_\epsilon^{t-t_0}d\tau\,\frac{F_c\cos(m\tau)-F_s\sin (m\tau)}{\tau}\,.\label{eqn-ABCD_with_back_react_Dc}
\end{align}
\end{subequations}
In the integrals in the right-hand-side of (\ref{eqn-ABCD_with_back_react_Ds}) and (\ref{eqn-ABCD_with_back_react_Dc}), which contain quickly oscillating functions, the effective value of $\tau$ is about $1/m$. Thus we can approximate
$F(t-\tau) \approx F(t)$ and take such factors out of the integrals. Let us analyse now for example equation (\ref{eqn-ABCD_with_back_react_Ds}) term by term. The analysis
of Eq.~(\ref{eqn-ABCD_with_back_react_Dc}) is similar. In what follows we neglect $\ddot D$ in comparison with $m^2 D$.

The dominant term in Eq.~(\ref{eqn-ABCD_with_back_react_Ds}), which is the coefficient multiplying $D_s$, determines the renormalisation of $m$:
\be
m_1^2 = m^2 + g\,m^2 \int_\epsilon^{t-t_0} \frac{d\tau}{\tau}\,\cos m\tau  \,.
\label{renorm-mass}
\ee
The next subdominant term, which is the coefficient in front of $\dot D_c$, determines the decay rate of $D_c$:
\be\label{dot-Dc}
\dot D_c=\frac{g m}{2}D_c\int_\epsilon^{t-t_0}\frac{d\tau}{\tau}\sin m\tau\approx \frac{\pi g m}{4}D_c\,.
\ee
We skipped here the term $g \dot D_c $, which leads to higher order corrections to the production rate.
Thus the decay rate is
\be
\Gamma_R = -\frac{\pi g m}{4} = \frac{m^3}{48m_{Pl}^2}\,.
\label{Gamma-R}
\ee
Correspondingly the oscillating part of $R$ or $H$ behaves as
\be
\cos m_1 t \to e^{-\Gamma_R t} \, \cos m_1 t\,.
\label{damping}
\ee
We will use this result in the next subsection in the calculation of the energy density influx of the produced particles into the primeval plasma.

\subsection{Particle Production Rate and Relic Energy Density of Produced Particles}

From equation (\ref{chi-diprime}) it follows that  the amplitude of gravitational production of two identical $\chi$
particles with momenta $p_1$ and $p_2$ in the first order in perturbation theory is given by \begin{equation}\label{particle_production_general}
A(p_1,p_2)\simeq \int d\eta\,d^3x\,\frac{a^2R}{6}\left\langle p_1,p_2\left|\chi\chi \right|0\right\rangle\,,
\end{equation}
where the final two-particle state is defined by
\[
 |p,q\rangle = \frac{1}{\sqrt 2}\,\hat a^\dagger_p\,\hat a^\dagger_q |0\rangle\,.
\]
The factor $1/\sqrt{2}$ is simply the correct normalisation of the two-particle state due to the Bose statistics. Using Eq.~(\ref{quantiz_chi}), we find
\begin{align}\label{chi2_p1_p2_amplitude}
 \langle p_1,p_2|\chi\chi|0\rangle &= \frac{\sqrt 2}{8(2\pi)^6}\int\frac{d^3k\,d^3k'}{E_k\,E_{k'}}e^{i(E_k+E_{k'})\eta-i(\mathbf k+\mathbf k')\cdot \mathbf x}\times \notag\\
&\qquad \times \langle 0|\,\hat a_{p_1}\,\hat a_{p_2}\,\hat a^\dagger_k\,\hat a^\dagger_{k'}\,|0\rangle\,,\notag\\
&=\sqrt 2\,e^{i(E_{p_1}+ E_{p_2})\eta-i(\mathbf p_1+\mathbf p_2)\cdot\mathbf x}\,.
\end{align}
Here $E_k^2 =\mathbf k^2$, and the function $a^2R$ has the form
\[
 a^2(\eta) R(\eta)=D(\eta)\sin(\tilde\omega\eta)\,,
\]
where $D(\eta)$ is a slowly-varying function of (conformal) time, $\tilde\omega $ is the frequency conjugated to conformal time. 
Under these approximations, the amplitude (\ref{particle_production_general}) becomes
\be\label{calc_part_prod_general}
\begin{aligned}
 A(p_1,p_2)&=\frac{i}{6\sqrt 2}\int d\eta\,d^3x\,D(\eta)\left(e^{i\tilde \omega \eta}-e^{-i\tilde\omega\eta}\right)\times \\
&\qquad  e^{i(E_{p_1}+ E_{p_2})\eta}\,e^{-i(\mathbf p_1+\mathbf p_2)\cdot \mathbf x}\,.
\end{aligned}
\ee
Taking $E_{p_i}\geq 0$ and neglecting at this stage variations of $D$ with time, we obtain
\begin{equation}\label{part_prod_2}
 A(p_1,p_2)\simeq -\frac{i}{6\sqrt 2}\,D(\eta)(2\pi)^4\,\delta^{(3)}(\mathbf p_1+\mathbf p_2)\,\delta(E_{p_1}+E_{p_2}-\tilde\omega)\,.
\end{equation}
In order to find the particle production
rate per unit comoving volume and unit conformal time, we need to integrate the modulus squared of this amplitude over the phase space, namely: 
\be\label{trans_rate}
\begin{aligned}
 n' &= \int\frac{d^2p_1\,d^3p_2}{(2\pi)^6\,4\,E_{p_1}E_{p_2}}\,\frac{|A(p_1,p_2)|^2}{V\,\Delta\eta} \\
 &\simeq\frac{1}{72}\,D^2(\eta)\int\frac{d^3p_1\,d^3p_2}{(2\pi)^6\,4\,E_{p_1}E_{p_2}}(2\pi)^4\,\delta^{(3)}
 (\mathbf p_1+\mathbf p_2)\times \\
&\qquad \times \delta(E_{p_1}+E_{p_2}-\tilde\omega) \\
&\simeq \frac{D^2(\eta)}{576\pi}\,,
\end{aligned}
\ee
where $V$ and $\Delta\eta$ are the total volume and conformal time, which of course go to infinity,
$n$ is the number density of the produced particles, and a prime denotes derivative with respect to conformal time.
Since the energy of the produced particles is equal to $\tilde\omega/2$,
we find for the rate of gravitational energy transformation into elementary particles:
\begin{equation}\label{rho_prime_part_prod}
\rho'= \frac{n' \tilde \omega}{2}=\frac{D^2(\eta)\tilde\omega}{1152\pi}
\end{equation}
and so the rate of variation of the physical energy density of the produced $\chi$-particles is
\begin{equation}\label{rho_dot_part_prod}
 \dot\rho_{\chi}= \frac{m\langle R^2\rangle}{1152\pi}\,.
\end{equation}
Here, $\langle R^2\rangle$ is the square of the amplitude of the oscillations of $R$ and we substituted $\tilde\omega=a m$. 
To obtain the  the total rate of the gravitational energy transformation into elementary
particles we should multiply the above result by the number of the produced particle species, $N_{eff}$,
so the total rate of production of matter is $\dot \rho_{PP} = N_{eff} \dot \rho_\chi $.\\
Note that our result (\ref{rho_dot_part_prod}) coincides with that of~\cite{Vilenkin:1985md},
although performed in a slightly different cosmological regime.

Now we can calculate the evolution of the cosmological energy density of matter,
which is determined by the equation:
\be
\dot\rho =-4H\rho + \dot\rho_{PP}\,.
\label{dot-rho-rel}
\ee
We assumed here that the produced matter is relativistic and so the first term in the r.h.s.
describes the usual cosmological red-shift, while the second term is the particle source from the
oscillations of $R$. Since $\rho$ is not oscillating but a smoothly varying function of time, its
red-shift is predominantly determined by the non-oscillating part of the Hubble parameter,
$H_{c} \simeq \alpha/2t$, see Eq.~(\ref{h_fit_func}).

Parametrizing the oscillating part of the Hubble parameter as $H_{osc} \simeq {\beta \cos mt}/{t}$, 
we find for the oscillating part of curvature:
\be
R\simeq -\frac{6\beta m\sin mt}{t}\, e^{-\Gamma_R t}\,.
\label{R-osc}
\ee
Here we took into account the exponential damping of $R$, which was for brevity omitted in 
the expression for $H$ just above.\\
Correspondingly the energy density of matter  obeys the equation:
\be
\dot \rho = -\frac{2 \alpha}{t} \rho + \frac{\beta^2 m^3 N_{eff}}{32\pi t^2}\,e^{-2\Gamma_R t}\,.
\label{dot-rho-2}
\ee
This equation can be explicitly integrated as it is, but for a simple analytical estimate we will use the instant decay approximation. Namely we neglect the exponential damping term, when $2\Gamma_R t <1$, and take $\alpha =\alpha_1 = 1.25$, according to the numerical estimate of sec.~\ref{ss-num-estim}. For $2\Gamma_R t  > 1$ we completely ignore the second (source) term in Eq.~(\ref{dot-rho-2}) and take $\alpha = \alpha_2 = 1$. This choice corresponds to the GR 
solution and we believe that it is realised when the oscillations disappear, as follows from the analytical estimates presented above. Thus at short times, $2 \Gamma_R t <1$, the energy density of matter would be:
\be
\rho = \rho_{in} \left(\frac{t_{in}}{t}\right)^{2\alpha_1} +
\frac{\beta^2 m^3 N_{eff}}{32 \pi (2\alpha_1-1) t} \left( 1 - \frac{t_{in}^{2\alpha_1 -1}}{t^{2\alpha_1 -1}} \right)\,.
\label{rho-short-t}
\ee
For large times, i.e. $2\Gamma_R t >1$, equation (\ref{dot-rho-2}) becomes homogeneous and its solution is simply the relativistically red-shifted energy density, whose initial value is to be determined from Eq.~(\ref{rho-short-t})
at $t= 1/(2\Gamma_R)$:
\be
\begin{aligned}
\rho &= \frac{ m^6 }{768 \pi\,m_{Pl}^2\, (2\Gamma_R t)^{2\alpha_2}}\,
\Bigg\{ \frac{\kappa}{8} \left(2 t_{in} \Gamma_R \right)^{2\alpha_1 - 2} + \\
&\qquad + \frac{\beta^2 N_{eff}}{2\alpha_1-1}\left[1-(2\Gamma_R t_{in})^{2\alpha_1-1}\right] \Bigg\} \,,
\end{aligned}
\label{rho-long-time}
\ee
where we parametrised the energy density of matter at the initial time $t_{in} $ as
\be
\rho_{in} =\frac{3 m_{Pl}^2 \kappa}{32 \pi t_{in}^2}\,.
\label{rho-in}
\ee
Parameter $\kappa$ is arbitrary, and depends upon the thermal history of the Universe before $t_{in}$.
In particular, $\kappa = 0$ is possible and does not contradict our picture, since the equations of motion
have non-trivial oscillating solutions even if $\rho =0$.

The first term in equation (\ref{rho-long-time}) is the contribution of normal thermalised relativistic matter,
while the second also describes relativistic matter, but this matter might not be thermalised, at least during
some cosmological period. Depending upon parameters the relative magnitude of non-thermalised matter
might vary from negligibly small up to being the dominant one.

\section{Discussion and Implications}

The characteristic decay time of the oscillating curvature is
\be
\tau_R = \frac{1}{2\Gamma_R} =  \frac{24 m^2_{Pl}}{m^3} \simeq 2
\left(\frac{10^5\text{ GeV}}{m}\right)^3 \text{ seconds}\,.
\label{tau-R}
\ee
The contribution of the produced particles into the total cosmological energy density reaches its maximum value at approximately this time. The ratio of the energy density of the newly produced energetic particles and that of those already existing in the plasma, according to Eq.~(\ref{rho-long-time}), is:
\be
\frac{\rho_{hi}}{\rho_{therm}} = \frac{8\beta^2 N_{eff}}{\kappa(2\alpha_1-1)}\,\frac{1-(2\Gamma_Rt_{in})^{2\alpha_1-1}}{(2\Gamma_R t_{in})^{2\alpha_1-2}}\,.
\label{ratio}
\ee
If we take $t_{in} \simeq 1/m$, then $t_{in} \Gamma_R \simeq m^2/m_{Pl}^2 \ll 1$ and the effects of non-thermalised matter may be negligible. However, for sufficiently large $\beta$ and possibly small $\kappa$ the non-thermal particles may play a significant role in the cosmological history.

The influx of energetic protons and antiprotons could have an impact on BBN. Thus this would either allow to obtain some bounds on $m$ or even to improve the agreement between the theoretical predictions for BBN and the measurements of primordial light nuclei abundances.

The oscillating curvature might also be a source of dark matter in the form of heavy supersymmetric (SUSY) particles. Since the expected light SUSY particles have not yet been discovered at LHC, supersymmetry somewhat lost its attractiveness. The contribution of the stable lightest SUSY particle into the cosmological energy is proportional to
\be
\Omega \sim{ m^2_{SUSY} }/m_{Pl}
\label{Omega-SUSY}
\ee
and for $m_{SUSY} $ in the range $10^2-10^3$ GeV the cosmological fraction of these particles would be of order unity. This is exactly what is necessary for dark matter. However, it excludes thermally produced LSP's if they are much heavier. If LSP's came from the decay of $R$ and their mass is larger than the scalaron mass $m$, the LSP production could be sufficiently suppressed to make a reasonable contribution to dark matter.

\chapter{Curvature Singularities from Gravitational Contraction
in \titlefr Gravity
}
\label{ch:curv_sing_grav_contr}

\chaptermark{\spacedlowsmallcaps{Singularities from Gravitational Contraction}}

\citazione{\footnotesize {\color{myhypercolor}\bf{L. Reverberi}}, \it{Phys. Rev. D} \bf{86}, 084005 (2013);\\
{\color{myhypercolor}\bf{L. Reverberi}}, \it{J. Phys. Conf. Ser.} \bf{442}, 012036 (2013).}{}

\section{Introduction}

We have already discussed the constraints for the cosmological viability of $f(\R)$ models recently thoroughly investigated \cite{Amendola:2006we,Sawicki:2007tf}, and the few models proposed which evade all such tests, therefore seeming to be good candidates for a gravitational theory of Dark Energy~\cite{Appleby:2007vb, Hu:2007nk,Starobinsky:2007hu} (see also below, Eq.~\ref{eq:models}).

Testing such modified gravity theories in astronomical/astrophysical systems is of paramount importance to constrain and possibly rule out models, and in general to improve our knowledge of the subject. Studies of the stability of spherically symmetric solutions have indicated the possibility of an infinite-$\R$ singularity developing inside relativistic, dense stars~\cite{Frolov:2008uf,Kobayashi:2008tq,Thongkool:2009js,Thongkool:2009vf}. Important steps forward in our understanding of static, spherically or axisymmetric astrophysical objects in $f(\R)$ gravity have recently been made (see e.g.\cite{Babichev:2009td,Babichev:2009fi,Capozziello:2011nr,Capozziello:2011gm}), and seem to point towards the existence of a rather general instability/singularity problem in these theories. Indeed, it has been shown that analogous problems occur in many different extended theories of gravity, not only $f(\R)$~\cite{Seifert:2007fr}.

Furthermore, similar results are obtained in the case of a less dense but contracting object~\cite{Arbuzova:2010iu,Bamba:2011sm}. In this case the singularity is not triggered by the large mass/energy density, but rather from its increase with time. One can write the trace of the modified Einstein equations as an oscillator equation for the additional gravitational scalar degree of freedom, which is sometimes dubbed \it{scalaron} and which we will denote with $\xi$ (see below), and it is easy to see that $\R$ oscillates around the GR solution $\R+\T=0$. The frequency and the amplitude of such oscillations usually grow along with the increasing density, and may eventually lead to a singularity. The key point is that $\xi$ moves in a matter- and therefore time-dependent potential, in which the ``energy'' corresponding to the point $\xi=\xi(|\R|\to\infty)$ may be finite, rendering this singular point, in principle, accessible by the field. This mechanism is strictly related to that responsible for the 
past cosmological singularities examined e.g. in~\cite{
Appleby:2008tv,Appleby:2009uf}. 

Singularity issues in infrared-modified $f(\R)$ theories could in principle be solved by the introduction of ultraviolet corrections, as investigated e.g. in~\cite{Arbuzova:2010iu,Bamba:2011sm}. Moreover, oscillations lead to gravitational particle production, and a large frequency/amplitude of the oscillating curvature could lead to a noticeable emission of cosmic rays~\cite{Arbuzova:2011fu}; in principle, this could severely affect the total cosmic ray flux, distort the power spectrum, and even serve as a possible mechanism to avoid the Greisen-Zatsepin-Kuzmin (GZK) cutoff~\cite{Arbuzova:2012su,Arbuzova:2013ina}.

In this Chapter, we focus again on the $f(\R)$ gravity models~\cite{Hu:2007nk,Starobinsky:2007hu} during the contraction of a nearly-homogeneous cloud of pressureless dust. Using a simplified approach, namely assuming spherical symmetry, homogeneity and low gravity, we work out simple expressions for the evolution of $\xi$ and hence $\R$, and in particular of the amplitude and frequency of their oscillations. We confirm the existence of a finite-time, future singularity, whose appearance depends on the duration of the contraction and on both model- and physical parameters, and derive estimates for the typical timescales for this process. 

Once we derive general results, we will apply them to two very similar models recently proposed and cited above:
\begin{subequations}\label{eq:models}
\begin{align}
 &F_{HS}(\R) = -\cfrac{\lambda \R_c}{1+(\R/\R_c)^{-2n}}\,,\quad\text{\cite{Hu:2007nk}}\\
 &F_S(\R) = \lambda \R_c\left[\left(1+\dfrac{\R^2}{\R_c^2}\right)^{-n}-1\right]\,.\quad \text{\cite{Starobinsky:2007hu}}
\end{align}
\end{subequations}
The subscripts stand for, respectively, Hu-Sawicki (HS) and Starobinsky (S). For both models, if $\lambda$ is of order unity $\R_c$ is of the order of the present cosmological constant (for details we refer the reader to the specific articles), which is much smaller than the typical values of $\R$ and $\T$ in astrophysical systems, such as pre-stellar, pre-galactic, and molecular clouds. Hence, in many cases we will take the limit $|\R_c/\T|\sim \R_c/\R \ll 1$ before presenting the final results.

For simplicity, we assume that the contraction of the system is stationary, i.e. the mass density grows linearly with time, on a typical timescale $t_{contr}$:
\be\label{eq:T_evol}
\T(t) = \T_0\left(1+t/t_{contr}\right)\,.
\ee
We must stress that this evolution law should not be regarded as accurate from a physical standpoint; the difficult task of computing the full dynamics of contraction of a self-gravitating system goes way beyond the scope of this Thesis (see e.g.~\cite{Cembranos:2012fd} and references therein).

Nevertheless, unless the contraction follows a radically different behaviour, and especially until ${t\sim t_{contr}}$, results obtained with this form should be qualitatively correct. We will find that a faster contraction contributes positively to the formation of a singularity, so we expect that contraction laws $\T\sim t^\gamma$ with $\gamma>1$ will lead to singularities even more effectively than what appears from our results. On the other hand, a slower contraction could help delaying ($\gamma<1$) or even avoiding, if the contraction would stop at some moment, the singularity. 

In this Chapter we will use the following dimensionless parameters characterising the physical properties of the system under scrutiny:
\be\label{eq:definitionsCSC}
\begin{aligned}
 \R_{29} &\equiv \frac{\mpl^2}{8\pi}\,\frac{(-\R_c)}{\rhoc}\,,\\
\rho_{29} &\equiv \frac{\rho_0}{\rhoc}\,,\\
t_{10} &\equiv \frac{t_{contr}}{10^{10}\text{ years}}\,.
\end{aligned}
\ee

\section{Curvature Evolution in Contracting Systems}\label{sec:CURV_EVOL}
\subsection{Field Equations}\label{sec:field_equations}
We start as usual from the gravitational action
\be\label{eq:grav_action_fR}
\begin{aligned}
A_{grav} &= -\frac{\mpl^2}{16\pi}\int d^4x\,\-g\,f(R)\,,\\
         &= -\frac{\mpl^2}{16\pi}\int d^4x\,\-g\,[R+F(R)]\,,
\end{aligned}
\ee
from which one obtains the field equations
\be\label{eq:field_equations}
f_{,R}(\R)\R_\mn-\frac{1}{2}f(\R)\,g_\mn + \left(g_\mn\D^2-\D_\mu\D_\nu\right)f_{,R}(\R)=\T_\mn\,.
\ee
The corresponding trace equation reads
\be\label{eq:traceCSC}
3\D^2F_{,R}+\R F_{,R}-2F-(\R+\T)=0\,.
\ee
We consider a nearly homogeneous and spherically symmetric cloud of pressureless dust, hence
\be
T^\mu_\mu \simeq \frac{8\pi}{\mpl^2}\,\rho_m\,.
\ee
where $\rho_m$ is the mass/energy density of the cloud. The homogeneity of the could allows us to neglect spatial derivatives, as intuitively clear and explicitly proved in~\cite{Arbuzova:2010iu}; assuming also low gravity, the D'Alambertian operator can be replaced by the second derivative in the time coordinate: ${\D^2\to\partial_t^2}$. Of course, a more careful investigation of the problem should take into account both time and spatial derivatives; this could be a subject for further research. Also notice that these and following arguments can also be applied to cosmology, if we consider the evolution of the Universe, backwards in time, during the matter-dominated epoch~\cite{Appleby:2008tv,Appleby:2009uf}. 

The approximation of low gravity and the formation of curvature singularities may seem utterly incompatible, but this is not true: in fact, ${R\sim \partial^2 g_\mn}$ may diverge even if $g_\mn$ is very close to $\eta_\mn$ (Minkowski). Details about this statement and about the dust assumption (${p/\rho\ll 1}$) can be found in Sec.~\ref{sec:approx}.

As mentioned in the Introduction, in a typical astrophysical situation we have
\be\label{eq:large_density_limit}
\frac{\T}{\R_c}\sim\frac{\R}{\R_c} \gg 1\,,
\ee
so that both models reduce to
\be\label{eq:model_approxCSC}
F_{HS}\simeq F_S\simeq-\lambda \R_c\left[1-\left(\frac{\R_c}{\R}\right)^{2n}\right]\,.
\ee
In this limit, these models are basically equivalent to adding a $\Lambda$-term, since $F$ is almost constant; in fact, we clearly have
\be\label{eq:Lambda}
\R+F\approx \R+2\Lambda\,,\qquad\Lambda = -\frac{\lambda \R_c}{2}\,.
\ee
See also~\cite{Hu:2007nk,Starobinsky:2007hu} for further details. Therefore, ${\R+\T+2F=0}$ is practically equivalent to GR with the addition of a cosmological constant $\Lambda$, that is the usual $\Lambda$CDM model\footnote{Of course, these $f(\R)$ models do \it{not} explain the nature of Dark Matter.}; in what follows, when referring to ``GR'' for brevity, we will mean precisely this. We also have ${F_{,R}\sim \lambda (\R_c/\R)^{2n+1}}$, thus
\be\label{eq:terms_com}
|\R| \gg |F| \gg |F_{,R}\R|\,.
\ee
Under these assumptions, Eq.~\rif{eq:traceCSC} becomes
\be\label{eq:traceCSC_approx}
3\partial_t^2F_{,R}-(\R+\T+2F)=0\,.
\ee
Since $F\sim\Lambda\ll T$, the only contribution of this effective cosmological constant is to offset the GR solution from $\R+\T=0$ to $\R+\T+2F=0$, which is a very small (and almost constant in absolute value) correction of order $|\R_c/\T|$. One should be careful, because it may appear from exact numerical results that $F$ is of the order of $\R+\T$ or even larger\footnote{But not of the order of $\R$ or $\T$ \it{individually}!}; nevertheless, its effect in~\rif{eq:traceCSC_approx} is completely trivial, unlike the dynamical term $\ddot F_{,R}$, because even large variations of $\R$, especially when $|\R|$ increases, result in extremely small variations of $F$ (see equation~\ref{eq:model_approxCSC}). From now on, we will include these corrections using
\be\label{eq:T_correction}
\tilde\T \equiv\T+2F \approx\T+4\Lambda\,,
\ee
However, we will still have to consider $\T$ alone because it is the quantity directly related to the \it{physical} energy/matter density at a given point and a given instant of time.\\
Defining the new scalar field
\be\label{eq:xi_def}
\xi\equiv -3F_{,R}\,,
\ee
which in the cases considered is approximately
\be\label{eq:xi_models_explicit}
\xi_{HS,\,S}\simeq 6n\lambda\left(\frac{\R_c}{\R}\right)^{2n+1}\,,
\ee
we rewrite~\rif{eq:traceCSC_approx} as an oscillator equation:
\be\label{eq:traceCSC_xi_KG}
\ddot\xi+\R+ \tilde\T=0\quad\Leftrightarrow \quad \ddot\xi+\frac{\partial U}{\partial\xi}=0\,.
\ee
Again, we stress the underlying assumption that ${\tilde\T=\tilde\T(t)}$. We are testing the behaviour of curvature with a simple, smooth \it{external} energy density evolution, which is arbitrarily chosen. As we have already mentioned, a more complete analysis could be subject of stimulating further research.

Usually, it is not possible to invert \rif{eq:xi_def} to obtain ${\R=\R(\xi)}$ and thus a simple form for $U(\xi)$, expect perhaps in some limit, but it is rather clear that solutions will oscillate around the solution ${\R+\tilde\T=0}$, with frequency roughly given by
\be
\omega_\xi^2\simeq \left.\frac{\partial^2 U}{\partial\xi^2}\right|_{\R+\tilde\T=0}\simeq \left.\frac{1}{\partial\xi/\partial \R}\right|_{\R+\tilde\T=0}\,.
\ee
If ${\omega^2 <0}$, one expects instabilities, and this is exactly the kind of instability of refs.~\cite{Dolgov:2003px,Faraoni:2006sy}. One can immediately see that for the two models considered we have
\be\label{eq:omega_models}
\omega^2\simeq -\frac{\R_c}{6n\lambda(2n+1)}\left(-\frac{\tilde\T}{\R_c}\right)^{2n+2}>0\,,
\ee
so there is no instability problem. We remind the reader that with our sign conventions ${\R_c<0}$, ${\tilde\T>0}$.\\
Even with ${\omega^2>0}$, however, we will show that if the model fulfils some requirements, then curvature \it{singularities} can develop. In particular, we need that
\begin{enumerate}
\item  there exist a certain value $\xi_{sing}$ corresponding to ${|\R|\to\infty}$,
\item the potential be finite in $\xi_{sing}$, i.e. ${U(\xi_{sing})<\infty}$.
\end{enumerate}

\subsection{Energy Conservation and the Scalaron Potential}
If the previous requirements are met, then in general it is possible that $\xi$ reach $\xi_{sing}$ and hence ${|\R|\to\infty}$. We can see this, for instance, from the ``energy'' conservation equation associated to \rif{eq:traceCSC_xi_KG}, that is
\be\label{eq:energy_conservation}
\frac{1}{2}\,\dot\xi^2+U(\xi,t)-\int^t dt'\,\frac{\partial \tilde\T}{\partial t'}\,\xi(t')=\text{ const}\,,
\ee
where
\be\label{eq:potential}
U(\xi,t) = \tilde\T(t)\,\xi + \int^\xi \R(\xi')d\xi'\,.
\ee
The last term in the l.h.s. of~\rif{eq:energy_conservation} is due to the explicit time-dependence of $\tilde\T$, and if ${\partial \tilde\T/\partial t>0}$, as is the case in contracting systems even without specifically assuming\footnote{Assuming \rif{eq:T_evol}, ${\partial\tilde\T/\partial t= \T_0/t_{contr}\neq \tilde\T_0/t_{contr}}$ (see~Eqs.~\ref{eq:T_evol} and~\ref{eq:T_correction}), so~\rif{eq:energy_conservation} is further simplified, with the last term simply being proportional to ${\int^t dt'\,\xi}$.}~\rif{eq:T_evol}, it will produce an increase in the ``canonical'' energy (kinetic + potential).

Note that this is true for $\xi>0$, whereas for $\xi<0$ it would give the opposite behaviour. However, it has been shown that the condition $F_{,R}<0$, corresponding to $\xi>0$, is crucial for the correct behaviour of modified gravity models at (relatively) low curvatures \cite{Appleby:2009uf}.

\begin{figure}[thb]
\centering
\includegraphics[width=\mywidthsingle]{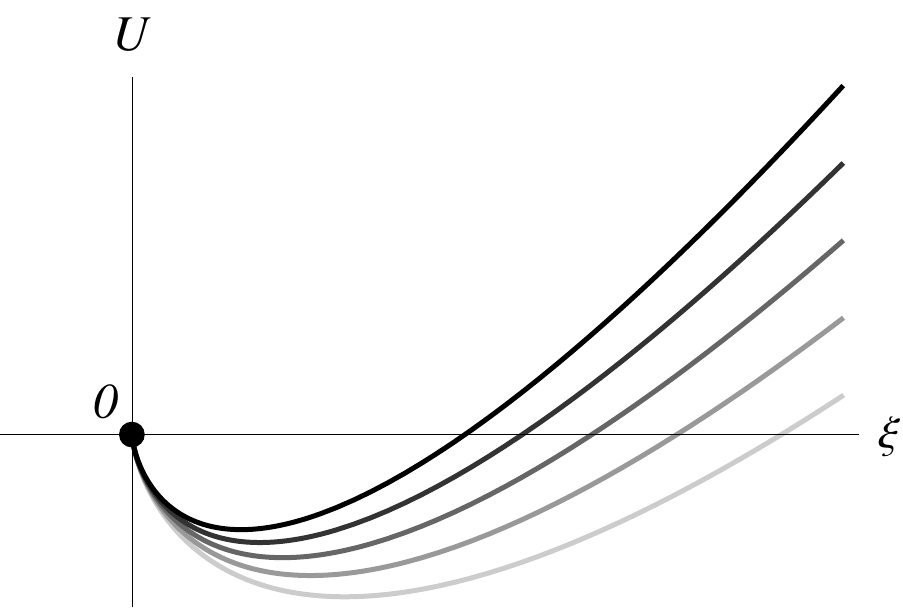}
\caption[Qualitative shape of potentials for models \rif{eq:models}, assuming ${\R/\R_c\gg 1}$.]{Qualitative shape of potentials for models \rif{eq:models}, assuming ${\R/\R_c\gg 1}$. For both models ${\xi=\xi_{sing}=0}$ (the black dot) corresponds to the singular point ${|\R|\to\infty}$. The typical time evolution is also shown, from light grey (earlier times) to black (later times).}
\label{fig:potentials}
\end{figure}

As we have previously mentioned, usually it not possible to invert the relation ${\xi=\xi(\R)}$ in order to obtain ${\R=\R(\xi)}$ and solve the integral in \rif{eq:potential} exactly. Nonetheless, for the two models considered and in the limit ${\R/\R_c\gg 1}$ this procedure is possible; apart from an additive constant, which we can put to zero, the approximate potentials are equal and read
\be\label{eq:potentials_approx}
U\simeq \tilde\T(t)\,\xi + 3\lambda \R_c(2n+1)\left(\frac{\xi}{6n\lambda}\right)^\frac{2n}{2n+1}\,.
\ee
The shape of this potential is shown in Fig.~\ref{fig:potentials}. The bottom of the potential, as expected from \rif{eq:traceCSC_xi_KG}, is in ${\R+\tilde\T=0}$. Moreover, ${\xi_{sing}=0}$ and ${U(\xi_{sing})=}$const.

\section{Adiabatic Region}\label{sec:adiabatic}
For simplicity, let us initially assume that the oscillations of $\xi$ in its potential are ``adiabatic'', in the sense that at each oscillation $\xi$ moves between two values
\be\label{eq:xi_min_xi_max}
\xi_{min}(t)\,,\quad\xi_{max}(t)\,,                                                                                                                                                                                                                                                                                                                           \ee
at roughly the same ``height'', that is with
\[
U(\xi_{min})=U(\xi_{max})\,.
\]
We should stress that $\xi_{min}$ and $\xi_{max}$ are considered to be slowly varying, so that it makes sense to compare these two values even though they do not correspond to the same instant of time, but are rather evaluated at different times with a $\delta t$ of order $\omega^{-1}$. 

The validity of this approximation can be understood as follows: the potential is roughly of the order of $\tilde\T\,\xi$, whereas the variation of the integral term in \rif{eq:energy_conservation} in one oscillation is of order $\dot{\tilde\T}\,\xi/\omega$. If $\omega$ is much larger than the inverse contraction time, that is
\[
\omega\gg \frac{\dot{\tilde\T}}{\tilde\T}\,,
\]
which is the case for the models considered below (see Eq.~\ref{eq:models}) provided that the contraction is sufficiently slow, then the integral term can be considered approximately constant over a large number of oscillations.\\
Assuming~\rif{eq:T_evol}, this basically results in the condition
\be\label{eq:fast-roll-condition}
\frac{\omega\,t_{contr}}{2\pi} \gg 1\,,
\ee
where the factor $2\pi$ only indicates that the period of the oscillations of $\xi$ is $2\pi\omega^{-1}$, not $\omega^{-1}$. For the models under investigation, this gives roughly
\be\label{eq:fast_roll_inequality}
\frac{\rho_{29}^{2n+2}\,t_{10}^2}{n\lambda(2n+1)\R_{29}^{2n+1}} \gg 142\,.
\ee
Later, we will relax this assumption and work in the opposite regime, where $\dot{\tilde \T}/\tilde\T \gtrsim \omega$.

Let us expand $\xi$ around the ``average'' value
\be\label{eq:xi_a_def}
\xi_a(t)\equiv \xi(\R=-\tilde\T)\,,
\ee
which corresponds to the value of $\xi$ if the behaviour of the system were described by the usual GR solution $\R+\tilde\T=0$.
We could be misled to infer from~\rif{eq:traceCSC_approx} that with this definition $\xi_a$ must exactly satisfy
\be\label{eq:xi_a_wrong}
\ddot\xi_a=0\,,
\ee
so that $\xi_a \sim t$. This is not true, because near the GR solution we can no longer neglect sub-leading terms in \rif{eq:traceCSC} and hence use \rif{eq:traceCSC_approx}. In some sense, \rif{eq:xi_a_wrong} remains true provided that we interpret it as the statement
\be
\left|\ddot\xi_a\right| \ll |\R|,\,\tilde\T\,.
\ee 
Ultimately, $\xi_a$ is the reference point for $\xi$ because it corresponds to the bottom of its potential (see Eq.~\ref{eq:traceCSC_xi_KG}). Nonetheless $\xi_a$ is not the solution of \rif{eq:traceCSC_xi_KG}, but merely a test function helping us quantifying how the behaviour of $\R$ in $f(\R)$ gravity differs from that of GR. After all, $\xi$ in GR is identically zero.

Thus we write
\begin{subequations}\label{eq:xi_expansionCSC}
\begin{align}
\xi(t) &= \xi_a(t)+\xi_1(t) \label{eq:xi_expansionCSC_xi1}\\
&\equiv \xi_a(t) + \alpha(t)\sin \Phi(t)\,, \label{eq:xi_expansionCSC_alpha}
\end{align}
where
\be
\Phi(t) \simeq \int^tdt'\,\omega\,.
\ee
\end{subequations}
The function $\alpha$ is also assumed to be relatively slowly-varying, that is
\be
\frac{\dot\alpha}{\alpha} \ll \omega\,.
\ee
In terms of the  quantities of~\rif{eq:xi_min_xi_max}, we have
\be\label{eq:xi_a_+-_alpha}
\xi_{min}\simeq \xi_a - \alpha\,,\quad \xi_{max} \simeq \xi_a+\alpha\,.
\ee

\subsection{Harmonic Regime}\label{sec:linearised}
We initially assume that the amplitude of oscillations is small enough that the potential can be approximated by a harmonic potential:
\be\label{eq:potential_approx_harmonic}
U(\xi,t) \simeq U_0(t)+\frac{1}{2}\,\omega^2(\xi-\xi_a)^2\,,
\ee
where $\omega$ was defined in~\rif{eq:omega_models} and as we can see from equations~\rif{eq:xi_models_explicit} and~\rif{eq:potentials_approx}
\be\label{eq:U0}
U_0(t) \equiv U\left(\xi_a(t)\right) = 3\lambda \R_c\left(-\frac{\R_c}{\tilde\T(t)}\right)^{2n}\,.
\ee
This is equivalent to considering the first-order approximation in $\xi_1$ (defined in Eq.~\ref{eq:xi_expansionCSC_xi1}).  Equation~\rif{eq:traceCSC_xi_KG} then reads
\be\label{eq:ddot_xi_1}
\ddot\xi_1+\omega^2\xi_1\simeq -\ddot\xi_a\,.
\ee
Using the expansion \rif{eq:xi_expansionCSC_alpha} and neglecting $\ddot\xi_a$ and $\ddot\alpha$ yields
\be\label{eq:alpha_general_sol}
\frac{\dot\omega}{\omega}\simeq -2\frac{\dot\alpha}{\alpha}\quad\so\quad \alpha(t)\simeq \alpha_0\sqrt\frac{\omega_0}{\omega(t)}\,.
\ee
As long as the approximations hold, this can be considered a rather general result, and the specific $F(\R)$ model will determine the behaviour of the oscillations. The value $\alpha_0$ in Eq.~\rif{eq:alpha_general_sol} is strictly related to the initial conditions, that is to the initial displacement from the GR behaviour. We will fix the initial values of $\R$ and $\dot \R$, and from those derive the initial values of $\xi$ and $\dot\xi$. Thus, $\alpha_0$ can be calculated differentiating Eq.~\rif{eq:xi_expansionCSC_alpha}, yielding
\be
\dot\xi_0(\R_0,\dot \R_0) \simeq \dot\xi_{a,0}+\alpha_0\omega_0\,\,\so\,\,\alpha_0\simeq \frac{\dot\xi_0-\dot\xi_{a,0}}{\omega_0}\,.
\ee
This corresponds to the explicit solution
\be\label{eq:sol_alpha_explicit}
\alpha(t) \simeq \left(\dot\xi_0-\dot\xi_{a,0}\right)\left[\omega_0\,\omega(t)\right]^{-1/2}\,.
\ee
Please note that, apparently, we have not made use of the assumption $U(\xi_{sing})<\infty$ considered before. Although not necessary to perform calculations, this condition is needed to ensure that the expansion \rif{eq:xi_expansionCSC} be reliable. In fact, oscillations are harmonic only if the potential is nearly quadratic; this assumption is usually quite reasonable, especially near the bottom of the potential, but loses validity, for instance, near points at which $U$ diverges. Therefore, models in which $U(\xi)$ is singular in ${\xi=\xi_{sing}=\xi(|\R|\to\infty)}$ cannot be discussed within the framework of this Chapter. 

Also, it is clear from Eq.~\rif{eq:sol_alpha_explicit} that if $\dot\xi_0=\dot\xi_{\alpha,0}$ the amplitude of oscillations would vanish at all times. This can be immediately proved to be wrong, for instance numerically. This is an unfortunate consequence of the approximations used to derive \rif{eq:sol_alpha_explicit}, particularly neglecting $\ddot\xi_a$ in \rif{eq:ddot_xi_1}; evidently, the source term $\ddot\xi_a\neq 0$ will produce oscillations regardless of the initial conditions. When $\alpha$ is initially very small, $\ddot\xi_a$ and in general terms proportional to $1/t_{contr}^2$ should be kept and the approximations used are no longer valid. Therefore, Eq.~\rif{eq:sol_alpha_explicit} is reliable when $(\dot\xi_0-\dot\xi_{\alpha,0})$ is ``large'' enough, say of the order of $\dot\xi_{\alpha,0}$.

In order to have simple and more or less reliable estimates, we will use the initial conditions
\be\label{eq:initial_conditions}
\begin{cases}
 \R_0=-\tilde\T_0 \\
 \dot \R_0 = -\kappa\,\dot{\tilde\T_0} = -\kappa\,\T_0/t_{contr}\,,
\end{cases}
\ee
where $\kappa$ is a free parameter quantifying the initial displacement from the GR behaviour $\R+\tilde\T=0$; in particular, $\kappa=1$ corresponds to the situation in which $\R$ initially behaves exactly as if there were no $F(\R)$ at all (but still a cosmological constant). For simplicity, only change the initial ``velocity'' $\dot \R_0$.\\
Because of the considerations made above and noting that with these initial conditions
\be
\dot\xi_0 = \kappa\,\dot\xi_{a,0}\,,
\ee
our results will be particularly reliable for values of $\kappa$ not too close to unity.

\begin{figure}[ht]
{
\includegraphics[width=\mywidthdouble]{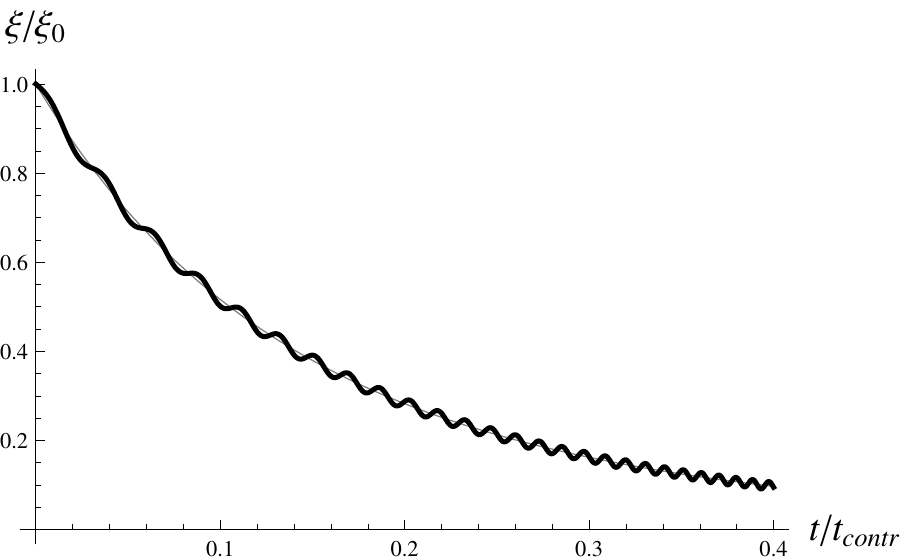}
\includegraphics[width=\mywidthdouble]{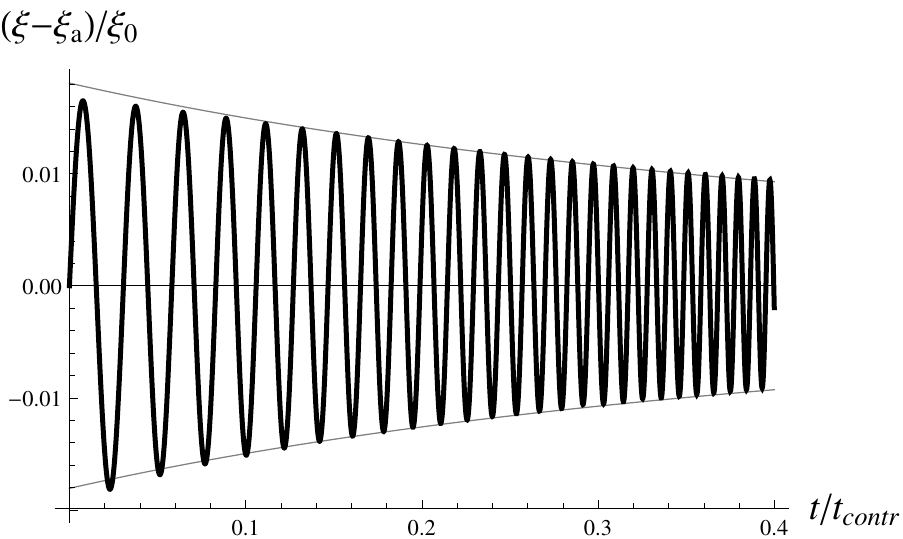}}\\
{\includegraphics[width=\mywidthdouble]{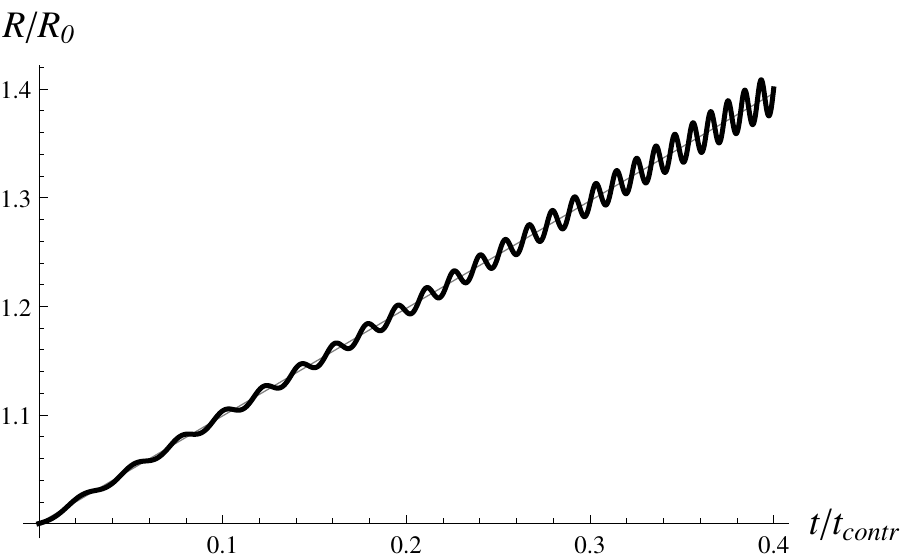}
\includegraphics[width=\mywidthdouble]{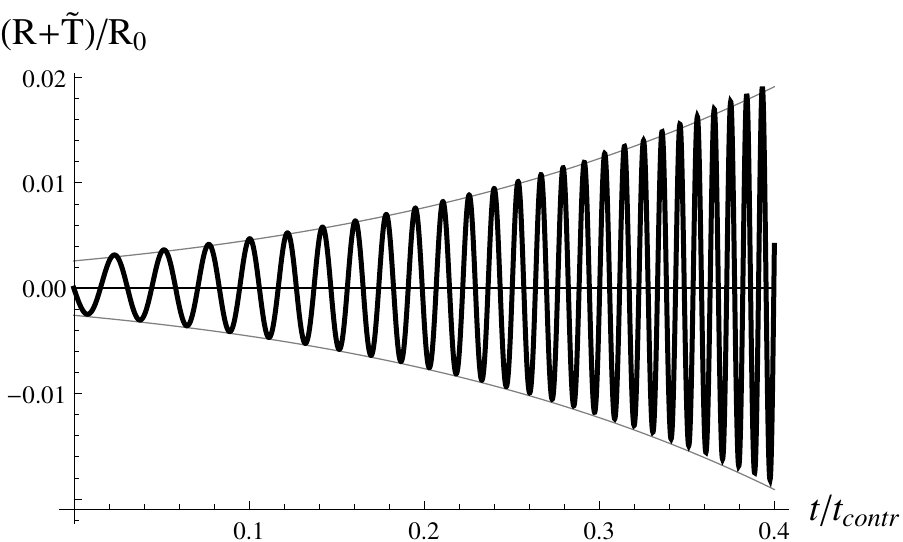}}
\caption[Comparison of the evolution of $\xi$ for the Hu-Sawicki model with the predicted result, see equations~\rif{eq:xi_expansionCSC} and \rif{eq:sol_alpha_explicit}. The values of parameters used are: ${n=3}$, ${\lambda =\R_{29}=1}$, ${\rho_{29}=2\cdot10^2}$ and ${t_{10}=1\cdot10^{-6}}$, ${\kappa=0.5}$.]{Comparison of the evolution of $\xi$ for the Hu-Sawicki model with the predicted result, see equations~\rif{eq:xi_expansionCSC} and \rif{eq:sol_alpha_explicit}. The values of parameters used are: ${n=3}$, ${\lambda =\R_{29}=1}$, ${\rho_{29}=2\cdot10^2}$ and ${t_{10}=1\cdot10^{-6}}$, ${\kappa=0.5}$. The value $n=3$ gives satisfactory result for these models in reproducing the known cosmological evolution. \it{Top left:} numerical solution for $\xi$ (black), compared to the ``average'' value $\xi_a$ (gray), defined in~\rif{eq:xi_a_def}, normalised in units of $\xi_0$. \it{Top right:} plot of $\xi_1$~\rif{eq:xi_expansionCSC}, compared to the expected evolution of the amplitude $\alpha$ \rif{eq:alpha_models_explicit}, normalised to $\xi_0$. \it{Bottom left:} evolution of $\R/\R_0$ with time (black), compared to the external energy/mass density (gray). Note that~\rif{eq:T_evol} is assumed. \it{Bottom right}: oscillations of $\R$ around its GR value $\R=-\tilde\T$; the amplitude grows with time following~\rif{eq:beta_fast_roll} quite accurately.}
\label{fig:xi_evol}
\end{figure}

\begin{figure}[tb]
\centering
 \includegraphics[width=\mywidthsingle]{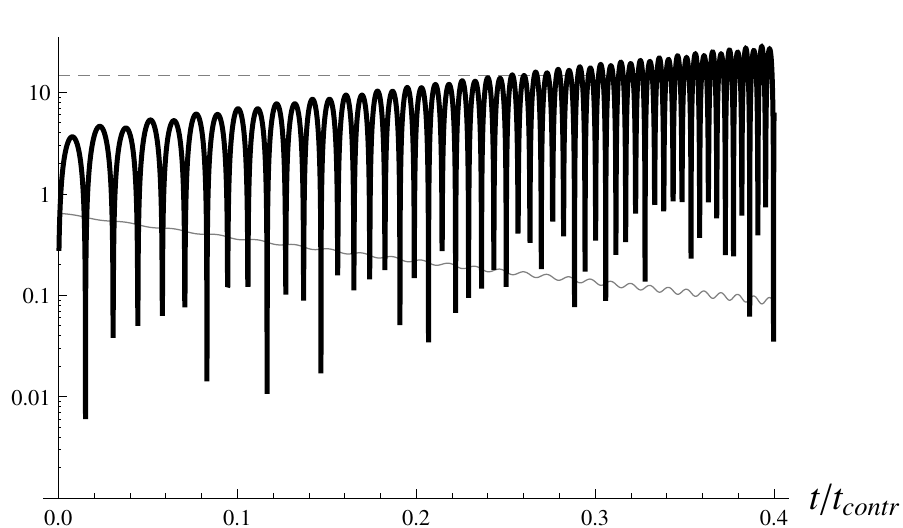}
\caption[Comparison of the different terms in~\rif{eq:traceCSC}; the lines are, respectively, $|\R+\T|$ (black), $|2F|$ (dashed) and $10^{12}|F_{,R}\R|$ (gray), and parameters are those of Fig.~\ref{fig:xi_evol}.]{Comparison of the different terms in~\rif{eq:traceCSC}; the lines are, respectively, $|\R+\T|$ (black), $|2F|$ (dashed) and $10^{12}|F_{,R}\R|$ (gray), and parameters are those of Fig.~\ref{fig:xi_evol}. Notice that, as expected, $F_{,R}\R$ is absolutely negligible, whereas $F$ is of the order of $\R+\T$. Still, since it only produces an almost constant offset in the GR solution, there is no appreciable effect on solutions as indicated by fig.~\ref{fig:xi_evol}. For details, see the discussion above and below Eq.~\ref{eq:terms_com}.}
\label{fig:terms_comparison}
\end{figure}

Using these initial conditions, the amplitude of the scalaron oscillations for models~\rif{eq:models} evolves as (see also Eq.~\ref{eq:omega_models}):
\begin{subequations}\label{eq:alpha_models_explicit}
\begin{align}
\alpha(t) &\simeq \frac{\left[6n\lambda(2n+1)\right]^\frac{3}{2}|1-\kappa||\R_c|^{3n+\frac{3}{2}}\,\dot{\tilde\T_0^2}}{\tilde \T_0^\frac{5(n+1)}{2}\,\tilde\T(t)^{(n+1)/2}}\,,\\
& = \frac{\left[6n\lambda(2n+1)\right]^\frac{3}{2}|1-\kappa||\R_c|^{3n+\frac{3}{2}}\,T_0^2}{\tilde \T_0^\frac{5(n+1)}{2}\,\tilde\T(t)^\frac{n+1}{2}\,t_{contr}^2}\,.
\end{align}
\end{subequations}
Accordingly, $\R$ oscillates around its GR value $\R=-\tilde\T$. We thus define
\be\label{eq:R_expansion}
\R(t) = -\tilde\T + \beta\,\R_0\,r_{osc}\,,
\ee
where $r_{osc}$ has maximum absolute value equal to 1 and contains all the information about the oscillations of curvature, whereas the dimensionless function $\beta$ contains the information about the amplitude of such oscillations. Using~\rif{eq:xi_models_explicit} and~\rif{eq:alpha_models_explicit}, and expanding at linear order in $\beta$, we find
\be\label{eq:beta_fast_roll}
|\beta| \simeq \frac{\tilde\T^{2n+2}\,|\alpha|}{6n\lambda(2n+1)\,\tilde\T_0\,|\R_c|^{2n+1}}\,.
\ee
In figure \ref{fig:xi_evol} we show a comparison between the numerical solutions of~\rif{eq:traceCSC_xi_KG} and our estimates, using the approximate model~\rif{eq:model_approxCSC}. With the chosen values of parameters, we have
\be\label{eq:omega_t_contr}
\frac{\omega_0\,t_{contr}}{2\pi}\simeq 30\,,
\ee
so the fast-roll condition~\rif{eq:fast-roll-condition} is satisfied. The agreement between our analytical estimates and numerical results is expected to improve as ${\omega_0\,t_{contr}}$ increases. In Fig.~\ref{fig:terms_comparison} we show a comparison of the various terms in~\rif{eq:traceCSC}.

\subsection{Approaching the Singularity: Anharmonic Oscillations}
As $\xi$ decreases and $|\R|$ increases, $\alpha$ grows so that eventually there appear anharmonic features in the oscillations of the scalaron. This is due to the fact that, when $\alpha$ becomes of the order of $\xi_a$, the higher-order terms in the potential, which had been neglected in~\rif{eq:potential_approx_harmonic}, become important. The different shape of the potential~\rif{eq:potentials_approx} on the left and on the right of the bottom (see also figure~\ref{fig:potentials}) determines an asymmetry of oscillations around the expected average value $\xi_a$. In particular, it is easy to infer that, redefining
\be
\xi_{min}\equiv \xi_a-\alpha_-\,,\qquad \xi_{max}\equiv \xi_a+\alpha_+\,,
\ee
we should have ${\alpha_-<\alpha_+}$, because the potential is steeper for ${\xi<\xi_a}$ than it is for ${\xi>\xi_a}$. Note that in the harmonic regime we assumed (see Eq.~\ref{eq:xi_a_+-_alpha})
\be
\xi_{max}-\xi_a = \xi_a-\xi_{min} =\alpha\,.
\ee
The variation of $\alpha$ is caused by the change in the shape of the potential with time and the increasing ``energy'' of the field, in the sense of equation~\rif{eq:energy_conservation}. In the harmonic region, using~\rif{eq:xi_min_xi_max} and~\rif{eq:potential_approx_harmonic} yields
\be
U(\xi_{max})\simeq U(\xi_{min}) \simeq U_0+\frac{1}{2}\,\omega^2\alpha^2\equiv U_0+\Delta U\,.
\ee
Note that all quantities involved are functions of time. The term $\Delta U$, if we neglect the integral term in~\rif{eq:energy_conservation}, corresponds to the maximum value of ${\dot\xi^2/2}$, that is the value this term has  when the field is at the bottom of the potential. Since we are basically considering a classical harmonic oscillator, this is an expected result. Substituting the explicit values, we find
\be\label{eq:delta_U}
\Delta U \simeq \frac{18\left[n\lambda(2n+1)(1-\kappa)\right]^2|\R_c|^{4n+2}\,\tilde\T^{n+1}\,\dot{\tilde\T_0}^2}{\tilde\T_0^{5n+5}}\,.
\ee
As mentioned before, this result depends essentially on the variation of the shape of the potential and on the increase of the energy of $\xi$, not on the assumption of harmonicity. Therefore, we will assume that $\Delta U$ continues to follow~\rif{eq:delta_U} even \it{away} from the harmonic region. In particular, we are interested in the region very close to the singularity, namely $\xi_a\simeq \alpha$. We will see numerically that this assumption is in good agreement with exact results.

Near the singularity, the term in the potential~\rif{eq:potentials_approx} linear in $\xi$ goes to zero more rapidly than the other term, so it can be neglected; therefore, the request that
\be
U(\xi_a-\alpha_-) = U_0 + \Delta U\,,
\ee
using equation~\rif{eq:xi_min_xi_max}, leads to the solution
\be\label{eq:alpha_singul_explicit}
\alpha_-(t) \simeq \xi_a(t)-6n\lambda\left[\frac{U_0(t)+\Delta U(t)}{3\lambda \R_c(2n+1)}\right]^\frac{2n+1}{2n}\,.
\ee
The explicit forms of $U_0$ and $\Delta U$ for the models considered are given, respectively, by equations~\rif{eq:U0} and~\rif{eq:delta_U}.

\begin{figure}[tb]
\centering
\includegraphics[width=\mywidthsingle]{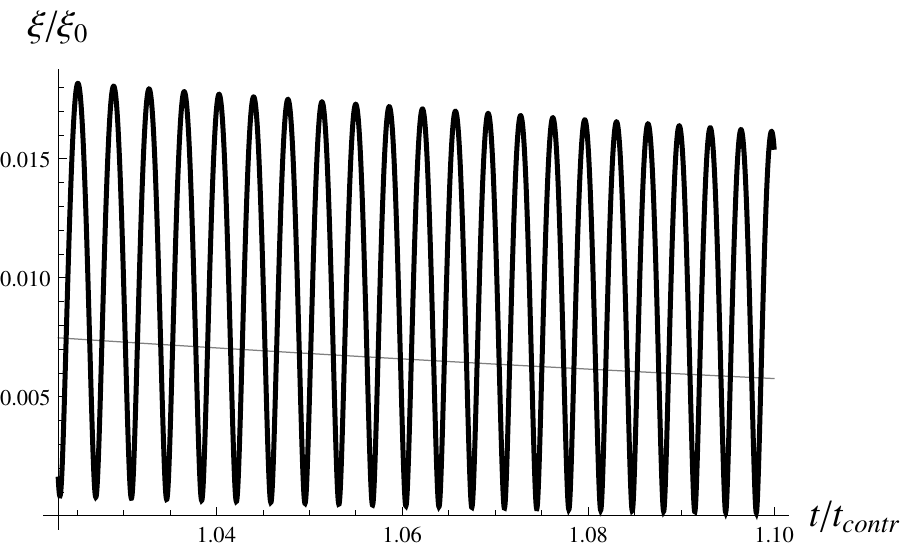}
\includegraphics[width=\mywidthsingle]{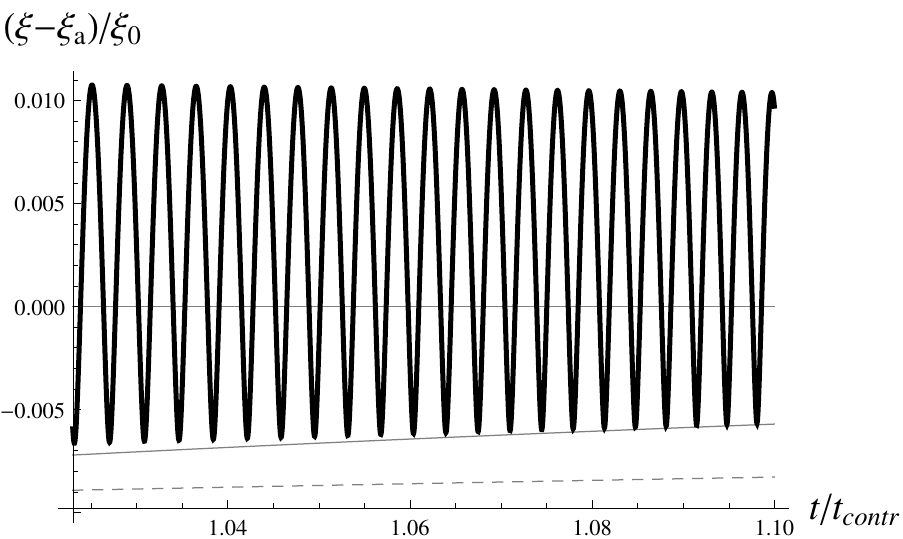}
\caption[Numerical solution for: $n=3$, $\lambda=\R_{29}=1$, $\rho_{29}=2\cdot 10^2$, $t_{10}=5\cdot10^{-7}$.]{Numerical solution for: $n=3$, $\lambda=\R_{29}=1$, $\rho_{29}=2\cdot 10^2$, $t_{10}=5\cdot10^{-7}$. \it{Top Panel:} as $\xi$ approaches the singularity, its oscillations around $\xi_a$ start showing anharmonic features.  \it{Bottom Panel:} the lower values of $\xi-\xi_a$ are clearly different from the naively predicted value~\rif{eq:alpha_models_explicit}, the dashed line in the picture, whereas the refined result of Eq.~\ref{eq:alpha_singul_explicit} (thin, solid line) is in very good agreement with the numerical solution.}
\label{fig:alpha_near_singularity}
\end{figure}

In figure~\ref{fig:alpha_near_singularity}, we show $\xi$ approaching the singularity, with ${\xi_{min}\ll \xi_a}$. As expected, the old estimate~\rif{eq:alpha_models_explicit} no longer reproduces the behaviour of numerical solutions, whereas the new result~\rif{eq:alpha_singul_explicit} works very well.

\subsection{Generation of the Singularity}\label{sec:fast_roll_singularity}
We are now ready to make the final calculations in order to derive the critical energy/mass density $\T_{sing}$ corresponding to the curvature singularity. We can either use~\rif{eq:alpha_singul_explicit} or equivalently the condition
\[
U_0+\Delta U = U(\xi_{sing}) = 0,
\]
to obtain
\begin{subequations}\label{eq:sing_fast_correct}
\be\label{eq:T_sing_fast_roll}
\begin{aligned}
\frac{\tilde\T_{sing}}{\tilde\T_0} &= \left[\frac{\tilde\T_0^{2n+4}}{6\lambda n^2(2n+1)^2(1-\kappa)^2\,|\R_c|^{2n+1}\,\dot{\tilde\T_0}^2}\right]^\frac{1}{3n+1}\\
&\simeq \left[0.28\,\frac{\rho_{29}^{2n+2}\,t_{10}^2\,(1+2\lambda \R_{29}/\rho_{29})^{2n+4}}{\lambda n^2(2n+1)^2(1-\kappa)^2\,\R_{29}^{2n+1}}\right]^\frac{1}{3n+1}\,.
\end{aligned}
\ee
The corresponding timescale for the formation of the singularity, using equation~\rif{eq:T_evol} and~\rif{eq:T_correction}, is simply
\be\label{eq:t_sing_fast_roll}
\frac{t_{sing}}{t_{contr}} = \frac{\T_{sing}}{\T_0}-1 = \frac{\tilde\T_{sing}-4\Lambda}{\tilde\T_0-4\Lambda}-1\,.
\ee
\end{subequations}
In table~\ref{tab:T_sing_fast_roll}, we show a comparison of a few analytical estimates with exact numerical results. The agreement increases with increasing $t_{contr}$ (and increasing $\T_{sing}$), which in fact corresponds to the situation in which the assumptions of adiabaticity are particularly reliable, see for instance Eq.~\ref{eq:fast-roll-condition}.\\
Even for the smallest values of $t_{contr}$ considered in Tab.~\ref{tab:T_sing_fast_roll}, the discrepancy between the numerical and analytical values is at most a few percent. Notice that this is a considerable and perhaps surprising result, since for the first value in table~\ref{tab:T_sing_fast_roll} ($t_{10}=1\cdot 10^{-6}$) we have
\[
\frac{\omega_0\,t_{contr}}{2\pi}\simeq 2\,,
\]
so that the condition~\rif{eq:fast-roll-condition} is actually barely fulfilled, and yet the analytical estimates work more than sufficiently well.
\begin{table*}[bt]
\centering
\begin{tabular}{c l l c c l l}
\toprule\noalign{\smallskip}
 &\multicolumn{2}{c}{$\tilde\T_{sing}/\tilde\T_0$} &  &  &\multicolumn{2}{c}{$\tilde\T_{sing}/\tilde\T_0$} \\
\noalign{\smallskip}
\cline{2-3}\cline{6-7}
\noalign{\smallskip}
$t_{10}$ & \multicolumn{1}{c}{Eq.~\rif{eq:sing_fast_correct}} & \multicolumn{1}{c}{exact} & & $t_{10}$ & \multicolumn{1}{c}{Eq.~\rif{eq:sing_fast_correct}} &  \multicolumn{1}{c}{exact}\\
\noalign{\smallskip}
\noalign{\smallskip}
$1\cdot 10^{-6}$ & 1.40877 & 1.56622 & & $1.0\cdot 10^{-5}$ & 2.23276  & 2.23157 \\
$2\cdot 10^{-6}$ & 1.61826 & 1.66162 & & $1.2\cdot 10^{-5}$ & 2.31567  & 2.3125 \\
$3\cdot 10^{-6}$ & 1.75495 & 1.77693 & & $1.4\cdot 10^{-5}$ & 2.38818  & 2.38475 \\
$4\cdot 10^{-6}$ & 1.85889 & 1.87618 & & $1.6\cdot 10^{-5}$ & 2.45282  & 2.44829 \\
$5\cdot 10^{-6}$ & 1.94373 & 1.95447 & & $1.8\cdot 10^{-5}$ & 2.51128  & 2.5066 \\
$6\cdot 10^{-6}$ & 2.01591 & 2.02037 & & $2.0\cdot 10^{-5}$ & 2.56476  & 2.55936 \\
$7\cdot 10^{-6}$ & 2.07903 & 2.08139 & & $2.5\cdot 10^{-5}$ & 2.68182  & 2.67566 \\
$8\cdot 10^{-6}$ & 2.1353  & 2.13753 & & $3.0\cdot 10^{-5}$ & 2.78141  & 2.77482 \\
$9\cdot 10^{-6}$ & 2.1862  & 2.18478 & & $5.0\cdot 10^{-5}$ & 3.0806   & 3.07293 \\

\noalign{\smallskip}\bottomrule
\end{tabular}
\caption{Critical energy/mass density $\T_{sing}$ obtained using~\rif{eq:T_sing_fast_roll}, compared to the exact numerical result. Parameters are $n=3$, $\lambda=\R_{29}=1$, $\kappa=0.5$, $\rho_{29}=10^2$, so results depend on the value of $t_{contr}$.}
\label{tab:T_sing_fast_roll}
\end{table*}

\begin{figure}[tb]
\centering
\includegraphics[width=\mywidthsingle]{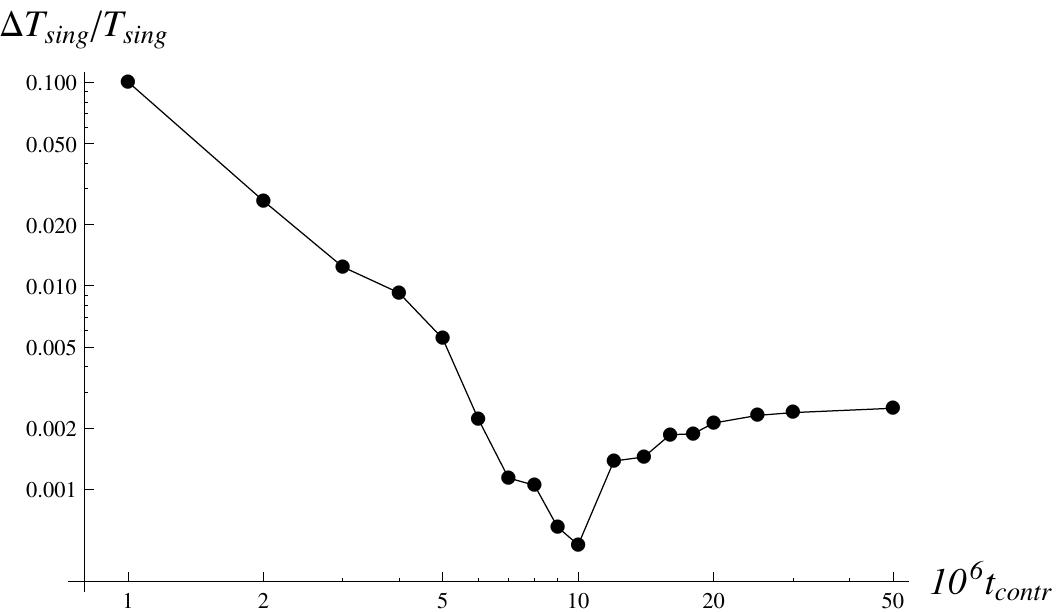}
\caption[Relative errors of table~\ref{tab:T_sing_fast_roll}.]{Relative errors of table~\ref{tab:T_sing_fast_roll}. The discrepancy between analytical estimates and exact values decreases as $t_{contr}$ increases, but tends to a constant value of about 0.2\%. The ``cusp'' at small errors at $t_{10}\simeq 10^{-5}$ corresponds to a change in sign of $\Delta \T_{sing}/\T_{sing}$. See the text for further details.}
\label{fig:sing_error_fast_roll}
\end{figure}

In general, the accuracy of our analytical estimate should increase with increasing $\omega_0\,t_{contr}$, as the adiabatic approximation is more and more accurate. Instead, the relative errors seem to tend asymptotically (mind the logarithmic scale) to a fixed, though quite small, value ${~0.2\%}$, for which we have not found an explanation; possibly, this feature could be due to numerical computation issues. Anyway, the accuracy of our analytical estimates is good enough for all practical purposes.

One could argue that we should have displayed results for longer $t_{contr}$ and perhaps larger $\rho_0$, which are physically more realistic. Unfortunately, exploring that range of parameters is almost prohibitive from a computational standpoint, due to the massive number of oscillations occurring until $t$ reaches $t_{sing}$. Using equations~\rif{eq:omega_models} and~\rif{eq:sing_fast_correct} and assuming for simplicity $t_{sing}/t_{contr}>1$, we obtain in fact
\be
N_{osc} \simeq \int_{t_0}^{t_{sing}}\omega\,dt \propto \left(\rho_{29}^{n+1}\,t_{10}\right)^\frac{5n+5}{3n+1}\,,
\ee
so even a small increase in $\rho_{29}$ and/or $t_{10}$, especially for large $n$, can lead to an enormous increase in the time required for computations.

Nevertheless, we have no reason to believe that the satisfactory agreement of our estimates and numerical results would not hold in the case of more realistic values of parameters.

\section{Slow-Roll Region}\label{sec:xi_stuck}
Let us relax the assumptions of adiabaticity of section \ref{sec:adiabatic}, and focus instead on the opposite regime, that is
\be\label{eq:slow_roll_condition}
\frac{\omega\,t_{contr}}{2\pi}\ll 1\,,
\ee
corresponding to Eq.~\rif{eq:fast_roll_inequality} with inverted inequality sign. This is a slow-roll regime, in which the initial ``velocity'' of the field dominates over the acceleration due to the potential. In first approximation, assuming that $\dot\xi_0\neq 0$, which is equivalent to $\kappa\neq 0$ in~\rif{eq:initial_conditions}, we have
\be\label{eq:xi_slowroll_linear}
\xi(t) \simeq \xi_0+\dot\xi_0\,t\,.
\ee
Notice that $\dot\xi_0 < 0$. This behaviour, i.e. the fact that $\xi$ is roughly linear in $t$, can also be understood as follows. Considering equation~\rif{eq:traceCSC_xi_KG}, we see that neglecting ${\R+\tilde\T}$ we are left with
\be\label{eq:traceCSC_xi_slowroll}
\ddot\xi = 0\,,
\ee
which has exactly the solution~\rif{eq:xi_slowroll_linear}. This does not mean that we are precisely sitting on the solution $\R+\tilde\T=0$, because $\ddot\xi(\R+\tilde\T=0)\equiv\ddot\xi_a\neq 0$; otherwise, we would not have any singularity since $\R$ would simply follow the smooth evolution of $\tilde\T$. Rather, it means that~\rif{eq:traceCSC_xi_slowroll} coincides with~\rif{eq:traceCSC_xi_KG} up to corrections of order $\R+\tilde\T$. Since we can estimate
\be
\ddot\xi \sim \frac{\xi}{t_{contr}^2}\,,
\ee
and
\be
\R+\tilde\T = \frac{\partial U}{\partial\xi} \sim \omega^2\xi\,,
\ee
we find that
\be
\frac{\ddot\xi}{\R+\tilde\T} \sim \frac{1}{\omega^2t_{contr}^2} \gg 1\,.
\ee
This means that~\rif{eq:traceCSC_xi_KG} and~\rif{eq:traceCSC_xi_slowroll}, in this regime, are equal provided that we neglect terms of order $(\omega\,t_{contr})^2$; this is a legitimate approximation when~\rif{eq:slow_roll_condition} holds.

The reader may compare this to the assumptions of Sec.~\ref{sec:adiabatic}, where instead we had neglected terms ${\propto t_{contr}^{-2}}$. In that regime, the dominant contribution to $\ddot\xi$ was oscillatory, with
\[
\ddot\xi_{adiab} \sim \omega^2\xi\,,
\]
because we had $\omega\,t_{contr}\gg 1$.

\begin{figure}[tb]
\centering
\includegraphics[width=\mywidthsingle]{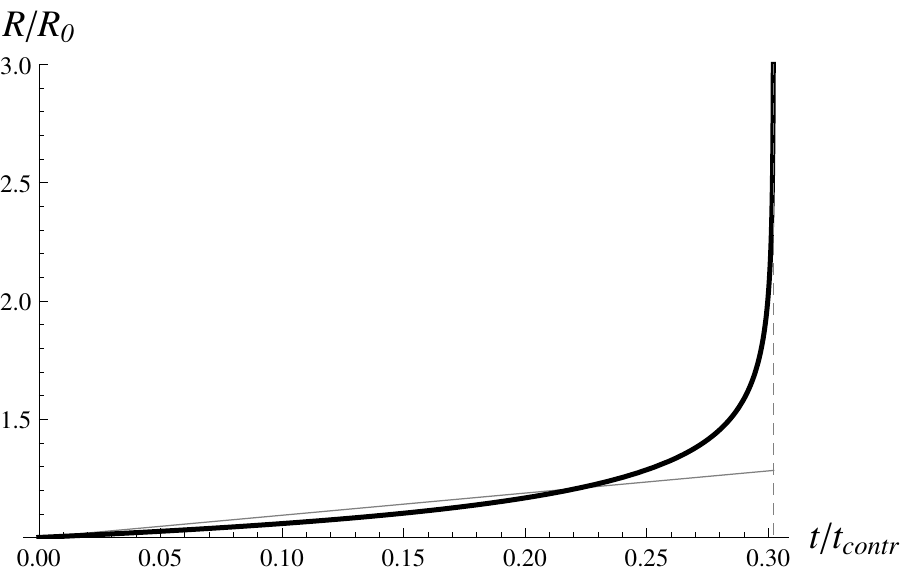}
\includegraphics[width=\mywidthsingle]{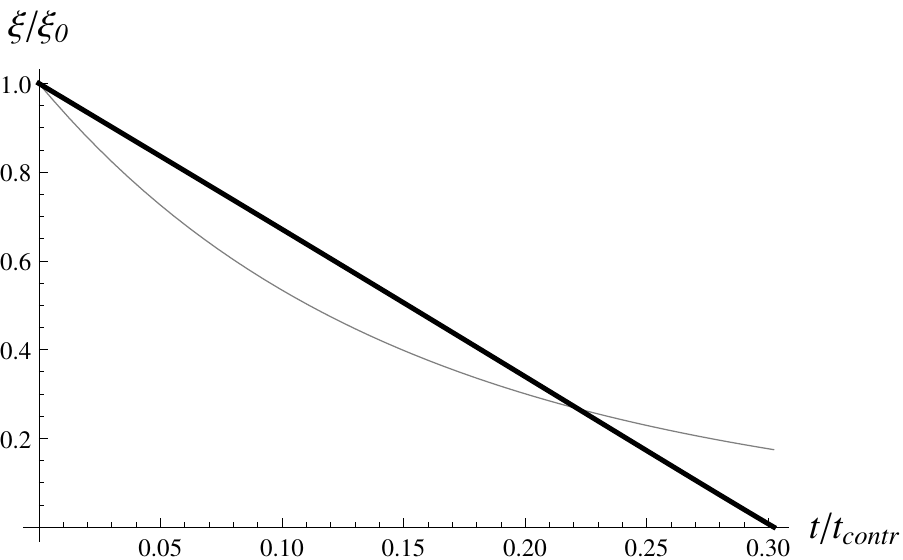}
\caption[Numerical results for parameters: ${n=3}$, ${\lambda=\R_{29}=1}$, ${\rho_{29}=30}$, ${t_{10}=1\cdot 10^{-5}}$, ${\kappa=0.5}$.]{Numerical results for parameters: ${n=3}$, ${\lambda=\R_{29}=1}$, ${\rho_{29}=30}$, ${t_{10}=1\cdot 10^{-5}}$, ${\kappa=0.5}$. Initially, ${\omega_0\,t_{contr}/2\pi\simeq 0.2}$, so the condition~\rif{eq:slow_roll_condition} is narrowly fulfilled. \it{Top Panel:} only a portion of $\R/\R_0$ (which diverges) is shown, in order to have a comparison with $\tilde\T/\tilde\T_0$ (thin, solid line); the singularity is reached at ${t_{sing}/t_{contr}\simeq 0.3}$ (dashed vertical line; see Tab.~\ref{tab:T_sing_slow_roll}). Initially, the slope of $\R/\R_0$ is different from that of $\tilde\T/\tilde\T_0$ because $\kappa\neq 1$. \it{Bottom panel}: $\xi$ does not follow $\xi_a$ (gray line) at all, but rather decreases roughly linearly with time, in qualitative agreement with~\rif{eq:xi_slowroll_linear}.}
\label{fig:singularity}
\end{figure}

\subsection{Generation of the Singularity}
With the simple solution~\rif{eq:xi_slowroll_linear}, it is straightforward to see that $\xi$ reaches the singularity $\xi_{sing}=0$ at
\be
t_{sing} \simeq -\frac{\xi_0}{\dot\xi_0}\,,\quad \tilde\T_{sing}\simeq \T_0\left(1-\frac{\xi_0}{\dot\xi_0\,t_{contr}}\right)+4\Lambda\,.
\ee
Using the explicit expressions for the models under consideration, we obtain the very simple expression
\begin{subequations}\label{eq:sing_slow_correct}
\be\label{eq:t_sing_slow_roll}
\frac{t_{sing}}{t_{contr}} \simeq \frac{1+4\Lambda/\T_0}{(2n+1)\kappa}\,,
\ee
or equivalently
\be\label{eq:T_sing_slow_roll}
\frac{\tilde\T_{sing}}{\tilde\T_0} \simeq 1 + \frac{1}{(2n+1)\kappa}\,.
\ee
\end{subequations}
In figure~\ref{fig:singularity}, we show the typical behaviour of $\xi$ and $R$ in this regime, until the singularity.
\hyphenation{initially}

\begin{figure}[tb]
\centering
\includegraphics[width=\mywidthsingle]{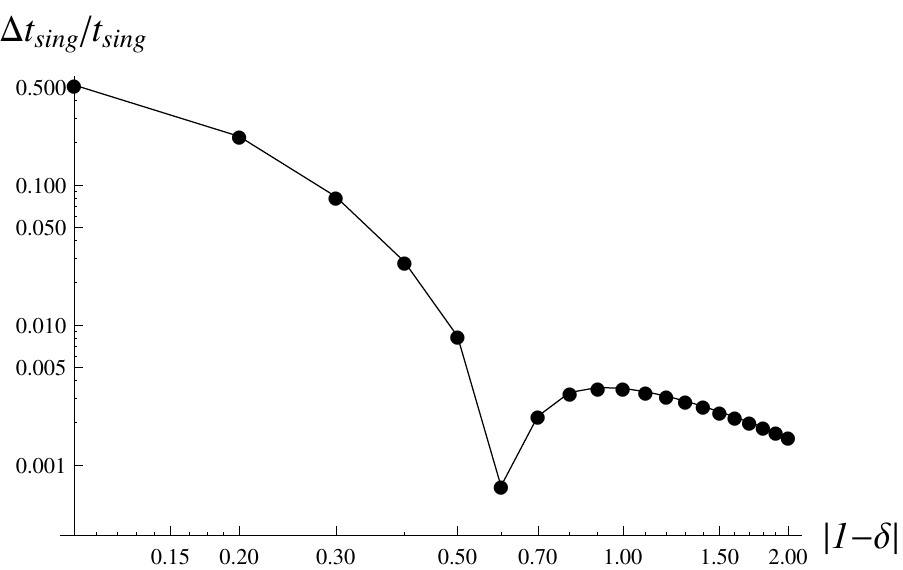}
\caption[Relative errors of table~\ref{tab:T_sing_slow_roll}.]{Relative errors of table~\ref{tab:T_sing_slow_roll}. Noticeably, only for very small values of $\kappa$, say less than a few percent, there is an appreciable difference between our estimate and the exact result.}
\label{fig:sing_error_slow_roll}
\end{figure}

Basically, we are assuming that the motion of $\xi$ is completely dominated by the initial conditions, and that the acceleration due to the potential is negligible. Of course, for $\kappa\to 0$ the approximation loses its validity because the initial velocity is practically zero, but this is not worrisome because the most physically sensible choices are those with $\kappa\simeq 1$. The theoretical estimates of Tab.~\ref{tab:T_sing_slow_roll} are in remarkable agreement with the exact numerical values, and as expected the two results differ significantly only when $\kappa\ll 1$. The relative errors are depicted in Fig.~\ref{fig:sing_error_slow_roll}.

As expected, errors decrease for increasing values of $\kappa$, except for the region $\kappa\simeq 0.6$, which is most likely a numerical feature and should have no physical meaning. Nonetheless, the agreement between analytical and numerical values is excellent, especially considering that the slow-roll condition~\rif{eq:slow_roll_condition} is barely fulfilled, in fact ${\omega_0\,t_{contr}/2\pi\simeq 0.2}$.\\
The particular choice of parameters was motivated by the requirement of a somewhat realistic value of $t_{contr}$, in particular not too small. Taking larger values of $\rho_{29}$ and tuning $t_{10}$ to have, say, ${\omega_0\,t_{contr}/2\pi<10^{-2}}$ yields estimates in outstanding agreement with numerical calculations, because the approximations~\rif{eq:large_density_limit} and~\rif{eq:slow_roll_condition} are all the more accurate. As an example, consider:
\be
\begin{cases}
n=3\\
\lambda=\R_{29}=1\\
\rho_{29}=10^2\\
t_{10}=10^{-9}\\
\kappa=1
\end{cases}
\quad \so \quad \frac{\omega_0\,t_{contr}}{2\pi}\simeq 0.2\%\,,
\ee
which gives the terrific value
\be
\frac{\Delta t_{sing}}{t_{sing}}\simeq  3\cdot 10^{-7}\,.
\ee
The price to pay, however, is to have unnaturally small contraction times, for instance ${t_{contr}=10}$ years in this case, therefore further similar results were not explicitly shown. Still, it is good to notice that the mathematical accuracy of our estimates improves as expected.

\begin{table*}[t!]
\centering
\begin{tabular}{c l l c c l l}
\toprule\noalign{\smallskip}
 &\multicolumn{2}{c}{$t_{sing}/t_{contr}$} &  &  & \multicolumn{2}{c}{$t_{sing}/t_{contr}$}\\
\noalign{\smallskip}
\cline{2-3} \cline{6-7}
\noalign{\smallskip}
\multicolumn{1}{c}{$\kappa$} & \multicolumn{1}{c}{Eq.~\rif{eq:sing_slow_correct}} & \multicolumn{1}{c}{exact} &   & \multicolumn{1}{c}{$\kappa$} & \multicolumn{1}{c}{Eq.~\rif{eq:sing_slow_correct}} & \multicolumn{1}{c}{exact}\\
\noalign{\smallskip}
\noalign{\smallskip}
0.1  & 1.52381   & 0.73706   &  & 1.1  & 0.138528  & 0.13899   \\
0.2  & 0.761905  & 0.592958  &  & 1.2  & 0.126984  & 0.127378  \\
0.3  & 0.507937  & 0.466093  &  & 1.3  & 0.117216  & 0.117551  \\
0.4  & 0.380952  & 0.370184  &  & 1.4  & 0.108844  & 0.109129  \\
0.5  & 0.304762  & 0.302237  &  & 1.5  & 0.101587  & 0.101831  \\
0.6  & 0.253968  & 0.253789  &  & 1.6  & 0.0952381 & 0.0954476 \\
0.7  & 0.217687  & 0.218171  &  & 1.7  & 0.0896359 & 0.0898169 \\
0.8  & 0.190476  & 0.191103  &  & 1.8  & 0.0846561 & 0.0848133 \\
0.9  & 0.169312  & 0.169917  &  & 1.9  & 0.0802005 & 0.0803378 \\
1.0  & 0.152381  & 0.152918  &  & 2.0  & 0.0761905 & 0.076311  \\


\noalign{\smallskip}\bottomrule
\end{tabular}
\caption{Comparison of analytical estimates and numerical results for $t_{sing}/t_{contr}$. The parameters used are: ${n=3}$, ${\lambda=\R_{29}=1}$, ${\rho_{29}=30}$, ${t_{10}=1\cdot 10^{-5}}$.}
\label{tab:T_sing_slow_roll}
\end{table*}

\section{Remarks on the Validity of the Approximations Used}\label{sec:approx}
\subsection{Low Gravity: $g_\mn\simeq \eta_\mn$}
At the beginning of section~\ref{sec:CURV_EVOL}, we made the substitution
\be
\D_\mu\D^\mu = g^\mn\D_\mu\D_\nu \to \square = \eta^\mn\partial_\mu\partial_\nu \to \partial_t^2\,,
\ee
assuming the homogeneity of the cloud and low gravity. The latter approximation is usually quite reasonable for astronomical densities, except for compact stars. However, one may be argue that if $\R\to\infty$, even with relatively low $\rho$, we are no longer in low-gravity regime, and thus the approximation fails. In order to show that $g_\mn\simeq \eta_\mn$ and $\R\to\infty$ are compatible, let us assume the simple homogeneous, isotropic line element
\be
ds^2 = dt^2 - \left[1+\psi(t)\right]d\mb x^2\,,\qquad |\psi|\ll 1\,,
\ee
with $\psi(t)$ parametrising the deviation from Minkowski. In this case, the scalar curvature is
\be\label{eq:R_psi}
\R = -\frac{3\ddot\psi}{1+\psi} \simeq -3\ddot\psi\,,
\ee
so we have
\begin{align}
\psi(t_{sing}) &\simeq -\frac{1}{3}\int^{t_{sing}}_{t_0}dt\int^{t}_{t_0}dt'\,\R(t') \notag\\
&\simeq \frac{1}{3}\int_{t_0}^{t_{sing}}dt\int^{t}_{t_0}dt'\,\tilde\T(t') \notag\\
&\simeq \frac{\tilde\T_0\,t_{sing}^2}{6}+\frac{\T_0\,t_{sing}^3}{18\,t_{contr}}\,\notag\\
&\simeq 0.28\,\rho_{29}\,t_{10}^2\,x^2\left(1 +\frac{x}{3}\right)\,.\label{eq:psi_sing}
\end{align}
where we have expanded $\R$ as in Eq.~\rif{eq:R_expansion}, neglected $\Lambda$ and defined
\be\label{eq:X_def}
x = \frac{t_{sing}}{t_{contr}}\,.
\ee
Low-gravity corresponds, roughly, to having~\rif{eq:psi_sing} smaller than unity. Firstly, we focus on the case $x\lesssim 1$, and assume that we are in the fast-roll (adiabatic) regime. Since in this case we can use Eq.~\rif{eq:sing_fast_correct}, the condition $x\lesssim 1$ becomes
\be
\frac{0.28\,\rho_{29}\,t_{10}^2}{\lambda n^2(2n+1)^2(1-\kappa)^2}\left(\frac{\rho_{29}}{\R_{29}}\right)^{2n+1}\lesssim 2^{3n+1}\,,
\ee
so that ${\rho_{29}\,t_{10}^2 \lesssim 1}$ as well, since ${\rho_{29}\gg \R_{29}}$. Therefore, 
\be
\psi \sim \rho_{29}\,t_{10}^2\,x < 1\,.
\ee
If on the other hand ${x\gtrsim 1}$, the condition ${\psi\lesssim 1}$ yields roughly
\be
\rho_{29}\,t_{10}^2\,x^3\lesssim 10\,,
\ee
that is
\[
\rho_{29}^{9n+7} \lesssim 4\cdot 10^{3n+1}\frac{\left[\lambda n^2(2n+1)^2(1-\kappa)^2\R_{29}^{2n+1}\right]^3}{t_{10}^{6n+8}}\,.
\]
It is easy to check that for all explicit numerical results presented in the text, this condition is very well satisfied. One should also keep in mind that when $t_{sing}/t_{contr}\gg 1$ the behaviour $\rho\sim t$ and thus the results of this Chapter are in any case expected to be less reliable.

The discussion of the slow-roll regime is even more straightforward, since $x<1$ (see Eqs.~\ref{eq:slow_roll_condition} and~\ref{eq:t_sing_slow_roll}) and
\be
\rho_{29}\,t_{10}^2 \sim \omega_0\,t_{contr}\left(\frac{\R_{29}}{\rho_{29}}\right)^{2n+1} \ll 1\,,
\ee
so $\psi$ is safely smaller than unity.

\subsection{Negligible Pressure: $p\ll\rho$}
Let us now consider the assumption of pressureless dust. We can combine the trace equation~\rif{eq:traceCSC_approx} with the time-time component of the modified Einstein equations~\rif{eq:field_equations} assuming the equation of state $p=w\rho$, obtaining:
\be\label{eq:sys_Einstein}
\begin{cases}
\ddot\xi + \R + 2F = \cfrac{8\pi}{\mpl^2}\,(\rho -3 p) = \cfrac{8\pi(1-3w)\rho}{\mpl^2}\,,\\
\left(1-\cfrac{\xi}{3}\right)\R_{tt}-\cfrac{f}{2} = \cfrac{8\pi\,\rho}{\mpl^2}\,.\\
\end{cases}
\ee
Then, as in GR, the space-space equation is automatically fulfilled\footnote{For simplicity we have also assumed isotropy, that is ${\R_{xx} = \R_{yy} = \R_{zz}}$ and ${p_x=p_y=p_z}$, but the result can be easily generalised.}, hence:
\be
\left(1-\frac{\xi}{3}\right)\R_{ii}+\frac{f}{2}+\frac{\ddot\xi}{3}=\frac{8\pi p}{\mpl^2} = \frac{8\pi w \rho}{\mpl^2}\,.
\ee
This is true for any equation of state $w$, including of course the non-relativistic case $w=0$, which corresponds to assuming
\be
\frac{p}{\rho}\ll 1\quad\so\quad \T^\mu_\mu \simeq \frac{8\pi}{\mpl^2}\,\rho\,.
\ee
The mathematical consistency of the Einstein equations is therefore guaranteed regardless of the assumed equation of state. Physically, we know from statistical mechanics that for non-relativistic particles
\be
\frac{p}{\rho} \sim \frac{v^2}{3\,c^2}\,,
\ee
where $v$ is the typical velocity of the dust particles and $c$ is the speed of light. Given the total mass of the cloud $M$ and its radius
\be
r \simeq \left(\frac{3M}{4\pi\rho}\right)^{1/3}\,,
\ee
the velocity of the particles at time $t$ should approximately be
\be
v \sim |\dot r| \simeq \frac{r_0}{3\,t_{contr}(1+t/t_{contr})^{1/3}}\,.
\ee
Ultimately, this yields
\be
\frac{p}{\rho} \sim 10^{-9}\,\frac{M_{11}^{2/3}}{t_{10}^2\,\rho_{29}^{2/3}(1+t/t_{contr})^{8/3}}\,,
\ee
where
\be
M_{11} \equiv \frac{M}{10^{11}\,M_\odot} \simeq \frac{M}{2\cdot 10^{44}\,M_{11}\text{ g}}\,.
\ee
In basically any conceivable astronomical situation, except for very massive, rarefied clouds with short, perhaps unnatural contraction times, this quantity is much smaller than one, so that $p$ is indeed negligible.

We should be completely honest and point out that for the smaller values of $t_{contr}$ shown in table~\ref{tab:T_sing_fast_roll}, taking $M_{11}\sim 1$ gives a ratio $p/\rho>1$, which seems to invalidate the initial assumptions. However, because of the considerations at the end of section~\ref{sec:fast_roll_singularity}, we can disregard these problems provided that we carefully choose physically realistic parameters. In other words, some of the values in table~\ref{tab:T_sing_fast_roll} are unlikely to describe existing physical systems, but are nonetheless a useful indication of the accuracy of our analytical estimates.

\section{Discussion and Conclusions}
The possibility of curvature singularities in DE $f(\R)$ gravity models has been confirmed and studied in a rather simple fashion. The trace of the modified Einstein equations has been rewritten, under the simplifying assumptions of homogeneity, isotropy and low-gravity, as an oscillator equation for the scalaron field $\xi$, which moves in a potential $U$ depending on the external energy/mass density and thus on time. In the two models considered~\cite{Hu:2007nk,Starobinsky:2007hu}, the potential is finite in the point corresponding to the curvature singularity $|R|\to\infty$, that is $\xi_{sing}=0$; the energy conservation equation associated with $\xi$ indicates that the development of the singularity can be triggered by an increase in the external energy/mass density.

The ratio between the typical contraction time and the inverse frequency of the scalaron determines two distinct regimes. In the adiabatic regime the oscillations of $\xi$ are very fast compared to relevant variations of $U$, and such oscillations are almost harmonic. Performing a linear analysis, we have estimated the scalaron amplitude and frequency analytically. The singularity is expected to be reached when the amplitude of the oscillations of $\xi$ exceeds the separation between the ``average'' value $\xi_a$, which corresponds at each instant to the position of the bottom of the potential, and the singular point $\xi_{sing}$.\\
In the slow-roll regime, the typical oscillation time of the field $\xi$ is much longer than the typical contraction time, which also determines the timescale for significant changes in the potential. Thus, $\xi$ is mainly driven by its initial conditions, and the slope of the potential is not enough to stop the field from reaching the singular point. This may occur on relatively short timescales.

In both regimes, our analytical estimates and numerical results are in remarkable agreement (see tables~\ref{tab:T_sing_fast_roll} and~\ref{tab:T_sing_slow_roll}, and figures~\ref{fig:sing_error_fast_roll} and~\ref{fig:sing_error_slow_roll}).

In principle, the results of this work could provide simple methods to constrain and possibly rule out models~\cite{Hu:2007nk,Starobinsky:2007hu}, and most likely the same technique could be applied to other models already proposed as well as to more sophisticated evolution laws different from~\rif{eq:T_evol}. The development of a curvature singularity could reveal unexpected consequences in a more detailed analysis of the models, and the mechanisms described herein may play a highly non-trivial r\^ole, for instance, for the study of Jeans-like instabilities and hydrodynamical stellar (non-) equilibrium~\cite{Babichev:2009td,Babichev:2009fi,Capozziello:2011nr,Capozziello:2011gm}. This goes beyond the scope of this Chapter, and could be subject of further research.

Two effects could on one hand hinder the development of singularities, and on the other hand provide additional methods to constrain models: ultraviolet gravity modifications and gravitational particle production.\\
Ultraviolet corrections to the gravitational action should start dominating at large $\R$, and set a limit to its growth; in turn, $\R$ would never reach the singularity (for details see e.g.~\cite{Arbuzova:2010iu,Appleby:2009uf}). Recently, a few works have investigated even more general ultraviolet aspects of (modified) gravity; a fully non-perturbative approach seems to point towards the altogether absence of singularities in gravity~\cite{Biswas:2011ar,Modesto:2011kw}.

Gravitational particle production, as is well known, is universal whenever curvature oscillates, and could in principle be a detectable source of high energy cosmic rays~\cite{Arbuzova:2012su,Arbuzova:2013ina}. The back-reaction on curvature is a damping of its oscillations, so this damping may prevent $\R$ from reaching infinity as well. This is particularly important in the adiabatic regime, where there can be very many oscillations before $\xi$ reaches $\xi_{sing}$ and therefore a large amount of energy could be released into SM particles. The produced cosmic rays would carry model-dependent signatures which could provide us valuable information to improve the constraints on the known models and maybe even suggest new gravitational theories.

\chapter{Gravitational Particle Production in \titlefr Gravity during Structure Formation}
\chaptermark{\spacedlowsmallcaps{Particle Production during Structure Formation}}
\label{ch:grav_part_prod_struct_form}

\citazione{\footnotesize E.V. Arbuzova, A.D. Dolgov, {\color{myhypercolor}\bf{L. Reverberi}}, \it{Eur. Phys. J. C} \bf{72}, 2247 (2012),\\
E.V. Arbuzova, A.D. Dolgov, {\color{myhypercolor}\bf{L. Reverberi}}, \it{Phys. Rev. D} \bf{88}, 024035 (2013).}{}

\section{Introduction}

The initial suggestion~\cite{Capozziello:2003tk,Capozziello:2003gx,Carroll:2003wy} of gravity modification with $F(R) \sim \mu^4/R$ suffered from strong instabilities in celestial bodies~\cite{Dolgov:2003px,Faraoni:2006sy}. Because of that, further modifications have been suggested~\cite{Hu:2007nk,Appleby:2007vb,Starobinsky:2007hu} which are free of these instabilities.

The suggested modifications, however, may lead to infinite-$R$ singularities  in the past cosmological history~\cite{Appleby:2009uf} and in the future in astronomical systems with rising energy/matter density~\cite{Frolov:2008uf,Thongkool:2009js,Thongkool:2009vf, Arbuzova:2010iu}. Some properties of the singularity found in \cite{Arbuzova:2010iu} were further studied in \cite{Bamba:2011sm,Bamba:2012cp}. These singularities can be successfully cured by the addition of an $R^2$-term into the action. Such a contribution naturally appears as a result of  
quantum corrections due to matter loops in curved space-time~\cite{Gurovich:1979xg,Starobinsky:1979ty,Starobinsky:1980te}.

Another mechanism which may in principle eliminate these singularities is particle production by the oscillating curvature. If the production rate is sufficiently high, the oscillations of $R$ are efficiently damped and the singularity could be avoided (see below).

The $R^2$ term may also have dominated in the early universe where it could lead to strong particle production. The process was studied long ago in~\cite{Zeldovich:1977, Starobinsky:1980te,Vilenkin:1985md}. Renewed interest to this problem arose recently~\cite{Arbuzova:2011fu,Motohashi:2012tt}, stimulated by the interest in possible effects of additional ultraviolet terms, $\sim R^2$, in infrared-modified $F(R)$ gravity models. 

In this paper we discuss the behaviour of a popular $F(R)$ model of dark energy in the case of a contracting system, discussing the evolution of the curvature scalar $R$ and the related effects of gravitational particle production. The calculations are done both numerically and analytically. For realistic values of the parameters, especially for extremely small coupling constant $g$, see Eq.~ (\ref{eq:definitionsPPSF}), numerical calculations are not reliable, so we have found an approximate analytical solution and compared it with numerical one with small but not too small values of $g$, for which numerical solutions are reliable. The comparison confirms the very good precision of the analytical solution.

\section{Basic Frameworks and Equations \label{s-basics}}

We consider the model proposed in~\cite{Starobinsky:2007hu}:
\be\label{eq:model}
F(R) = -\lambda R_c\left[1-\left(1+\frac{R^2}{R_c^2}\right)^{-n}\right]-\frac{R^2}{6m^2}\,,
\ee
where $n$ is an integer, $\lambda>0$, and $| R_c |$ is of the order of $8\pi \rho_c / m_{Pl}^2$, where
$\rho_c$ is the present day value of the total cosmological energy density. More precisely the
value of $R_c$ is determined by equation (\ref{eq-Rc}) below.
The $R^2$-term, absent in the original formulation, has been included to prevent curvature singularities in the presence of 
contracting bodies \cite{Arbuzova:2010iu}, and is relevant only at very large curvatures, because we need $m\gtrsim 10^5$ GeV in order to preserve the successful predictions of the standard BBN \cite{Arbuzova:2011fu}.

The evolution of $R$ is determined from the trace of the modified Einstein equations:
\be\label{eq:tracePPSF}
3\D^2 F_{,R} -R+RF_{,R} -2F=T\,,
\ee
where $\D^2\equiv \D_\mu\D^\mu$ is the covariant D'Alambertian operator, $F_{,R} \equiv d F/ d R$,  $T~\equiv~8\pi T^\mu_\mu/\mpl^2$, and $T_{\mu\nu}$ is the energy-momentum tensor of matter.

To describe the accelerated cosmological expansion, the function $F(R)$ is chosen in such a way that equation (\ref{eq:tracePPSF}) has a non-zero
constant curvature solution, $R=\bar R $, in the absence of matter. Observational data demand
\be
\bar R = - \frac{32\pi \Omega_\lambda \rho_c}{m_{Pl}^2},
\label{R-c}
\ee
where $\Omega_\lambda \approx 0.75$ is the vacuum-like cosmological energy density, deduced from the observations under the assumption of validity of the usual General Relativity (GR) with non-zero cosmological constant. Using this condition we can determine
$R_c$ from the solution of the equation:
\be
\bar R-\bar R F_{,R}  (\bar R) + 2 F(\bar R) = 0.
\label{eq-Rc}
\ee
This equation has two different limiting solutions for sufficiently large $\lambda$, roughly speaking $\lambda > 1$, namely $\bar R/R_c = 2\lambda$ and
$\bar R/R_c =1/ [n(n+1)\lambda]^{1/3}$. Following~\cite{Starobinsky:2007hu}, we should consider only the \it{maximal} root $\bar R\lesssim 2\lambda R_c$. Moreover, for the sake of simplicity and definiteness, we will neglect these subtleties and assume $\lambda \sim 1$ and
\be
R_c \simeq \bar R \simeq 1/t_U^2,
\label{R-0}
\ee
where $t_U \approx 4\cdot 10^{17}$ s is the universe age. Still, for a more detailed study of the parameter space of the model, it could be necessary to consider the full numerical solution of Eq.~\rif{eq-Rc} for all values of $\lambda$.

We are particularly interested in the regime $|R_c|\ll|R|\ll m^2$, in which $F$ can be approximated by
\be\label{eq:model_approxPPSF}
F(R)\simeq -R_c\left[1-\left(\frac{R_c}{R}\right)^{2n}\right]-\frac{R^2}{6m^2}\,.
\ee
We consider a nearly-homogeneous distribution of pressureless matter, with energy/mass density rising with time but still relatively low (e.g. a gas cloud in the process of galaxy or star formation). In such a case the spatial derivatives can be neglected and, if the object is far from forming a black hole, 
the space-time would be almost Minkowski. Then equation \rif{eq:tracePPSF} takes the form
\be\label{eq:tracePPSF_approx}
3\partial_t^2F_{,R} -R-T = 0\,.
\ee
Let us introduce the dimensionless quantities\footnote{The parameter $g$ should not be confused with $\det g_\mn$.}
\begin{subequations}\label{eq:definitionsPPSF}
\begin{align}
& z\equiv \frac{T(t)}{T(t_{in})}\equiv \frac{T}{T_0}= \frac{\rho_m(t)}{\rho_{m0}}\,,\\
& y\equiv -\frac{R}{T_0}\,, \\
& g\equiv \frac{T_0^{2n+2}}{6 n(-R_c)^{2n+1}m^2}= \frac{1}{6 n (m t_U)^2} \,\left( \frac{\rho_{m0}}{\rho_c}\right)^{2n+2}\,,\\
& \tau\equiv m\sqrt g\,t\,,
\end{align}
\end{subequations}
where $\rho_c \approx 10^{-29} $ g/cm$^3$ is the cosmological energy density at the present time,
$\rho_{m0}$ is the initial value of the mass/energy density of the object under scrutiny,
and $T_0 = 8\pi \rho_{m0}/m_{Pl}^2$. Next let us introduce the new scalar field:
\be\label{eq:xi_definition}
\xi\equiv  \frac{1}{2 n}\left(\frac{T_0}{R_c}\right)^{2n+1}F_{,R}  = \frac{1}{y^{2n+1}}-gy\,,,
\ee
in terms of which Eq.~\rif{eq:tracePPSF_approx} can be rewritten in the simple oscillator form:
\be\label{eq:xi_evol}
\xi''+z-y=0\,,
\ee
where a prime denotes derivative with respect to $\tau$. The potential of the oscillator is defined by:
\be
\frac{\partial U}{\partial \xi}= z - y(\xi).
\label{U-prime}
\ee
The substitution (\ref{eq:xi_definition}) is analogous to that done in \cite{Arbuzova:2010iu} but now $y$ cannot be 
analytically expressed through $\xi$ and we have to use approximate expressions.

It is clear that~\rif{eq:xi_evol} describes oscillations around $y=z$ (the ``bottom'' of the potential), 
which corresponds to the usual GR solution $R+T=0$. So we can separate solutions into an average and an oscillatory part. 
For small deviations from the minimum of the potential, solutions take the form:
\be
\begin{aligned}
\xi(\tau) &=\left[\frac{1}{z(\tau)^{2n+1}}-gz(\tau)\right]+\alpha(\tau)\sin F(\tau)\\
&\equiv \xi_a(\tau)+\xi_1(\tau)\,,
\end{aligned}
\label{eq:xi_expansionPPSF}
\ee
where
\be
F (\tau) \equiv \int^\tau_{\tau_0}d\tau'\,\Omega (\tau')\,,
\ee
and the dimensionless frequency $\Omega$ is defined as
\be\label{eq:frequency_U}
\Omega^2 = \frac{\partial^2 U}{\partial \xi^2}\,,
\ee
taken at $y=z$. From \rif{eq:xi_evol}, we find that it is equal to
\be\label{eq:frequency_2}
\Omega^2 = -\left.\frac{\partial y}{\partial\xi}\right|_{y=z} = -\left.\frac{1}{\partial\xi/\partial y}\right|_{y=z} = \left(\frac{2n+1}{z^{2n+2}}+g\right)^{-1}\,.
\ee
The conversion into the physical frequency $\omega$ is given by
\be\label{eq:Omega}
\omega = \Omega\,m \sqrt g\,.
\ee
It is assumed that initially $\xi (\tau_0)$ sits at the minimum of the potential, otherwise we would need to add a cosine term in~\rif{eq:xi_definition}. 
If initially $\xi (\tau_0)$ was shifted from the minimum, the oscillations would generally be stronger and the effect of particle 
production would be more pronounced.

\subsection{Potential for $\xi$}
\begin{figure}
\centering
\includegraphics[width=\mywidthsingle]{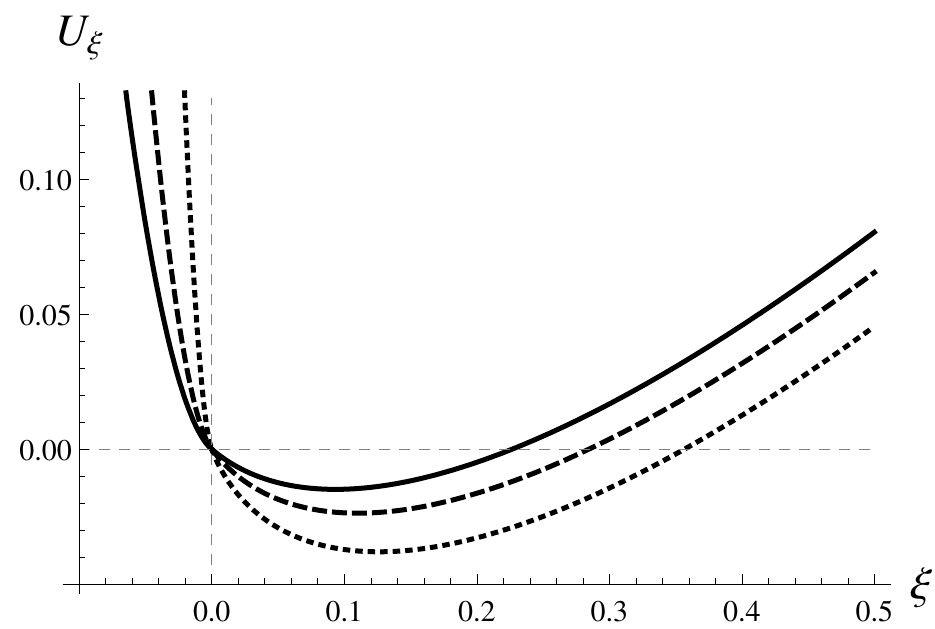}
\includegraphics[width=\mywidthsingle]{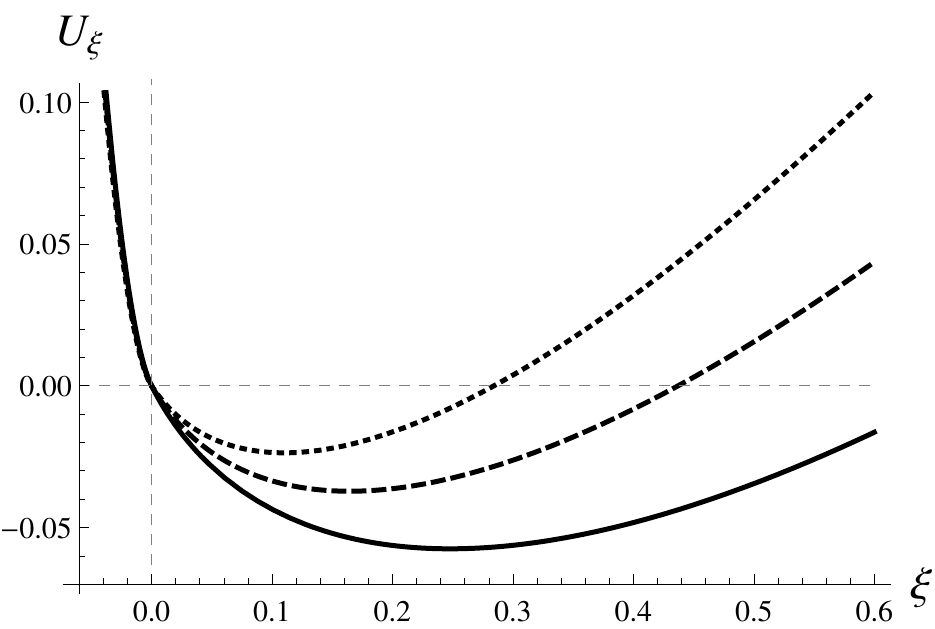}
\caption[Examples of the variation of potential~\rif{eq:xi_potential} for different values of parameters.]{Examples of the variation of potential~\rif{eq:xi_potential} for different values of parameters. \it{Left panel} $(n=2,\,z=1.5)$ solid line: $g=0.02$, dashed line: $g=0.01$, dotted line: $g=0.002$; the part of the potential at $\xi<\xi_a$ is increasingly steeper as $g$ decreases; the bottom of the potential also moves. \it{Right panel} $(n=2,\,g=0.01)$ solid line: $z=1.3$, dashed line: $z=1.4$, dotted line: $z=1.5$; the bottom of the potential moves to higher values of $U$ and lower values of $\xi$, as $z$ increases.}
\label{fig:potential}
\end{figure}

One cannot analytically invert Eq.~\rif{eq:xi_definition} to find the exact expression for $U(\xi)$. However, we can find an approximate expression for $gy^{2n+2}\ll 1$ ($\xi>0$) and $gy^{2n+2}\gg 1$ ($\xi<0$). The value $\xi=0$ separates two very distinct regimes, in each of which $\Omega$ 
has a very simple expression [see Eq.~\rif{eq:frequency_2}] and $\xi$ is dominated by either one of the two terms in the r.h.s. 
of Eq. \rif{eq:xi_definition}. Hence, in those limits the relation $\xi=\xi(y)$ can be inverted giving an explicit expression for $y=y(\xi)$, and therefore the following form for the potential:
\begin{subequations}\label{eq:xi_potential}
\be\label{eq:xi_pot_theta}
U(\xi) = U_+(\xi)\Theta(\xi) + U_-(\xi)\Theta(-\xi)\,,
\ee
where
\be\label{eq:potential_+_-}
\begin{aligned}
&U_+(\xi) = z\xi - \frac{2n+1}{2n}\left[\left(\xi+g^\frac{2n+1}{2n+2}\right)^\frac{2n}{2n+1}-g^\frac{2n}{2n+2}\right]\,,\\
&U_-(\xi) = \left(z-g^{-\frac{1}{2n+2}}\right)\xi+\frac{\xi^2}{2g}\,.
\end{aligned}
\ee
\end{subequations}
By construction $U$ and $\partial U/\partial\xi$ are continuous at $\xi=0$. The shape of this potential is shown in Fig.~\ref{fig:potential}.
We can write a conservation equation for a quantity which is analogous to the ``energy'' of the field $\xi$:
\begin{align}
\text{const.} &= \frac{1}{2}\,\xi'^2 + U(\xi) - \int^\tau_{\tau_0}d\eta \frac{\partial U}{\partial\eta}\notag \\
& = \frac{1}{2}\,\xi'^2 + 
U(\xi) - \int^\tau_{\tau_0}d\eta \frac{\partial z}{\partial\eta}\,\xi(\eta)
\label{eq:conserved}
\end{align}
where $\xi$ and $\xi'$ are taken at time moment $\tau$ coinciding with the upper integration bound.
The oscillating part of $\xi$ in the last integral term in~\rif{eq:conserved} would be integrated away for  fast harmonic oscillations of $\xi$. However, since the oscillations 
at late time become  strongly asymmetric, this term rises with time, see Fig. 3 and Sec. 3.3.1 below.

The bottom of the potential, as it is obvious from Eq.~\rif{U-prime}, corresponds to the GR solution $R=-T$, or $y(\xi) = z$, and its depth (for $gz^{2n+2}<1$) is
\be\label{eq:potential_bottom}
U_0(\tau) \simeq -\frac{1}{2n\,z(\tau)^{2n}}\,.
\ee

We will use a very simple form for the external energy density $z$, namely
\begin{subequations}\label{eq:z_linear}
\begin{align}
z(\tau) &= 1+\kappa(\tau-\tau_0)\\
\rho(t)&= \rho_0\left(1+\frac{t-t_0}{t_{contr}}\right)\\
\kappa^{-1} &\equiv m\sqrt g\,t_{contr}\,.
\end{align}
\end{subequations}
Here, $\kappa^{-1}$ and $t_{contr}$ are respectively the dimensionless and physical timescales of the contraction of the system; analogously, $\tau_0$ and $t_0$ are respectively the dimensionless and physical initial times, which for simplicity and without loss of generality will be taken equal to 0.
This evolution law may not be accurate when $t/t_{contr}>1$, but results obtained with more sophisticated functions describing the contraction of the system are most likely in qualitative agreement with our results, provided that $\dot\rho$ remains positive at all times.

It is also useful to express physical parameters such as $m$,  the initial energy density $\rho_{m0}$, etc., in terms of their respective ``typical'' values. Let us define
\begin{subequations}\label{eq:typ_param}
\begin{align}
 &\rho_{29}\equiv \frac{\rho_{m0}}{\rho_c}\,,\\
&m_5 \equiv \frac{m}{10^5\text{ GeV}}\,,\\
&t_{10}\equiv \frac{t_{contr}}{10^{10}\text{ years}}\,,
\end{align}
\end{subequations}
where $\rho_c=10^{-29}\text{ g cm}^{-3}$ is the present (critical) energy density of the Universe. In terms of these quantities, we can rewrite $g$ and $\kappa$ as
\begin{subequations}
\begin{align}
g &\simeq 1.2\times 10^{-94}\,\frac{\rho_{29}^{2n+2}}{n 
\,m_5^2}\,,\\
\kappa &\simeq 1.9 \,\frac{\sqrt{n}
}{\rho_{29}^{n+1}\,t_{10}}\,.
\label{g-kappa}
\end{align}
\end{subequations}

\section{Solutions}

\subsection{Oscillations of $\xi$}
\begin{figure}
\centering
\includegraphics[width=\mywidthsingle]{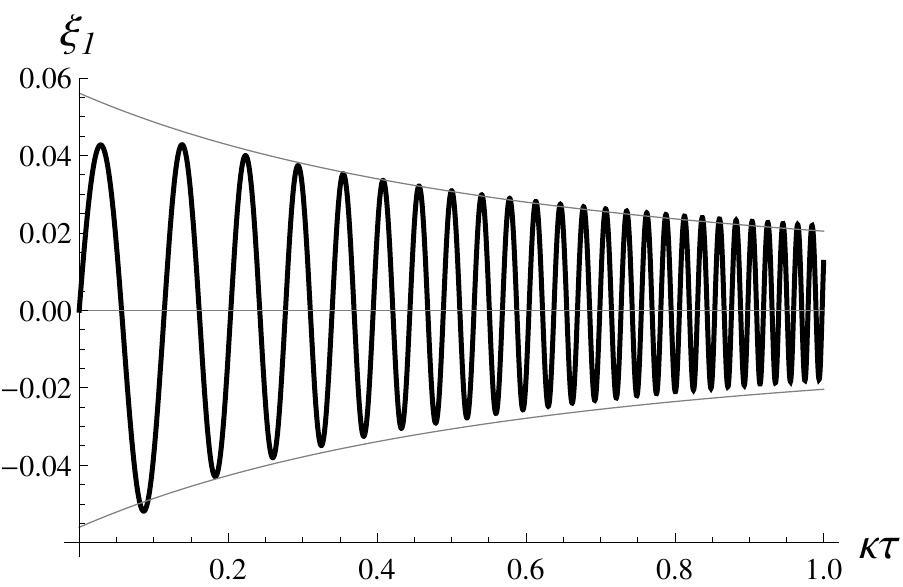}
\caption{Numerical results for $\xi_1(\tau)$, in the case $n=2$, $\kappa=0.01$, $g=0.01$ and initial conditions $y_0=1$, 
$y'_0=\kappa/2$. The amplitude and the frequency of the oscillations are in very good agreement with the analytical result~\rif{eq:xi_amplitude_solution}.}
\label{fig:xi_amplitude}
\end{figure}

At first order in $\xi_1$, equation \rif{eq:xi_evol} can be written  as
\be\label{eq:xi_evol_approx}
\xi_1''+\Omega^2\xi_1= -\xi_a'' \,,
\ee
with $\Omega$ given by~\rif{eq:frequency_2}. The term $\xi_a''$ is proportional to $\kappa^2$, which is usually 
assumed small, so in first approximation it can be neglected, though an analytic solution for constant $\Omega$ or in the limit of large $\Omega$ can be obtained considering this term as well. Using~\rif{eq:xi_expansionPPSF} and neglecting $\alpha''$, we obtain
\be
2\,\frac{\alpha'}{\alpha} \simeq -\frac{\Omega'}{\Omega}\,,
\ee
so
\be\label{eq:alpha_omega}
\alpha\simeq \alpha_0\,
\sqrt\frac{\Omega_0}{\Omega}= \alpha_0\left(\frac{1}{z^{2n+2}}+\frac{g}{2n+1}\right)^{1/4}
\left({1}+\frac{g}{2n+1}\right)^{-1/4}.
\ee
Here and in what follows sub-0 means that the corresponding quantity is taken at initial moment $\tau = \tau_0$. We impose the following initial conditions
\be
\begin{cases}
y(\tau=\tau_0)=z(\tau=\tau_0)=1\,,\\
y'(\tau=\tau_0)=y'_0\,,
\end{cases}
\label{init-y}
\ee
which correspond to the GR solution at the initial moment. In terms of $\xi$  it means that ${\xi_1(\tau_0)=0}$. 
The initial value of the derivative $\xi'_1 (\tau_0)$ can be expressed through $y_0'$, which we keep as a free parameter; 
according to~\rif{eq:xi_definition}: $\xi'_0 = -y'_0 (2n+1+g) $. Differentiating Eq.~(\ref{eq:xi_expansionPPSF}) with respect to $\tau$ and using \rif{eq:alpha_omega} we find:
\be 
\alpha_0 = (\kappa - y'_0) (2n+1+g)^{3/2}.
\label{alpha-0}
\ee
Correspondingly:
\be\label{eq:xi_amplitude_solution}
\left|\alpha(\tau)\right| = |y'_0-\kappa|(2n+1+g)^{5/4}\left(\frac{2n+1}{z^{2n+2}}+g\right)^{1/4}\,.
\ee
Because of the assumptions made to obtain \rif{eq:alpha_omega}, we expect this result to hold when $|y'_0-\kappa|\sim\kappa$ or slightly less. 
In this regime the numerical results, shown in Fig.~\ref{fig:xi_amplitude}, are in excellent agreement with the analytical 
estimate~\rif{eq:xi_amplitude_solution}. 
We remark that the agreement improves for larger $g$  and/or smaller $\kappa$, while for small $g$ and ``large'' $\kappa$ it may become 
significantly worse (see paragraph \ref{sec:spikes}).

For $ y'_0 = \kappa$ and $ \xi_1 (\tau_0)=0$, it would seem from Eq.~\rif{eq:xi_amplitude_solution} that oscillations are not excited. However, this is an artifact of the approximation used. In fact, the "source" term in the r.h.s. of Eq.~\rif{eq:xi_evol_approx} produces oscillations and hence deviations from GR
with any initial conditions.

\subsection{Oscillations of $y$}
We shall now exploit this result to evaluate the amplitude of the oscillations of $y$. We first expand $y$ as it was done for $\xi$ in Eq.~\rif{eq:xi_expansionPPSF}:
\be\label{eq:y_expansion}
y(\tau) = z(\tau) + \beta(\tau)\,\sin F(\tau) \equiv y_a(\tau) + \beta(\tau)\,\sin\left(\int_{\tau_0}^\tau d\tau'\,\Omega \right)\,, 
\ee
where it is easy to prove that $\Omega$ must coincide with that given by Eq.~\rif{eq:frequency_2}. For $|\beta/z| < 1$, we expand $\xi$ as
\begin{align}\label{eq:xi_exp_small_beta}
\xi&=\frac{1}{z^{2n+1}\left[1+({\beta}/{z})\sin F(\tau)\right]^{2n+1}}-g\left(z+\beta\sin F(\tau)\right)\notag\\
 &\simeq \frac{1}{z^{2n+1}}-gz - \left(\frac{2n+1}{z^{2n+2}}+g\right)\beta\sin F(\tau)
\end{align}
Comparing this expression with Eqs.~\rif{eq:xi_expansionPPSF} and \rif{eq:frequency_2}, we find that
\be
|\beta| = |\alpha|\left(\frac{2n+1}{z^{2n+2}}+g\right)^{-1} = |\alpha|\Omega^2\,.
\ee
Accordingly, $\beta$ evolves as:
\be\label{eq:y_amplitude_evolution}
\beta(\tau) \simeq \left|y'_0-\kappa\right|\left(2n+1+g\right)^{5/4}\left(\frac{2n+1}{z^{2n+2}}+g\right)^{-3/4}\,.
\ee
This is in reasonable agreement with numerical results, especially in both limiting cases ${gz^{2n+2}\ll 1}$ and ${gz^{2n+2}\gg 1}$, as expected.

\subsection{``Spike-like'' Solutions}\label{sec:spikes}
We have found simple analytical solutions for $\xi$ and $y$ in two separate limits: $gz^{2n+2}\ll 1$ and $gz^{2n+2}\gg 1$. 
However, in the intermediate case numerical calculations show interesting features which are worth discussing.
\begin{figure}[tb]
\centering
\includegraphics[width=\mywidthsingle]{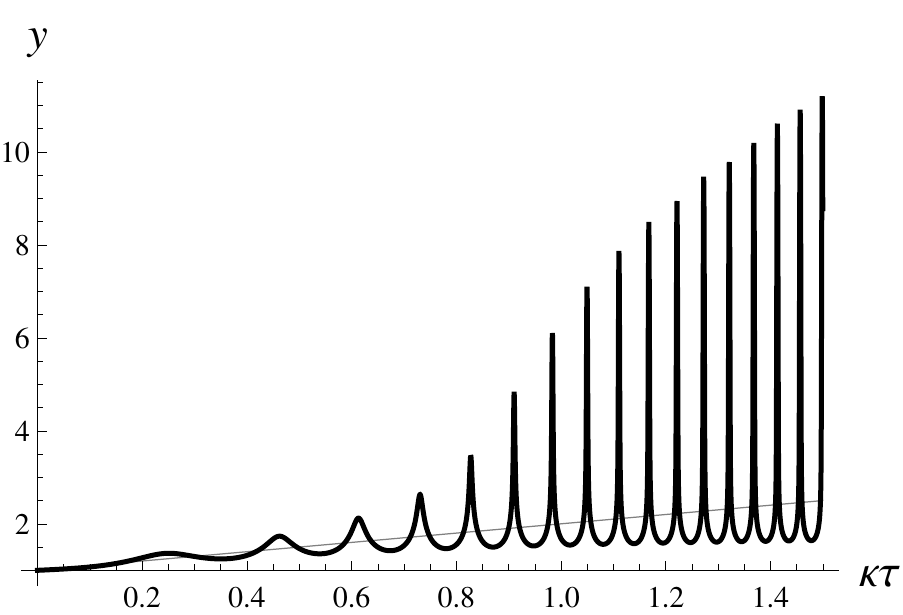}
\includegraphics[width=\mywidthsingle]{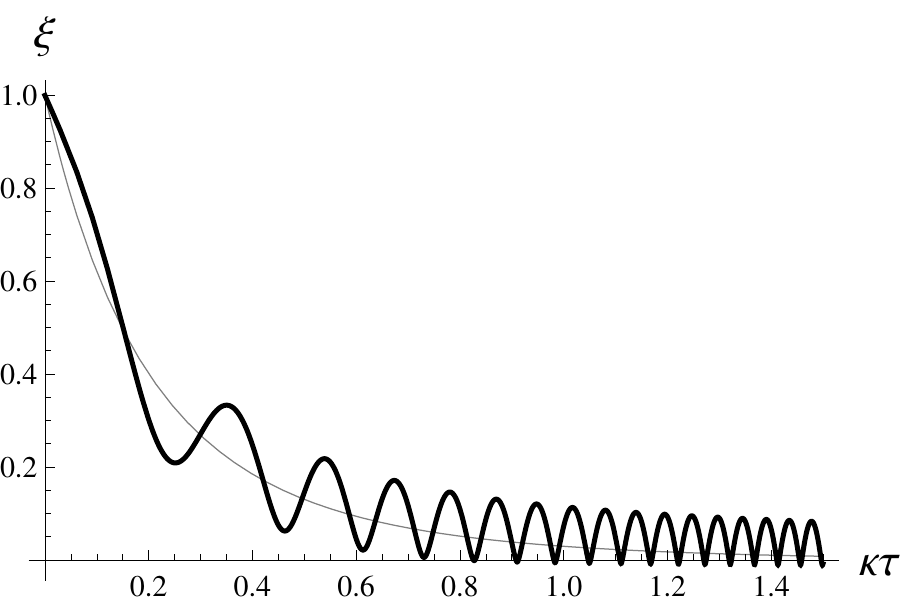}
\caption[``Spikes'' in the solutions. The results presented are for $n=2$, $g=0.001$, $\kappa=0.04$, and $y'_0=\kappa/2$.]{``Spikes'' in the solutions. The results presented are for $n=2$, $g=0.001$, $\kappa=0.04$, and $y'_0=\kappa/2$. Note the asymmetry of the oscillations of $y$ and $\xi$ around $y=z$ and their anharmonicity. See the text for further details.}
\label{fig:spikes}
\end{figure}
As shown in figure \ref{fig:spikes}, when $\xi$ approaches (and even crosses) zero but $\xi_a$ does not, that is when $gy^{2n+2}\simeq 1$ but $gz^{2n+2}<1$, we have the largest deviations from the harmonic, symmetric oscillations around $y=z$. This happens especially when $g$ is very small and $\kappa$ is not too small.\\
The reason for this behaviour is qualitatively explained by the following considerations. Inspecting  Eq.~\rif{eq:xi_potential} and/or figure \ref{fig:potential}, we see that when $\xi<\xi_a$ the potential becomes increasingly steep, hence reducing the time spent in that region. Moreover, a given variation $\delta\xi$ in this region corresponds to a large variation of $y$. Thus there appear high, narrow ``spikes'' in $y$. On the other hand, for $\xi>\xi_a$ the potential is much less steep, and the oscillation in that region lasts longer, yielding slow ``valleys'' between the spikes of $y$.\\
Please note that in the region with spikes, the assumption $|\beta/z| \ll 1$ is no longer accurate, and we have deviations from the analytical estimate~\rif{eq:y_amplitude_evolution}, which is usually smaller than the exact numerical value.

\subsubsection{Estimate of $\beta$}\label{sec:beta_estimate}

In order to obtain an estimate of $\beta$ in this region we use an analogous approach to that of~\cite{Reverberi:2012ew}. Let us first introduce some new notations:
\be 
U(\tau) = U[\xi(\tau), z(\tau)], \,\,\,\, U_a(\tau) = U[\xi_a (\tau), z(\tau)].
\label{z-of-tau}
\ee
Remember that according to Eq.~(\ref{eq:xi_expansionPPSF}) $\xi_a (\tau)  =  z(\tau)^{-(2n+1)} - g z(\tau)$ is the value of $\xi$ where the potential has its minimum value. Let us denote by $\tau_a$ the time at which $\xi$ hits  the minimum of the potential, i.e. 
$\xi (\tau_a) = \xi_a (\tau_a)$ and correspondingly $\xi_1 (\tau_a) = 0$. Note that $U(\tau_a) =U_a(\tau_a)$, while they are evidently different
at other values of $\tau$. We will examine Eq.~\rif{eq:conserved} choosing the initial value of time at the moment when $\xi$ passes through the
minimum of the potential. Correspondingly $U(\tau_0) = U_a (\tau_0)$ and hence we denote $\tau_0$ as $ \tau_{a0}$. 
With this initial value of $\tau_0$ we rewrite Eq.~(\ref{eq:conserved}) as
{\small
\be\label{new_conserved}
\frac{1}{2} [\xi'(\tau)]^2+U(\tau)-\int^\tau_{\tau_{a0}} d\eta\,z'(\eta) \left[ \xi_a(\eta) + \xi_1(\eta)  \right]= U_a(\tau_{a0}) + \frac{1}{2} [\xi'(\tau_{a0})]^2\,.
\ee
}
We start by noticing that the first term under the integral in the l.h.s. of this equation can be explicitly integrated and is equal to 
\be\label{eq:int_source_potential}
\int^\tau_{\tau_{a0}} d\eta\,z' (\eta) \,\xi_a (\eta)= U_a(\tau) - U_a(\tau_{a0})\,.
\ee
So $U_a(\tau_{a0})$ disappears from this equation.

Assume that the upper limit of integration, $\tau$, is sufficiently high, such that the nearest minimum value of $\xi$ is negative.
Let us first take $\tau$ exactly equal to $\tau_m$, i.e. to one of the numerous  values when $\xi$ hits a minimum, $\xi = \xi_{min}<0$.
At this point $\xi'(\tau_m) = 0$ and we obtain:
\be
U(\tau_m) = U_a(\tau_m) + \frac{1}{2}\, [\xi'(\tau_{a0})]^2 + \int^{\tau_m}_{\tau_{a0}} d\eta z'(\eta) \xi_1(\eta).
\label{U-of-tau-m}
\ee
Now we need to estimate the last term (integral) in this equation. To this end let us use again Eq.~(\ref{new_conserved}) but now take the upper integration limit equal to the moment when $\xi$ happens to be at the minimum of $U$, at the nearest point from $\tau_m$ taken above, so that $\tau = \tau_a$. According to Eq.~(\ref{eq:int_source_potential}) the potential terms and the integral of $z' \xi_a$ cancel out and we are left with
\be\label{max_kin_estim}
 \int^{\tau_a}_{\tau_{a0}}   d\eta\,z'(\eta) \,\xi_1 (\eta)  =  \frac{1}{2}\,[\xi'(\tau_a)]^2 -  \frac{1}{2}\,[\xi'(\tau_{a0})]^2 .
\ee
When $\xi$ passes through the minimum of the potential its velocity reaches maximum value for a given oscillation
and $[\xi (\tau_a)']^2  = \Omega^2 \alpha^2 $, where $\alpha$ and $\Omega$ are given by~\rif{eq:xi_amplitude_solution} and~\rif{eq:frequency_2}
respectively. Notice that $[\xi (\tau_a)']^2 $ rises with time and thus for large time  $[\xi'(\tau_{a0})]^2$ may be neglected.
It worth noting that in the limit of harmonic oscillations symmetric with respect to the minimum of $U$ the integral in the
r.h.s. of Eq.~(\ref{max_kin_estim}) does not rise with time but, as we have seen above, the oscillations are
strongly asymmetric with respect to $\xi_a$ and because of that the integral rises with time and Eq.~(\ref{max_kin_estim})
is self-consistent.

The upper integration limits in Eqs.~(\ref{U-of-tau-m}) and (\ref{max_kin_estim}) are slightly different: $\tau_m$ corresponds to the moment
when $\xi$ reaches its minimum value (since $\xi<0$, maximum in absolute value), while $\tau_a$ is the nearest time moment when $\xi$ passes through
the minimum of the potential. So they differ by about a quarter of a period as calculated when $\xi$ is on the left side of the potential minimum.
Since this time interval is quite short the difference between these two integral can be neglected.

So we finally obtain
\be
U(\tau_m) = U_a(\tau_m) + \frac{1}{2} [\xi'(\tau_a)]^2 \approx \frac{1}{2}\left (\alpha^2\Omega^2-\frac{1}{n\,z^{2n}}\right).
\label{U-of-tau-m2}
\ee
Using expression (\ref{eq:potential_+_-}) for $U_- (\xi)$ and relation between $\xi$ and $y$: $y\simeq -\xi/g$, we find
\be
y^2 - 2zy - \frac{1}{g}\left[\alpha^2\Omega^2-\frac{1}{n\,z^{2n}}\right] = 0\,,
\ee
which yields the amplitude
\be\label{eq:beta_spikes}
\beta_{spikes} = \sqrt{\frac{1}{g}\left(\alpha^2\Omega^2-\frac{1}{n\,z^{2n}}\right)+z^2}\,.
\ee
Quite remarkably, this result is exactly equivalent to~\rif{eq:y_amplitude_evolution} in the limit $gz^{2n+2}\gg 1$, so we will assume that from the moment the harmonic approximation fails to be accurate, $\beta$ will follow~\rif{eq:beta_spikes} up to the asymptotic harmonic regime where $gz^{2n+2}\gg 1$. In particular, the moment of transition from harmonic to spike regime is roughly the time at which
\be
\beta_{harm} = \beta_{spikes}\,.
\ee


\section{Gravitational Particle Production}
As is well known, an oscillating curvature gives rise to gravitational particle production. Basically, the energy stored in oscillating gravitational degrees of freedom is released into pairs of elementary particles/antiparticles. As shown in~\cite{Arbuzova:2011fu} in the case of a minimally-coupled scalar field, the energy released into particles per unit volume and unit time is
\be\label{eq:PP_harmonic}
\dot\rho_{PP}\simeq \frac{\Delta_R^2\,\omega}{1152\pi}\,,
\ee
where $\Delta_R$ is the amplitude of the oscillations of $R$, and $\omega$ is their (physical) frequency. In our case,
\be
\Delta_R \equiv \beta\,T_0\,.
\ee
Moreover, the lifetime of oscillations of $R$ is
\be\label{eq:lifetime}
\tau_R = \frac{48\,\mpl^2}{\omega^3}\,.
\ee
Rigorously speaking, this result is valid when the oscillations of $R$ are perfectly harmonic, or at least when $R$ can be separated in a slowly-varying and an oscillating part, which is supposed to have a constant (or almost constant) frequency. As we have seen in the previous section, solutions of Eq. \rif{eq:xi_evol} show spikes, which are very far from being harmonic oscillations. Thus, we must find a more general result than \rif{eq:PP_harmonic}.

We consider the gravitational particle production of pairs of massless scalar particles $\varphi$, quantised in the usual way:
\be
\begin{aligned}
&\hspace{-1cm}\hat\varphi(x)= \int \frac{d^3 k}{(2\pi)^32 E_k}\left[\hat a_k\,e^{-ik\cdot x}+\hat a_k^\dagger\,e^{ik\cdot x}\right]\,,\quad (x\cdot k\equiv \omega t-\mb k\cdot \mb x)\\
&\hspace{-1cm}\left[\hat a_k,\hat a_{k'}^\dagger\right] = (2\pi)^3\,2E_k\,\delta^{(3)}\left(\mb k-\mb{k'}\right)\,.
\end{aligned}
\ee
At first order in perturbation theory, the amplitude for the creation of two particles of 4-momenta $p_1$ and $p_2$ is equal to~\cite{Arbuzova:2011fu}:
\begin{align}
A_{p_1,p_2}&\simeq \frac{1}{6}\int dt\,d^3x\,R(t)\langle p_1,p_2 | \hat\varphi^2 |0\rangle\notag\\
&= \frac{(2\pi)^3}{3\sqrt 2}\,\,\delta^{(3)}\left(\mb p_1+\mb p_2\right)\int dt\,R(t)\,e^{i(E_1+E_2)t}\,.
\label{eq:amplitude_1}
\end{align}
In terms of the Fourier transform of $R$, defined by
\be
R(t) \equiv \frac{1}{2\pi}\int d\omega\,\tilde{\mc R}(\omega)\,e^{-i\omega t}\,,
\label{R-of-t-Fur}
\ee
we can recast Eq. \rif{eq:amplitude_1} as
\be
A_{p_1,p_2}\simeq \frac{(2\pi)^3}{3\sqrt 2}\,\,\delta^{(3)}(\mb p_1+\mb p_2)\,\tilde{\mc R}(E_1+E_2)\,.
\label{A-of-p-2}
\ee
In order to calculate the number of particles produced per unit time and unit volume, we need to integrate $|A_{p_1,p_2}|^2$ over all phase space and to divide by the 3-dimensional volume and time duration of the process, $\Delta t$. This yields
\be\label{eq:n_PP}
\dot n_{PP}\simeq \frac{1}{288\pi^2\Delta t}\int d\omega \left|\tilde{\mc R}(\omega)\right|^2\,,
\ee
{and correspondingly, because each particle is produced with energy $E=\omega/2$,}
\be\label{eq:rho_PP}
\dot\rho\simeq \frac{1}{576\pi^2\Delta t}\int d\omega\,\omega\left|\tilde{\mc R}(\omega)\right|^2\,.
\ee
The time duration of the process $\Delta t$ may be considered infinitely large if the characteristic frequency satisfies
the condition $\omega_{ch} \Delta t \gg 1$. Correspondingly, the square of the delta functions in $\tilde{\mc R}$ would be proportional 
to $\delta (0) \sim \Delta t$. For instance, when $R$ is a perfect sine or cosine of frequency $\omega$, one has 
\be\label{eq:deltas}
\tilde{\mc R}(\varepsilon) \sim \delta(\varepsilon-\omega)+\delta(\varepsilon+\omega)\,,
\ee 
and
\be
\left|\delta(\varepsilon\pm \omega)\right|^2 = \frac{\Delta t}{2\pi}\,\delta(\varepsilon\pm \omega)\,.
\ee
Thus $\Delta t$ does not appear in the probability of particle production per unit time.
The physical cut-off of $\Delta t$ in the considered case is given roughly by $\tau_R$ in Eq.~(\ref{eq:lifetime}), therefore for $\omega \gg 1/\tau_R$ the approximation used here is accurate enough. Moreover, since frequencies must be positive, only the first delta function in~\rif{eq:deltas} gives a non-vanishing contribution.

\subsection{Regular Region}
First, let us concentrate on the ``regular'' region (see subsections 3.1 and 3.2). 
Substituting $\Delta_R\equiv \beta\,T_0$ in~\rif{eq:PP_harmonic}, using~\rif{eq:y_amplitude_evolution}
 and \rif{eq:Omega}, and taking $y'_0\simeq 0$ for simplicity, we find
\be
\begin{aligned}
\dot \rho_{PP,\,reg} &= \frac{\pi \sqrt{6 n} }{18}\, \frac{t_U \rho_c^{n+1}}{m_{Pl}^4 t_{contr}^2 \rho_{m0}^{n-1}}\,
\frac{ \left( 2n+1+g \right)^{5/2} }{\left(\frac{2n+1}{z^{2n+2}} + g \right)^2}\\
&= C \left(\frac{t_U}{t_{contr}}\right)^2\, \frac{ \rho_c^2}{m_{Pl}^4 t_U}, 
\end{aligned}
\label{dot-rho-reg}
\ee
where the coefficient $C$ has the following  expressions in the two limits:\\
\begin{subequations}
\begin{align}
&\hspace{-1cm}C(g z^{2n +2} < 1) =  \frac{\pi \sqrt{ 6 n (2n+1) } }{18} \left( \frac{\rho_c}{\rho_{m0} }\right)^{n-1}\, z^{4n+4}
\label{C-small-g}\\
&\hspace{-1cm}C(g z^{2n +2} > 1)= \frac{ \pi  \left[ 6 n (2n+1+g)  \right]^{5/2}\,(m t_U)^4}{18} \left( \frac{\rho_c}{\rho_{m0}} \right)^{5n+3}.
\label{C-large-gz}
\end{align}
\end{subequations}
The last factor in Eq.~(\ref{dot-rho-reg}) is extremely small. Since $\rho_c^2 \sim m^4_{Pl}/ t_U^4$, this factor
is about $1/t_U^5$. So unless $C$ is very large, particle production in the regular region would be negligible.
The most favorable possibility would be small $g$ and large $z$, but keep in mind that $g \sim \rho_{m0}^{2n+2}$.
We present an estimate of the flux in the conventional units as:
\be \label{eq:part_prod_rate}
\begin{aligned}
\frac{\dot\rho_{PP,\,reg}}{\text{GeV\,s}^{-1}\text{m}^{-3}} &\simeq 3.6\times 10^{-141}\, \frac{C_1(n,g,z)\,z^{4n+4}}{\rho_{29}^{n-1} t_{10}^2 }\\
& \simeq  2.5\times 10^{47}\, \frac{C_2 (n,g,z)\,m_5^4}{\rho_{29}^{5n+3} t_{10}^2}.
\end{aligned}
\ee
The coefficients $C_1$ and $C_2$ are convenient to use when $gz^{2n+2}\ll 1$ and $gz^{2n+2}\gg 1$. 
They are respectively:
\begin{subequations}\label{C1-C2}
\begin{align} 
C_1 &= \sqrt{n(2n+1 +g)} \left(\frac{2n+1+g}{2n+1 + g z^{2n+2}}\right)^2\notag \\
& \approx \sqrt{n(2n+1)}\,,\\
C_2 &= n^{5/2} \sqrt{2n+1+g}\,g^2 \left(\frac{2n+1+g}{2n+1 + g z^{2n+2}}\right)^2 z^{4n+4}\notag \\
& \approx {[n(2n+1)]^{5/2}}.
\end{align}
\end{subequations}

\subsection{Spike Region}
In the spike region the particle production rate would be strongly enhanced due to much larger amplitude of the
oscillations of $R$. We parametrise the solutions in this region
as a sum of gaussians with slowly varying amplitude $\Delta_R(t)$, 
superimposed on the smooth power-like background, $-T(t)$:
\be\label{eq:gaussians}
R(t) = -T(t)+\Delta_R(t)\sum_{j=1}^{N}\exp\left[-\frac{(t-jt_1)^2}{2\sigma^2}\right]\,.
\ee
Here $t_1$ is the time-shift between the spikes and $\sigma$ is the width of the spikes. 
The values of these parameters are determined from the solution obtained above.
The slow variation of the functions $T(t)$ and $\Delta_R(t)$ means that  $\dot T\ll T/t_1$ and $\dot \Delta_R\ll \Delta_R/t_1$.\\
In principle, $N$ could be infinitely large, which for $\Delta_R =$ const. corresponds to an infinitely long duration of the process. As we have mentioned before, this does not have an essential impact on the probability of particle production per unit time.

In accordance with the solutions of the equations determining the evolution of curvature, we consider the case 
$\sigma\ll t_1 $, that is the spacing between the spikes is much larger than their width. 
At high frequencies the Fourier transform of \rif{eq:gaussians} is dominated by the contribution of the quickly-varying gaussians, i.e.
\be
\tilde{\mc R}(\omega)\simeq \sqrt{2\pi}\,\Delta_R|\sigma|e^{-\omega^2\sigma^2/2+i\omega t_1}\,\,\frac{e^{iN\omega t_1}-1}{e^{i\omega t_1}-1}\,.
\ee
When squared, this gives
\be\label{eq:tilde_R_2}
\left|\tilde{\mc R}(\omega)\right|^2 \simeq 2\pi \Delta_R^2\,\sigma^2e^{-\omega^2\sigma^2}\,\frac{\sin^2 N\omega t_1/2}{\sin^2 \omega t_1/2}
\ee
The dominant part of this expression comes from $\omega=\omega_j = 2j \pi/t_1$, where it is equal to $N^2$. Around these points we have
\be
\frac{\sin^2 N\omega t_1/2}{\sin^2\omega t_1/2}\underset{\omega\simeq 2j\pi/t_1}{\simeq} \left(\frac{\sin \left[N(\omega t_1/2-j\pi)\right]}{\omega t_1/2-j\pi}\right)^2\,.
\ee
We take the limit $N\to\infty$ and use a representation of Dirac's delta function to write
\be\label{eq:sin2}
\frac{\sin^2 N\omega t_1/2}{\sin^2\omega t_1/2} \simeq \sum_j\left|\pi\,\delta\left(\frac{\omega t_1}{2}-j\pi\right)\right|^2 = 
\sum_j\left|\frac{2\pi}{t_1}\,\delta\left(\omega-\omega_j\right)\right|^2\,,
\ee
which yields
\be
\left|\tilde{\mc R}(\omega)\right|^2\simeq \frac{4\pi^2 \Delta_R^2\,\sigma^2e^{-\omega^2\sigma^2}\Delta t}{t_1^2}\sum_j\delta\left(\omega-\frac{2\pi j}{t_1}\right)\,.
\ee
Each particle produced by the oscillation component of frequency $\omega$ has energy $\omega/2$, 
so the gravitational particle production rate is
\begin{align}\label{eq:PP_general}
\notag \dot\rho_{PP} &= \frac{1}{288\pi^2\Delta t}\int\,d\omega\,\frac{\omega}{2}\left|\tilde{\mc R}(\omega)\right|^2\\
 \notag &= \frac{\pi \Delta_R^2\,\sigma^2}{72\,t_1^3}\sum_{j} j\exp\left[-\left(\frac{2\pi j\,\sigma}{t_1}\right)^2\right]\\
&\simeq \frac{\Delta_R^2}{576\pi\,t_1}\,.
\end{align}
In the last step, we have again assumed that $\sigma/t_1\ll 1$, so that the summation over $j$ can be replaced by an integral. Remarkably, the dependence on $\sigma$ has disappeared from the result.
Lastly, we use $\Delta_R=\beta\,T_0$ and~\rif{eq:beta_spikes}, obtaining
\be
\dot \rho_{PP,\,sp} = \frac{ (6 n)^{3/2} (2n+1+g)^{5/2} }{ 18 (2n+1 + gz^{2n+2}) }\,\frac{ m^2 t_U^3 z^{2n+2} }{t_{contr}^2 m_{Pl}^4}
\frac{ \rho_c^{3n+3} }{\rho_{m0}^{3n+1}}.
\label{dot-rho-sp}
\ee
or in conventional units:
\be\label{eq:part_prod_spikes}
\frac{\dot\rho_{PP,\,sp}}{\text{GeV\,s}^{-1}\text{m}^{-3}} \simeq 3.0 \times 10^{-47}\,\frac{C_3(n,g,z)\,m_5^2\,z^{2n+2}}{t_{10}^2\,\rho_{29}^{3n+1}}
\ee
where
\be
C_3 = \frac{(2n+1+g)^{5/2}n^{3/2}}{(2n+1+gz^{2n+2})} \approx [n(2n+1)]^{3/2}\,.
\label{C3}
\ee
All elementary particles couple to gravity, so in order to get an order-of-magnitude estimate of the overall particle production 
one should multiply \rif{eq:part_prod_rate} and \rif{eq:part_prod_spikes} by the number of elementary particle species,  $N_s$,
with masses bound from above by $m \lesssim  2\pi/\sigma $.

\subsection{Backreaction on Curvature and Mode-dependent Damping}
So far we have not taken into account that the oscillation amplitude should be damped due to the back reaction of particle production. Neglecting such damping 
would be an accurate approximation up to $t/\tau_R \simeq 1$; for larger times, however, the damping should be
taken into consideration. In the regular region oscillations are practically harmonic, so only a single
frequency mode is involved and one simply needs to add the exponential damping 
factor $\exp[-2\Gamma(\omega)t]$ to \rif{eq:PP_harmonic} and \rif{eq:part_prod_rate}. In the spike region the problem is more complicated because, according to Eq.~(\ref{eq:lifetime}), the damping depends upon the frequency, so different modes are  damped differently and this can noticeably distort the form of the initial $R(t)$.
A simple approximate way to take into account this damping is to introduce the factor $\exp [-t/\tau_R (\omega)] $ into the integrand of Eq.~(\ref{R-of-t-Fur}). After a sufficiently long time, only the modes with the lowest frequency survive and one may naively expect that the lowest frequency modes give the dominant contribution to particle production. However, one should keep in mind that the time duration is finite: in fact, it is equal to the time of stabilisation of the collapsing system and is surely shorter than the cosmological time $t_c \approx 4\cdot 10^{17} $~s. Thus the the energy is predominantly  emitted with the frequencies determined by the condition  $\Delta t /\tau_R(\omega)\simeq 1 $, see discussion 
below Eq.~(\ref{rho-fin}).

So to take into account the damping of $R$-oscillations we introduce into the amplitude (\ref{eq:amplitude_1}) 
the damping factor $\exp [-\Gamma (\omega) t ]$, where $\Gamma (\omega) = 1/\tau_R(\omega) = \omega^3/48 m_{Pl}^2$ and integrate
over time up to a finite upper limit:
\begin{align}
\hspace{-1cm}A_{p_1,p_2} &\simeq \frac{(2\pi)^3}{3\sqrt 2}\,\,\delta^{(3)}\left(\mb p_1+\mb p_2\right)
\int \frac{d\omega}{2\pi} \, \tilde{ R}(\omega) \int_0^{t} dt' \, e^{i(E_1+E_2 - \omega)t' -\Gamma(\omega) t'} \notag \\
&= \frac{(2\pi)^3}{3\sqrt 2}\,\,\delta^{(3)}\left(\mb p_1+\mb p_2\right) 
\int \frac{d\omega}{2\pi} \, \tilde{ R}(\omega)\, 
\frac{e^{i(2E -\omega)t - \Gamma(\omega) t}-1}{i ( 2E - \omega) -\Gamma(\omega)}\,,
\label{A-damped}
\end{align}
where $E= E_1 = E_2$ {and $t$ has here the same meaning as $\Delta t$ in \rif{eq:n_PP}}.\\
As we have seen above, $\tilde R (\omega) $ can be written as:
\be
\tilde R (\omega) \simeq \sqrt{2}\,\pi^{3/2} \Delta_R\,\sigma e^{-\omega^2 \sigma^2/2}\, e^{i\Phi} 
\sum_j \delta\left(\frac{\omega t_1}{2} - \pi j \right),
\label{tilde-R}
\ee
where $\exp (i\Phi)$ is a phase factor of modulus unity. Substituting this expression in Eq.~(\ref{A-damped}) and integrating over $\omega$, we find up to a phase factor:
\be
\hspace{-1cm}
A_{p_1,p_2} \simeq  \frac{(2\pi)^{7/2} \Delta_R\,\sigma }{3\sqrt 2\, t_1}\,\,\delta^{(3)}\left(\mb p_1+\mb p_2\right)
\sum_j e^{-\sigma^2\omega_j^2/2}\,\frac{1-e^{(i \epsilon_j -\Gamma_j)t}}{\epsilon_j +i\,\Gamma_j}  ,
\label{A-fin}
\ee
where $\omega_j = 2\pi j /t_1$, $\epsilon_j = 2E - \omega_j$, and $ \Gamma_j = \Gamma (\omega_j)$.\\
The energy density of the produced particles is:
\begin{subequations}\label{rho-damped}
\begin{align}
&\hspace{-.8cm}\rho = \frac{\Delta_R^2\,\sigma^2 }{72\pi t_1^2}\,\int_0^\infty dE\,E \sum_j \,e^{-\sigma^2 \omega_j^2}\,\frac{1 + e^{-2\Gamma_jt} -2\,e^{-\Gamma_jt} \cos (\epsilon_jt)}{\epsilon_j^2 +\Gamma_j^2} \\
&\hspace{-.6cm}=\frac{\Delta_R^2\,\sigma^2}{288\pi t_1^2}\sum_j e^{-\sigma^2\omega_j^2}
\int_{-\omega_j t}^\infty\,d\eta_j(\eta_j+\omega_j t)\,\frac{1 + e^{-2\Gamma_jt} -2\,e^{-\Gamma_jt} \cos \eta_j}{\eta_j^2 +(\Gamma_jt)^2}\,.\qquad  
\end{align}
\end{subequations}
We have introduced here  a new integration variable
$\eta_j\equiv \epsilon_j t$. It was also assumed that the diagonal terms with $j=k$ dominate in the double sum over $j$ and $k$ in the expression for $|A (p_1,p_2)|^2$; this is a good approximation for small $\Gamma$, in particular, $\Gamma \ll \omega$. 
Assuming $\omega_jt\gg 1$, the integral is easily taken at the poles of the denominator and we finally obtain:
\be
\rho = \frac{\pi \Delta_R\, \sigma^2}{144\,t_1^3} \,\sum_j j\,e^{-\sigma^2\omega_j^2}\,\frac{1 - e^{-2\Gamma_j t}}{\Gamma_j}.
\label{rho-fin}
\ee
For $\Gamma_j t \ll 1$ this result coincides  with (\ref{eq:PP_general}) after dividing by the total elapsed time $t$. The summation over $j$ can be separated into two regions of large and small $\Gamma_j t$. The boundary value of $\omega_j$ is given by the condition $2\Gamma_b t=\omega_b^3 t /24 m_{Pl}^2 =1$. Correspondingly
\be
\omega_b \simeq  {180} \,{\rm MeV} \left( t_c / t\right)^{1/3}\,,
\label{omega-b}
\ee
where $t_c = {4}\cdot 10^{17}$ s is the cosmological time. The boundary value of $j$ is 
\be
j_b  = \frac{ \omega_b t_1 }{2\pi} = \frac{\omega_b}{ \Omega m \sqrt{g}}, 
\label{j-b}
\ee
where we took $t_1 = 2\pi/\omega$ with  $\omega$ defined in eqs.~\rif{eq:frequency_2} and \rif{eq:Omega}.
In particular, for small $g$ we have $\Omega \simeq z^{n+1}/\sqrt{2n+1}$ and using Eq.~\rif{g-kappa} we find
\be
j_b \simeq {1.6} \cdot 10^{41}\, \frac{\sqrt{(2n+1) {n}}}{(z\, \rho_{29})^{n+1}}\,\left(\frac{t_c}{t}\right)^{1/3}.
\label{j-b2}
\ee

Accordingly, the time $t_1$ can be estimated as
\be
t_1 \approx 4\cdot 10^{18}\,{\rm  s}\,\frac{\sqrt{(2n+1)n}}{(z\, \rho_{29})^{n+1}}.\ee
Separating the summation over $j$ into two intervals of small and large $\Gamma_j t$, we obtain:
\be\label{eq:rho-sum-separ}
{\rho \approx \frac{\pi \Delta_R^2\,\sigma^2}{144\,t_1^3} \,\left[ 2t \sum_{j=1}^{j_b}  j\,e^{-(\sigma \omega_j)^2} + \sum_{j=j_b}^{\infty} 
 \frac{j\,e^{-(\sigma \omega_j)^2}}{\Gamma_j}\right]\,.}
\ee
If $\sigma \sim 1/m$, the exponential suppression factor $\exp (-\sigma \omega_j)^2$ is weak near $j = j_b$ and the sums over $j$ can be easily evaluated:
\be
{\rho \approx \frac{\pi \Delta_R^2 \sigma^2}{144\,t_1^3} \,\left( j_b^2 t + \frac{6 m_{Pl}^2 t_1^3}{\pi^3 j_b}\right) }\,.
\label{rho-fin-sum}
\ee
Alternatively, using the fact that $\sigma/t_1\ll 1$, we can replace the summation with an integral, obtaining the similar, more general result:
\begin{align}
\rho &\simeq \frac{\pi \Delta_R^2\,\sigma^2}{144\,t_1^3}\left[2t\int_0^{j_b}dj\,j\,e^{-\sigma^2\omega_j^2}+\int_{j_b}^\infty dj\,\frac{j\,e^{-\sigma^2\omega_j^2}}{\Gamma_j}\right] \notag\\
&\simeq \frac{\Delta_R^2\,t}{576\pi\,t_1}\left[1-e^{-\sigma^2\omega_b^2}\right]+\notag\\
&\qquad + \frac{\Delta_R^2\,\mpl^2\,\sigma^2}{12\pi\,\omega_b\,t_1}\left[e^{-\sigma^2\omega_b^2}-\sqrt\pi\,\sigma \omega_b\,\text{erfc}(\sigma\omega_b)\right]\,,
\label{eq:rho-fin-integral}
\end{align}
where $\text{erfc}(x)$ is the complementary error function:
\[
 \text{erfc}(x) = \frac{2}{\sqrt\pi}\int_x^{\infty} dt\,e^{-t^2}\,.
\]
Note that for $\sigma\omega_b\ll 1$, as was assumed before, this gives
\be
\rho\simeq \frac{\Delta_R^2\,\omega_b^2\,\sigma^2}{576\pi\,t_1}\left(t+\frac{48\mpl^2}{\omega_b^3}\right)= \frac{\Delta_R^2\,\omega_b^2\,\sigma^2\,t}{192\pi\,t_1}\,,
\label{eq:rho-fin-sum-equiv}
\ee
which is exactly equivalent to \rif{rho-fin-sum}. For $\sigma\omega_b\gg 1$, instead, we recover \rif{eq:PP_general}. 
This makes sense, because $\sigma\omega_b\gg 1$ corresponds to $j_b\to\infty$ and hence to the limit in which the time elapsed is not long enough for particle production to have had a noticeable back-reaction on curvature. Nonetheless, during this time particles may have been effectively produced.

These estimates are valid even in the spike region when $g z^{2n+2} \ll 1$ but $gy^{2n+2}$ may reach values much larger than unity.

\subsection{Damping of Oscillations \label{ss-damping}}
As we have seen, having a wide frequency spectrum of 
the Fourier transform of $R$ makes it impossible to simply use an exponential damping $R\to R\,e^{-\Gamma t}$ to include the effects of particle production. Moreover, the increasing energy density acts as a source term and increases the amplitude of the oscillations of $R$, which makes things even more complicated. However, the picture for the field $\xi$ is relatively simple, because its oscillations are almost harmonic. We still have a source component, given by the increasing $z$, but we can once again use the energy conservation equation to determine the time at which oscillations basically stop due to the damping. The effect of particle production on the evolution equation for $\xi$ (see Eq.~\ref{eq:xi_evol_approx}) is to transform it into
\be
\xi_1'' + 2\gamma(\Omega)\,\xi_1' + \Omega^2\xi_1 = -\xi_a''\,,
\ee
where
\be
\gamma(\Omega)\equiv \frac{\Gamma(m\,\Omega \sqrt g)}{m\sqrt g} = \frac{\Omega^3m^2g}{48\,\mpl^2}\,,
\ee
which in fact for $\gamma/\Omega\ll 1$ and $\xi_a''/\xi_1''\ll 1$ generates the wanted behaviour
\be
\xi_1 \sim e^{-\gamma\tau}\,\sin\Omega\tau\,.
\ee
Correspondingly, the energy conservation~\rif{eq:conserved} becomes
\be
\frac{1}{2}\,\xi'^2 + U(\xi) + 2\int_{\tau_0}^\tau d\eta\,\gamma\,\xi_1'^2 - \int^\tau_{\tau_0} d\eta \,z'\,\xi = \frac{1}{2}\,\xi_0'^2+ U(\xi_0)
\label{balance-gamma}
\ee
Let us consider values of $\tau$ in which $\xi(\tau) = \xi_a(\tau)$. In this case the potential disappears from this 
equation [see Eq.~\rif{eq:int_source_potential}], which can be now symbolically written as an equality between 
the variation of the kinetic energy and two integral terms:
\begin{subequations}\label{eq:balance_PP}
\be
\Delta K = I_{source} - I_\gamma\,,
\ee
where
\begin{align}
&\Delta K = K-K_0 \equiv \frac{1}{2}(\xi'^2 - \xi_0'^2)\,,\\
&I_{source} \equiv \int^\tau_{\tau_0}d\eta\,z'\,\xi_1\,,\\
&I_\gamma \equiv 2\int^\tau_{\tau_0}d\eta\,\gamma\,\xi_1'^2\,.
\end{align}
\end{subequations}
By definition $\xi = \xi_a$ is the position of $\xi$ at the minimum of the potential. So when
$\xi=\xi_a$, the kinetic energy, $K$, reaches one of the local maxima (in time).
This picture is particularly clear if we compare the system with a classical oscillator: when the field passes through the equilibrium point, the potential has minimal value and the velocity is maximal.

We estimate the effect of damping perturbatively applying the energy balance (\ref{balance-gamma})
with unperturbed functions $\xi (\tau)$ for which the effects of damping are neglected. In the absence of damping, the condition (\ref{eq:balance_PP}) turns into $\Delta K = I_{source}$, which is essentially Eq. (\ref{max_kin_estim}). Evidently the impact of damping on the oscillations of $\xi$ starts to be important when $I_\gamma$ becomes of the order of $\Delta K$. Though the damping coefficient is small, \bf{i.e.} $\Gamma \ll \omega$, the integral $I_\gamma$ rises with time faster than $\Delta K$ and ultimately it will overtake it. So we need to check when the condition 
\be 
\alpha^2 (\tau)\Omega^2 (\tau)  \sim 2\int^\tau d\eta\, \gamma (\eta) \alpha^2 (\eta) \Omega^2 (\eta)
\label{K-equal-gamma}
\ee
starts to be fulfilled with $\alpha$ and $\Omega$ given by~\rif{eq:xi_amplitude_solution} and~\rif{eq:frequency_2}.

Keeping in mind that $\gamma \tau = \Gamma t$ and using Eq.~(\ref{eq:lifetime}) for $\Gamma = 1/\tau_R$
with $\omega  = \Omega m \sqrt{g}$, we find that the equality~\rif{K-equal-gamma} is satisfied when the energy density is equal to
\begin{align}
z_\gamma^{3n+4} &= \frac{24(2n+1)^{3/2}(4n+5)\kappa\,\mpl^2}{g\,m^2}\notag \\
&= \frac{24(2n+1)^{3/2}(4n+5)}{g^{3/2}}\left(\frac{\mpl}{m}\right)^3\frac{1}{\mpl\,t_{contr}}\notag \\
&\simeq 6\times 10^{123}\,\frac{[n(2n+1)]^{3/2}(4n+5)}{t_{10}\,\rho_{29}^{3n+3}}
\label{t-gamma}
\end{align}
The corresponding boundary value of $t$ is $t_\gamma = z_\gamma\,t_{contr}$. 

For times smaller than $t_\gamma$ the effects of damping are negligible and particle production 
can be very effective giving rise to substantial production of cosmic rays, as we shall see in the next section.
Particle production is especially pronounced in the spike region when the amplitude of $R$ is very large.

As we have seen, the spikes' width is $\sigma\sim m^{-1}$ while their spacing, which determines the effective frequency, is $t_1\sim \omega^{-1}$. High frequency oscillations of $R$ should be damped very rapidly, since $\Gamma \sim \omega^3$, but  due to the non-harmonicity of the potential and the non-linearity of the relation between $\xi$ and $y$ or $R$, the low frequency oscillations are efficiently transformed into high frequency spikes of small amplitude in $\xi$ but of very large amplitude in $R$. It is worth noting in this connection that the bulk of the energy density associated with the oscillations of $R$ is concentrated at low frequencies, $\rho_{osc} \sim \Delta_R^2 m_{Pl}^2 /\omega^2 $, so the energy reservoir at low frequencies is deep enough to feed up the spikes.

The physical frequency, $\omega= \Omega m \sqrt{g}$, depends upon the product $Q \equiv g z^{2n+2}$. If $Q \ll 1$ the frequency may be rather low, of the order of hundred MeV, while for $Q \gg 1$ the frequency reaches 
the maximum value $\omega = m$. In the first case the lifetime of harmonic oscillations could be larger than
the universe age, while in the second case  it would be shorter than a second.

\section{Estimate of Cosmic Ray Emission}
\subsection{Regular Region}

Let us consider a cloud (e.g. a protogalaxy) with total mass $M$ and density $\rho$. Particles
would be uniformly produced over its whole volume, which is equal to:
\be\label{eq:volume}
V=\frac{M}{\rho} = 2\times 10^{73} \,{\rm cm}^3 \,\frac{M_{11}}{z\,\rho_{29}}\,,
\ee
where the mass of the cloud, $M$, is expressed in terms of the solar mass $M_\sun$:
\be
M_{11}\equiv \frac{M}{10^{11}M_\sun} = \frac{M}{2\times 10^{44}\text{ g}}\,.
\ee
In the regular case the oscillations of $R$ are almost harmonic, so we can rely on the adiabatic approximation and use Eq.~(\ref{eq:PP_harmonic}) 
for the particle production rate or Eq.~(\ref{eq:part_prod_rate}) 
corrected by the damping factor ${\exp[-2\Gamma (\omega) t]}$. To be more precise, in the exponent we should take the integral 
of $\Gamma$ over time to take into account the (slow) variation of $\omega$.

The total luminosity relative to gravitational particle production is obtained multiplying the rate of energy production per unit time and 
volume \rif{eq:part_prod_rate} by the total volume \rif{eq:volume}, that is $L= V(t)\dot\rho_{PP}(t)$, or
{\small
\be\label{eq:luminosity_regular}
\frac{L_{reg}}{\text{GeV s}^{-1}}\simeq 7.3 \times 10^{-74}\,N_s\,\frac{\sqrt{(2n+1+g)n}\,M_{11}\,z^{4n+3}}
{\left(1+\cfrac{gz^{2n+2}}{2n+1}\right)^2\rho_{29}^n\,t_{10}^2}\, e^{-2\int^t_{t_0}  dt'\,\Gamma (\omega)}\,,
\ee
}
where $\omega$ depends upon time due to the variation of $z(t)$, see eqs.~\rif{eq:frequency_2}, \rif{eq:Omega}. The initial time $t_0$ should be taken at the onset of structure formation, when the energy density locally started to rise.

For $gz^{2n+2}\ll 1$, the luminosity is negligible with respect to the luminosity
in the spike region (see below), so we will not consider this case further. When $gz^{2n+2}$ becomes larger than unity and $\omega \sim m$, the lifetime of oscillations turns out to be at most a few seconds and an explosively fast particle production takes place. The integrated luminosity can be approximately obtained from Eq.~(\ref{eq:luminosity_regular}) dividing it by $\Gamma (m)$ and taking the exponential factor equal to 1. 
The time duration of the production process, of the order of a few seconds, is close to that of some Gamma Ray Bursts but the characteristic particle energies are much higher, instead of MeV it is of the order of the scalaron mass $m\gtrsim 10^5$ GeV.

If $g>1$, oscillations are very mildly excited and particle production is negligible. This can be seen from \rif{eq:y_amplitude_evolution} with $g>1$, keeping in mind that $\kappa\sim g^{-1/2}$. Alternatively one may use Eq.  \rif{eq:luminosity_regular}
with  $g>1$ taking into account that a large value of $g$ corresponds to large values of $\rho_{29}$ and/or of $n$.
 
\subsection{Spike Region}
There remains to consider the spike region, where we need to use~\rif{eq:part_prod_spikes} instead of \rif{eq:part_prod_rate}, or \rif{rho-fin} and \rif{rho-fin-sum}. Equation \rif{eq:part_prod_spikes} yields
\be
\label{eq:luminosity_spike}
\frac{L_{sp}}{\text{GeV s}^{-1}} \simeq 6.0\times 10^{20}\,\frac{C_3\,N_s\,M_{11}\,m_5^2\,z^{2n+1}}{t_{10}^2\,\rho_{29}^{3n+2}}
\ee
where $C_3$ is given by Eq.~(\ref{C3}). 
This result is valid when $\Gamma_j t < 1$ for all essential values of $j$, see Eq.~\rif{rho-fin}, that is when the damping due to particle production is negligible. See also the discussion in Sec. \ref{ss-damping}.

In the opposite case, we cannot use the approximate account of damping made above (see Eq.~\ref{eq:luminosity_regular}) 
because oscillations are strongly anharmonic. If modes with both $\Gamma_j t$ greater and smaller than unity are essential, we 
have to use eqs.~\rif{rho-fin} or {\rif{eq:rho-fin-integral}}.
We start by rewriting Eq.~\rif{eq:rho-fin-sum-equiv} using $\omega_b^3=24\mpl^2/t $, $\Delta_R=\beta_{spikes}\,T_0$ and $t_1=2\pi/\omega$, where $\beta_{spikes}$ and $\omega$ are given, respectively, by \rif{eq:beta_spikes} and \rif{eq:frequency_2}-\rif{eq:Omega}. This yields, assuming $gz^{2n+2}\ll 1$ and $t\simeq z\,t_{contr}$,
\be
\frac{\rho}{\text{GeV m}^{-3}} \simeq 1.1\times 10^{-40}\,\frac{C_3\,(\sigma m)^2\,z^{2n+7/3}}{t_{10}^{5/3}\,\rho_{29}^{3n+1}}
\ee
and
\be\label{eq:spikes_rhodot_damped}
\frac{\dot\rho}{\text{GeV\,s}^{-1}\text{m}^{-3}} \simeq 6.9\times 10^{-58}\,\frac{\tilde C_3\,(\sigma m)^2\,z^{2n+4/3}}{t_{10}^{8/3}\,\rho_{29}^{3n+1}}\,,
\ee
where
\be
\tilde C_3 = \frac{(2n+1)(n+1)}{(2n+1+gz^{2n+2})}\,C_3\,.
\ee
From~\rif{rho-fin} it is clear that for each mode labelled by $j$ we have
\be
\dot\rho_j \simeq \frac{\pi\,\Delta_R^2\,\sigma^2j\,e^{-\sigma^2\omega_j^2}}{72\,t_1^2}\,e^{-2\Gamma_j t}\,
\ee
which has the predicted behaviour $\dot\rho\sim e^{-2\Gamma t}$. However, this behaviour is not obvious in the complete solution~\rif{eq:spikes_rhodot_damped}. This means that in this case the overall effect is more complicated than a simple exponential damping. The anharmonicity of the oscillations and the dependence of both $\dot\rho$ and $\Gamma$ on the frequency give non-trivial results which were impossible to predict without performing explicit calculations. 

When the damping due to particle production is relevant, the total luminosity becomes
\be
\label{eq:luminosity_spikes_damped}
\frac{L_{sp}}{\text{GeV s}^{-1}} \simeq 1.4\times 10^{10}\,\frac{\tilde C_3\,N_s\,M_{11}(\sigma m)^2z^{2n+1/3}}{t_{10}^{8/3}\,\rho_{29}^{3n+2}}
\ee
This value, though smaller than~\rif{eq:luminosity_spike}, might not be completely negligible, especially for short contraction times and relatively small initial densities. This means that even with the damping of oscillations taken into account, the produced cosmic rays could in principle be detectable.

\section{Discussion and Conclusions}

We have shown that in contracting astrophysical systems with rising energy density, powerful oscillations of curvature scalar $R$ are induced. Initially harmonic, these oscillations evolve to strongly anharmonic ones with high frequency and large amplitude, which could be much larger than the value of curvature in standard General Relativity.

Such oscillations result in efficient particle production in a wide energy range, from a hundred MeV up to the scalaron mass, 
$m$, which could be as large as $10^{10} $ GeV (and maybe even larger). Such high frequency oscillations could be a 
source~\cite{Arbuzova:2012su} of ultra high energy cosmic rays (UHECR) with $E \sim  10^{19}-10^{20}$ eV, see e.g. the review in~\cite{PhysRevD.86.010001}, which might avoid the GZK cutoff~\cite{Greisen:1966jv,Zatsepin:1966jv}, and may even have implications for the so-called ``ankle'' problem~\cite{Abbasi:2005ni, Abbasi:2007sv, Tsunesada:2011mp, Abreu:2011pj}.

Possibly the considered mechanism would give too large a fraction of high energy photons in UHECRs, see e.g.~\cite{Aloisio:2004ap,Kalashev:2008dh}, if no special care is taken, because gravity couples to all elementary particles with the same intensity. However, direct photon production may be suppressed due to the conformal invariance of electrodynamics.
To avoid a too strong indirect photon production one may need to introduce "photo-fobic" heavy particles predominantly created by the oscillating curvature.

It is tempting to explain gamma bursts by these curvature oscillations, but the emitted particle energy seems to be
much above the MeV range. To this end some modification of the model or a mechanism of energy depletion would be necessary, and could be an interesting subject of future research.

The oscillations considered here may also have an essential impact on the gravitational (Jeans) instability in $F(R)$ gravity studied for instance in~\cite{Capozziello:2011nr,Capozziello:2011et,Capozziello:2011gm}, where this effect was not taken into consideration.

The efficiency of particle production strongly depends upon the system under scrutiny, the values of the parameters
of the theory, and upon the explicit form of the function $F(R)$. These problems deserve further study, but the framework presented in this paper can be applied to many possible cases.

\chapter{Spherically Symmetric Solutions in \titlefR Gravity and Gravitational Repulsion}
\chaptermark{\spacedlowsmallcaps{Sph. Symm. Solutions and Grav. Repulsion}}
\label{ch:spher_symm_grav_repuls}

\citazione{\footnotesize E.V. Arbuzova, A.D. Dolgov, {\color{myhypercolor}\bf{L. Reverberi}}, \it{Astropart. Phys.} \bf{54}, 44 (2014).}{}

\section{Introduction}\label{s-introduction}

Popular $F(R)$ models phenomenologically acceptable for cosmology have been suggested in~\cite{Starobinsky:2007hu,Hu:2007nk,Appleby:2007vb}. They are more or less equivalent, particularly the former two, and in what follows we will use the specific $F(R)$ of~\cite{Starobinsky:2007hu}:
\be
F(R) =  -\lambda R_0\left[1-\left(1+\frac{R^2}{R_0^2}\right)^{-n}\right] - \frac{R^2}{6m^2},
\label{F-of-R}
\ee
where $R_0$ is a constant parameter with dimensions of curvature and similar in magnitude to the cosmological curvature at the present day universe, $\lambda$ is a dimensionless constant of order unity and the power $n$ is usually taken to be an integer (though not necessarily so).

The last term is introduced to avoid infinite $R$ singularities in the past cosmology~\cite{Appleby:2009uf} or in the future in astronomical systems with rising energy density~\cite{Frolov:2008uf,Arbuzova:2010iu,Reverberi:2012ew}.  

The corresponding field equations are
\be\label{eq:eq-of-mot}
(1+F_{,R})\,R_\mn - \frac{R+F}{2}\,g_\mn + (g_\mn\square - \D_\mu\D_\nu)F_{,R} = \frac{8\pi}{\mpl^2}\,T_\mn\,,
\ee
whose trace is
\be\label{D2-R}
3\square F_{,R} - R + RF_{,R} - 2F = \tilde T\,,
\ee
with $\tilde T = \tilde{T}^\mu_\nu$ and $\tilde T_\mn = 8\pi\,T_\mn/\mpl^2$. 

A detailed study of the solutions of the modified gravity equations in the present day universe was performed in~\cite{Arbuzova:2012su,Arbuzova:2013ina} for finite-size astronomical objects. It was found that if the energy density rises with time, fast oscillations of the scalar curvature are induced, with an amplitude possibly much larger than the usual GR value $R=-\tilde T$. The solution has the form:
\be
R = R_{GR} (r) y (t),
\label{R-of-tSSR}
\ee
where $R_{GR} = - \tilde T(r)$ is the would-be solution in the limit of GR, while the quickly oscillating function $y(t)$ may be much larger than unity. According to~\cite{Arbuzova:2013ina} the maximum value of $y$ in the so-called spike region is:
\be
y(t) \sim 6 n (2n+1)  m t_u \left(\frac{t_u}{t_{contr} }\right) \left[\frac{\rho_m (t)}{\rho_{m0}} \right]^{(n+1)/2}
\left(\frac{\rho_c}{\rho_{m0}} \right)^{2n+2} ,
\label{y-of-t}
\ee 
where $t_u$ is the universe age, $t_{contr} $ is the characteristic contraction time, so the energy density of the contracting cloud behaves as $\rho_m (t) = \rho_{m0} (1 + t/t_{contr})$, with $\rho_{m0} $ being the initial energy density of the cloud, and $\rho_c~=~10^{-29}$~g/cm$^3$ being the present day cosmological energy density. According to~\cite{Arbuzova:2011fu}, the mass parameter $m$ entering eq.~(\ref{F-of-R}) should be larger than about $10^5$ GeV to avoid a conflict with BBN. So the factor $m t_u $ is enormous: $ m t_u \geq 10^{47}$ and $y$ can  reach a very high value, if not suppressed by a small ratio $(\rho_c/\rho_{m0})^{2n+2}$, when $n$ is  large.

As shown in~\cite{Arbuzova:2013ina}, such spikes of high amplitude are formed if 
\be
6 n^2 (2n+1)^2 \left(\frac{t_u}{t_{contr}}\right)^2 \left[\frac{\rho_m (t)}{\rho_{m0}} \right]^{3n+1}
\left(\frac{\rho_c}{\rho_{m0}} \right)^{2n+2} > 1.
\label{spike-region}
\ee
The values of the densities $\rho_{m0}$ and $\rho_m (t)$ depend upon the objects under scrutiny. If we speak about formation of galaxies or clusters thereof, the following ratios can be expected: $\rho_{m0}/\rho_c = 1 - 10^3$ and $\rho_{m} (t) / \rho_{m0}  $ varying  in the range $1 - 10^5$.  Indeed the oscillations of  curvature in such systems are excited if their mass density rises with time. For large scale structures this process began when they decoupled from the overall Hubble flow, which mostly took place at redshifts in the interval $z = 10 - 0$, and could result in creation of galaxies with energy density 5 orders of magnitude higher than the present day cosmological one. If we consider the formation of stellar or planetary objects from the intergalactic gas with initial density $10^{-24} $ g/cm$^3$, then $\rho_{m0}/\rho_c = 10^5$ and $\rho_m (t) / \rho_{m0} $ can vary in the range $ 1 - 10^{24}$ or even larger.

If the condition (\ref{spike-region}) is not fulfilled and the spiky solution with high amplitude is not excited, still as calculations of refs.~\cite{Arbuzova:2012su,Arbuzova:2013ina} show, both numerically and analytically, the amplitude of $y(t)$ would be larger than unity, which is essential for the result presented below about gravitational repulsion inside systems with rising energy density.

\section{Spherically Symmetric Solutions in \titlefR Gravity}

The analysis in~\cite{Arbuzova:2012su,Arbuzova:2013ina} has been done under the assumption that the background 
space-time is nearly flat and so the background metric is almost Minkowsky. However, the large deviation of curvature from its GR value, found in these works, may invalidate the assumption of an approximately flat background and should be verified.

In what follows we consider a spherically symmetric bubble of matter, e.g. a gas cloud or some other astronomical object, which occupies a finite region of space of radius $r_m$, and study spherically symmetric solution of the field equations~\ref{eq:eq-of-mot} assuming that the metric has the Schwarzschild form:
\be
ds^2 =  A (r,t) dt^2 - B(r,t) dr^2 - r^2 (d\theta^2 + \sin^2 \theta\,d\phi^2) .
\label{ds2}
\ee
A metric of this type in $F(R)$ theories was analysed e.g. in~\cite{delaCruzDombriz:2009et,Cembranos:2012fd} but the curvature oscillations, which are in the essence of our work (see below), were not taken into account there. If these
oscillations are not taken into account our results agrees with the papers cited above.

We assume that the metric coefficients $A$ and $B$ weakly deviate from unity and check when this is true. The nonzero components of the Ricci tensor corresponding to the metric~(\ref{ds2}) are:
\begin{subequations}
\begin{align}
& R_{00}=\frac{A''-\ddot B}{2B}+\frac{(\dot B)^2-A'B'}{4B^2}+\frac{\dot A \dot B - (A')^2}{4AB}+\frac{A'}{rB}\, , 
\label{R00}\\
& R_{rr}=\frac{\ddot B -A''}{2A}+\frac{(A')^2-\dot A \dot B}{4A^2}+\frac{ A' B' - (\dot B)^2}{4AB}+\frac{B'}{rB}\, , 
\label{R11}\\
& R_{\theta \theta }=-\frac{1}{B}+\frac{rB'}{2B^2}-\frac{rA'}{2AB}+1\, , 
\label{R22} \\
& R_{\varphi \varphi }=\left(-\frac{1}{B}+\frac{rB'}{2B^2}-\frac{rA'}{2AB}+1\right )\sin^2\theta  = R_{\theta \theta} \sin^2\theta \, ,
\label{R33} \\
& R_{0r} =\frac{\dot B}{rB}\, .
\label{R01}
\end{align}
\end{subequations}
Here a prime and an overdot denote differentiation with respect to $r$ and $t$, respectively. The corresponding Ricci scalar is equal to:
\begin{align}
R&=\frac{1}{A}R_{00}-\frac{1}{B}R_{rr}-\frac{1}{r^2}R_{\theta \theta}-\frac{1}{r^2\sin^2\theta }R_{\varphi \varphi}  \notag \\
&=\frac{A''-\ddot B}{AB}+\frac{(\dot B)^2-A'B'}{2AB^2}+\frac{\dot A \dot B - (A')^2}{2A^2B}+\notag \\
&\qquad\qquad + \frac{2A'}{rAB}-\frac{2B'}{rB^2}+\frac{2}{r^2B}-\frac{2}{r^2}\notag \\
&= \frac{2}{A}R_{00}-\frac{2B'}{rB^2}+\frac{2}{r^2B}-\frac{2}{r^2} \, .
\label{Rscalar}
\end{align}
We assume that the metric is close to the flat one, i.e.
\be
A_1 = A - 1 \ll 1\,\,\,{\rm and} \,\,\, B_1 = B-1 \ll 1
\label{A1SSR-B1}
\ee
and study if and when this assumption remains true for the solutions with very large values of $R$ found in our previous works~\cite{Arbuzova:2012su,Arbuzova:2013ina}. It is convenient to use equations~\rif{eq:eq-of-mot} in the following form:
\bea
R_{00}  - R/2 &=& \frac{ \tilde T_{00} + \Delta F_{,R} + F/2 - RF_{,R}/2}{1+F_{,R}},
\label{R-00} \\
R_{rr}  + R/2 &=&  \frac{ \tilde T_{rr} + (\partial_t^2 +\partial_r^2 - \Delta) F_{,R} - F/2 + RF_{,R}/2}{1+F_{,R}} ,
\label{R-rr}
\eea
because their  left hand sides contain only first derivatives of the metric coefficients. In the weak field limit, when derivatives of $A(r,t)$ and $B(r,t)$ are sufficiently small so that their square can be neglected, we obtain the following expressions for the $R_{00}$ and $R_{rr}$ components of the Ricci tensor and for the Ricci scalar $R$:
\begin{subequations}\label{R_ij_weak}
\begin{align}
&R_{00} \approx \frac{A''-\ddot B}{2}+\frac{A'}{r}\, , 
\label{R00-weak}\\
&R_{rr} \approx \frac{\ddot B -A''}{2}+\frac{B'}{r}\, , 
\label{R11-weak}\\
&R \approx A''-\ddot B+\frac{2A'}{r}-\frac{2B'}{r}+\frac{2(1-B)}{r^2}\, .
\label{R-weak}
\end{align}
\end{subequations}
If the energy density of matter inside the the cloud, i.e. for  $ r<r_m$, is much larger than the cosmological energy density,
the following restrictions are fulfilled: 
\be
F_{,R} \ll 1\,\,\,\,\, {\rm and}\,\,\,\,\,\, F \ll R
\label{bounds-on-F}
\ee 
For static solutions the effects of gravity modifications in this limit are weak and, as we will see in what follows, 
the solution is quite close to the standard Schwarzschild one in agreement with other works 
on this subject. We assume that the spatial derivatives 
of $F'_R$ are small in comparison with the time derivatives. This assumption is justified a posteriory because we use in what follows the 
solution with quickly oscillating $R$ found in refs. [8,9]. The characteristic time variation of this
solution is microscopically small, while the space variation scale is macroscopically large.
So from eq.~(\ref{D2-R}) it follows that $(\partial_t^2 - \Delta ) F_{,R} = (\tilde T + R)/3 $ and we find:
\begin{subequations}
\begin{align}
&\hspace{-.8cm}B_1' + \frac{B_1}{r} = r \tilde T_{00},
\label{B1-prime} \\
&\hspace{-.8cm}A''_1-\frac{A_1'}{r} = -\frac{3B_1}{r^2} +\ddot B_1 +\tilde T_{00}-2\tilde T_{rr} + \frac{\tilde T_{\theta\theta}}{r^2} +
 \frac{\tilde T_{\varphi\varphi}}{r^2\sin^2\theta}
\equiv S_A\, .
\label{A1SSR-two-prime}
\end{align}
\end{subequations}
Since we assumed small deviations from the Minkowsky metric, we neglected the corresponding corrections in $T_{\mu \nu}$. The validity of this assumption is precisely what we have to check.

Equation~(\ref{B1-prime}) has the solution:
\be 
B_1(r,t) = \frac{C_B(t)}{r}+\frac{1}{r}\,\int_0^r dr' r'^2 \tilde T_{00} (r',t)\, .
\label{B1-new}
\ee
To avoid a singularity at $r=0$ we have to assume that $C_{B}(t) \equiv 0$. Then this expression for $B_1$ formally coincides with the usual Schwarzschild solution, while the equation determining the metric coefficient $A_1$ allows for an additional freedom:
\be 
A_1(r,t)=C_{1A}(t)r^2+C_{2A}(t)+\int^{r_m}_r dr_1\,r_1 \int^{r_m}_{r_1} \frac{dr_2}{r_2}\, S_A(r_2,t)\, .
\label{A1SSR-common}  
\ee
The integration limits are chosen in such a way that the singularity at $r_2=0$ is avoided. Using equation (\ref{B1-new}) with $C_{B}=0$ we can rewrite $S_A$ as:
\begin{align}
\notag & S_A = -\frac{3}{r^3}\,\int_0^r dr' r'^2 \tilde T_{00} (r',t) 
+\frac{1}{r}\,\int_0^r dr' r'^2 \ddot {\tilde {T}}_{00} (r',t) \, + \\
&\qquad\qquad + \tilde T_{00}-2\tilde T_{rr} + \frac{\tilde T_{\theta\theta}}{r^2} +
 \frac{\tilde T_{\varphi\varphi}}{r^2\sin^2\theta}\, .
\label{S-A}
\end{align}
Accordingly we obtain the following expression for $A_1(r,t)$:
\begin{align}
 A_1(r,t) &= C_{1A}(t)r^2 + C_{2A}(t) + \notag \\
&\quad +\int_{r}^{r_m} dr_1\,r_1\int_{r_1}^{r_m} \frac{dr_2}{r_2}\, \left[\tilde T_{00} (r_2,t) - \right.\notag\\
&\quad\left. -2\tilde T_{rr}(r_2,t) +  \frac{\tilde T_{\theta\theta}(r_2,t)}{r^2} + \frac{\tilde T_{\varphi\varphi}(r_2,t)}{r^2\sin^2\theta}\right]
-\notag \\
&\quad - \int_{r}^{r_m} dr_1\,r_1\int_{r_1}^{r_m} \frac{dr_2}{r_2} \, \left[\frac{3}{r_2^3}\,\int_0^{r_2} dr' r'^2 \tilde T_{00} (r',t) - \right.\notag \\
&\quad \left. - \frac{1}{r_2}\,\int_0^{r_2} dr' r'^2 \ddot {\tilde {T}}_{00} (r',t)\right].
\label{A-1}
\end{align}

\subsection{The Schwarzschild Case}

It is instructive to check how solutions (\ref{B1-new}) and (\ref{A-1}) reduce to the vacuum Schwarzschild solution in GR. The mass of matter inside a radius $r$ is defined in the usual way:
\be
M (r,t) =\int_0^{r} d^3r\,  T_{00} (r,t)=4\pi \int_0^{r} dr\,  r^2\,  T_{00} (r,t)
\label{M}
\ee
If all matter is confined inside a radius $r_m$, the total mass is $M \equiv M(r_m)$ and due to mass conservation it does not depend on time. Since $\tilde T_{00}=8\pi T_{00}/m_{Pl}^2$, we obtain for $r>r_m$, as expected, $B_1=r_g/r$, where $r_g=2M/m_{Pl}^2$ is the usual Schwarzschild radius. 
 
Let us turn now to the calculation of $A_1$ (\ref{A-1}). Evidently, for $r>r_m$ the first integral term vanishes because $r_2$ is also larger than $r_m$, in fact in this region we have $T_{\mu \nu}=0$. The integral containing $\ddot {\tilde {T}}_{00} $ is also zero due to total mass conservation. The remaining integral can be easily taken:
\be
\int_{r}^{r_m} dr_1\,r_1\int_{r_1}^{r_m} \frac{dr_2}{r_2} \, \frac{3}{r_2^3}\,\int_0^{r_2} dr' r'^2 \tilde T_{00} (r',t) = 
\frac{r_g}{r}+\frac{r_g\,r^2}{2r_m^3}-\frac{3r_g}{2r_m}\, .
\label{int-A}
\ee
Thus the metric coefficient outside the source is:
\be 
A_1 = -\frac{r_g}{r}+\left[C_{1A}(t)-\frac{r_g}{2r_m^3}\right] r^2 +\left[C_{2A}(t)+\frac{3r_g}{2r_m}\right] \, .     
\label{A-1-Sch}
\ee
Choosing $C_{1A}=r_g/(2r_m^3)$, to eliminate the $r^2$-term at infinity, and $C_{2A}=-3r_g/(2r_m)$ we obtain the usual Schwarzschild solution. Note that it is not necessary to demand that the space-independent constant in $A_1$ must vanish, because it can be removed by a redefinition of the time variable.

\subsection{Modified Gravity Solutions}

In the modified theory the internal solution remains of the same form (\ref{B1-new}) and (\ref{A-1}), where the coefficient $C_{1A}$, however, may depend non-trivially on time. This coefficient can be found from eq. (\ref{R-weak}) if the curvature scalar is known. As previously mentioned, we have shown in papers \cite{Arbuzova:2012su, Arbuzova:2013ina} in systems with rising energy density that the curvature scalar may be much larger than its value in GR. Using eqs. (\ref{B1-new}) and (\ref{A-1}) and comparing them to eq. (\ref{R-weak}) we can conclude that the dominant contribution into such form of the curvature is given by $A''+2A'/r$, i.e.
\be\label{eq:C1A_approx}
C_{1A}(t) \approx \frac{R(t)}{6}\,,
\ee
where $R(t)$ is given by eqs. (\ref{R-of-tSSR},\ref{y-of-t}).

There is an essential difference between the modified and the standard solutions in vacuum. 
In the standard case the term proportional to $r^2$ appears both at $r<r_m$ and $r>r_m$ with the same coefficient and 
hence it must vanish. On the other hand, for modified gravity such condition is not applicable and the $C_{1A} r^2$-term may be present at $r<r_m$ and absent at $r \gg r_m$. The vacuum solution for $R$ is presumably $R\sim R_c$, where $R_c$ is the (small) cosmological curvature, plus possible oscillating terms.

Thus to summarise, the metric functions inside the cloud are equal to:
\begin{subequations}\label{eq:metric_sol}
\begin{align}
&A(r,t) = 1 + \frac{R(t )\,r^2}{6} + A_1^{(Sch)} (r,t) \label{A-of-r-t}\,,\\
&B (r, t) = 1 + \frac{2M(r,t)}{\mpl^2r} \equiv  1+ B_1^{(Sch)}\,.
\label{B-of-r-t}
\end{align}
\end{subequations}
In other words we construct the internal solution assuming that it consists of two terms: the Schwarzschild one and the oscillating part generated by the rising density as is shown in our works~\cite{Arbuzova:2012su, Arbuzova:2013ina}. The expression for $A_1^{(Sch)} (r,t) $ can be found from (\ref{A-1}) with constant $C_{A1}=r_g/2r_m^3$ and $C_{A2}=-3r_g/r_m$, as determined from eq.~(\ref{A-1-Sch}). As for the integrals in eq.~(\ref{A-1}), we calculated them assuming that matter is nonrelativistic, so the space components of $T_{\mu\nu}$ are negligible in comparison to $T_{00}$, and that the matter/energy density, $T_{00} \equiv \rho_m (t)$, is spatially constant but may depend on time. 
The first two integrals in eq. (\ref{A-1}) cancel out and only the integral containing the second time derivative of the mass density survives. So for the Schwarzschild part of the solution we find: 
\be
A_1^{(Sch)} (r,t) = \frac{r_g r^2}{2r_m^3} -\frac{3r_g}{2r_m}+
 \frac{ \pi \ddot\rho_m}{3 m_{Pl}^2}\, ( r_m^2 - r^2)^2\, .
\label{A-1-2}
\ee

As we noted at the end of Sec.~\ref{s-introduction}, $R(t)$ is typically larger that the GR value: $|R_{GR}|=8 \pi \rho_m /m_{Pl}^2$, so the second term in eq. (\ref{A-of-r-t}),  $R(t)r^2/6$, gives the dominant contribution into $A_1$ at sufficiently large $r$. Indeed, $r^2 R(t) \sim r^2 y R_{GR}$ with $y > 1$, while the canonical Schwarzschild terms are of the order of $r_g/r_m \sim \rho_m r_m^2/m_{Pl}^2 \sim r_m^2 R_{GR} $.

As already mentioned, the solution with large oscillating $R(t)$ was obtained~\cite{Arbuzova:2012su,Arbuzova:2013ina}  under the assumption that the background metric weakly deviates from the flat Minkowsky one. Though this is certainly true for the Schwarzschild part of the solution~(\ref{A-1-2}), this may be questioned for the $r^2 R(t)/6$ - term. Evidently the flat background metric is not noticeably distorted if $ r ^2<  6/R(t)$. If the initial energy density of the cloud is of the order of the cosmological energy density, i.e. $R_{GR}\sim 1/t_u^2$,  then the metric would deviate from the Minkowsky one for clouds having radius $r_m > t_u/\sqrt y$,  where the maximum value of $y$ is given by eq. (\ref{y-of-t}). For systems where very large values of $y$ are reached, the flat space approximation may be broken already for non-interestingly small $r$. However, at the stage of rising $R(t)$ when $y>1$ but not too large, the flat space approximation would be valid over all the volume of the collapsing cloud. For large objects or large $y$, such that $R r^2 /6 \sim 1$, the approximation of flat background metric becomes inapplicable and one has to solve the exact non-linear equations~(\ref{R00}-\ref{R01}); this situation will be studied elsewhere. If $A_1$ becomes comparable with unity, the evolution of $R(t)$ may significantly differ from that found in~\cite{Arbuzova:2012su, Reverberi:2012ew, Arbuzova:2013ina}, but it seems evident that once a large $y>1$ is reached, it would remain larger than unity despite a possible back-reaction of the non-flat metric.

One more comment is in order here. Above we presented the solution which tends to the flat one at large distances, though strictly speaking this should not be the case, since in the considered $F(R)$ theory the metric of the empty space has the De Sitter form. Nevertheless our is approximation is good enough when the deviation of the metric from the flat one is large in comparison with the cosmological part of the metric, i.e. $A_1$ or $B_1$ are large in comparison with the De Sitter part. It is completely analogous to the case of the usual GR, when the gravitational field of an isolated body is close to the Schwarzchild one for sufficiently small distances from the center despite the cosmological FRW metric at large distances.

In the lowest order in the gravitational interaction the motion [the geodesic equation in metric (\ref{ds2})] of a non-relativistic  test particle is governed by the equation:
\be
\ddot r = - \frac{A'}{2} = -\frac{1}{2}\left[ \frac{R(t) r}{3} + \frac{r_g r}{r_m^3} \right],
\label{ddot-r}
\ee
where $A$ is given by eq. (\ref{A-of-r-t}). Since $R(t)$ is always negative and large, the modifications of GR considered here lead to anti-gravity inside a cloud with energy density exceeding the cosmological one. Gravitational repulsion dominates over the usual attraction if
\be\label{eq:antigrav_cond}
\frac{|R|r_m^3}{3r_g} = \frac{|R|r_m^3m_{Pl}^2}{6 M} = \frac{|R|r_m^3m_{Pl}^2}{8\pi \rho\,r_m^3} = 
\frac{|R|}{\tilde T_{00}} \equiv y > 1\,.
\ee
This condition may be misleading, in that it seems that whenever $R$ oscillates, so that its absolute value exceeds the GR value $\simeq \tilde T_{00}$ (for non-relativistic matter), the $f(R)$ dynamics should lead to repulsive gravity. This is of course not the case, and the reason is the following: to obtain~\rif{eq:antigrav_cond}, and particularly in deriving~\rif{eq:metric_sol}, we have assumed that $R$ oscillates with large amplitude so that
\be\label{eq:R_dominance}
|R|\gg T_{00}
\ee
at the ``top'' of oscillations, as found in Chapters~\ref{ch:curv_sing_grav_contr} and~\ref{ch:grav_part_prod_struct_form}, see~\cite{Arbuzova:2012su, Reverberi:2012ew, Arbuzova:2013ina}. Clearly, the condition~\rif{eq:antigrav_cond} is implied by~\rif{eq:R_dominance}, so it fails in giving us additional information as to when we may expect antigravity. Ultimately, the true condition for antigravity is thus~\rif{eq:R_dominance}, which ensures that~\rif{eq:C1A_approx} holds. It is probably possible to derive a more precise condition, we plan to address this issue in the future.

So, we have seen that in modified gravity and in systems with rising energy density, the curvature scalar would typically exceed the GR value $R_{GR}$, i.e. $y>1$, and thus the gravitational repulsion would dominate over the usual Schwarzschild attraction. The back-reaction of this repulsion would slow down the contraction but evidently not stop it. Moreover, the repulsion could overtake the contraction at sufficiently large radius. As a result shell type structures could be formed.
Hence the gravitational repulsion found here might be responsible for the formation of cosmic voids but the 
lengthy analysis of realistic scenarios is outside the framework of this Chapter. 

\section{Conclusions}
As it was shown in~\cite{Arbuzova:2012su,Arbuzova:2013ina}, the time evolution of curvature exhibits a periodic succession of high narrow spikes with $R>R_{RG}$ over some smooth background with relatively low $R$ -- see e.g. eq.~(4.17) of~\cite{Arbuzova:2013ina}. These oscillations are damped due to gravitational particle production but the corresponding life-time could be comparable or even larger than the cosmological time. So structure formation in modified gravity would be different from that in the standard GR. Sufficiently large primordial clouds would not shrink down to smaller and smaller bodies with more or less uniform density but could form thin shells empty (or almost empty) inside, except possibly for some central mass.  At least for some types of objects this result would modify the recent studies of the formation and stability of astronomical structures in $F(R)$ gravity~\cite{Capozziello:2011nr,Capozziello:2011gm}. We should however stress that those works neglected time derivatives, so the contrast with our results is mainly due to the different physical phenomena involved.

\chapter*{Conclusions}
\addcontentsline{toc}{chapter}{Conclusions}
\titolino{Conclusions}

Among modified theories of gravitation, $f(R)$ theories are possibly the most straightforward and ``natural'' purely geometric extension of GR. The first appearance of modified gravity theories dates back to the 1920's~\cite{Weyl:1919fi, Eddington:1922}, although their relevance and popularity vastly increased about 40 years ago, when pioneering works~\cite{Gurovich:1979xg, Starobinsky:1979ty, Starobinsky:1980te} showed the possibility of generating the early inflationary period with quadratic theories, which arise naturally from quantum corrections in curved spacetime.

After the discovery of the cosmic acceleration~\cite{Riess:1998cb,Perlmutter:1997zf,Perlmutter:1998np,Riess:2004nr}, new life was infused into $f(R)$ theories, sparked by the early works~\cite{Capozziello:2002rd, Capozziello:2003tk, Capozziello:2003gx, Carroll:2003wy, Nojiri:2003ft}. These models were soon realised to suffer from severe instabilities~\cite{Dolgov:2003px, Faraoni:2006sy}, because the additional scalar degree of freedom acquires an imaginary mass. During the following years, an impressive amount of work was directed to determining the cosmological viability conditions of $f(R)$ models~\cite{Carloni:2004kp, Clifton:2005aj, Abdelwahab:2007jp, Amendola:2006we, Amendola:2007nt, Sawicki:2007tf, de_Souza:2007fq, Tsujikawa:2007xu, Li:2007xn, Bean:2010xw}. Clearly, despite the conceptual simplicity of $f(R)$ theories, the additional dynamics and the higher-order equations make it difficult --  and fun -- to come up with ``good'' models to model dark energy.

Furthermore, these models must be tested in a variety of cosmological and astrophysical situations, and may lead to important detectable signatures which could in principle be observed soon. As important as inventing new models is, finding new ways to constrain and even exclude them is no smaller task. It is believed that $f(R)$ models can be considered as low-energy phenomenological limits of some more fundamental theory such as string theory, etc.~\cite{Birrell:1982ix, Buchbinder:1992rb, Vilkovisky:1992pb}. Every step towards the ``right'' $f(R)$ model may very well be a step towards the ``right'' theory of quantum gravity, so there is no overestimating the relevance of any result in this direction. This has been precisely the intent of my work.\\ 

Chapter~\ref{ch:vacuum_energy_puzzle} is devoted to introducing the vacuum energy problem. After a brief review of the standard cosmological scenario and of the main observational indications for a vacuum energy component, I presented a few of the most important theoretical models proposed to explain the cosmic acceleration, from both the modified matter (dark energy) and modified gravity standpoints.

In Chapter~\ref{ch:cosm_evol_R2_grav}, I studied the radiation-dominated epoch in $R+R^2$ gravity, discussing the modified curvature dynamics analytically and numerically. The curvature scalar exhibits fast oscillations around some power-law behaviour which may or may not correspond to the standard GR solution. These curvature oscillations are however damped due to gravitational particle production effects, so that eventually the solutions relax to the GR ones, but with an additional relic density of gravitationally produced particles which can in principle give some imprint on the cosmological evolution and perhaps even make up an effective mechanism to produce dark matter.

In Chapter~\ref{ch:curv_sing_grav_contr}, I investigated the formation of curvature singularities inside astronomical contracting systems within the framework of two recently proposed $f(R)$ models~\cite{Hu:2007nk, Starobinsky:2007hu}, studying the problem analytically and comparing my estimates with exact numerical results. I showed that such infinite-$R$, finite-$\rho$ singularities can arise in a number of physically reasonable systems, and derived the time scales for this to happen.

Naturally, as $R$ approaches infinity, one expects high-curvature effects to come into play. In Chapter~\ref{ch:grav_part_prod_struct_form}, I studied the curvature evolution in the models~\cite{Hu:2007nk, Starobinsky:2007hu} with the addition of an $R^2$ term (irrelevant for cosmology, but important for large $R$). I showed that this term prevents the formation of the curvature singularity while still allowing $R$ to reach very large values, and in turn may lead to strong particle production. I calculated the particle production rate, which depends on both physical properties of the system and on model parameters. These high-energy cosmic rays could in principle be detectable and, if observed, would represent a unique model-dependent signature. Some unexplained features in the cosmic ray spectrum, e.g. the so-called ``ankle''~\cite{Abbasi:2005ni, Abbasi:2007sv, Tsunesada:2011mp, Abreu:2011pj}, might find a fascinating explanation in this framework.

In Chapter~\ref{ch:spher_symm_grav_repuls}, I discussed another interesting and unexpected consequence of these high-$R$ solutions, namely the possibility of gravitational repulsion in contracting systems. The modified Einstein equations lead to new solutions for the metric in spherically symmetric systems, and the new geodesic equation for a test particle essentially shows repulsive behaviour if curvature is large compared to its GR value. The phenomenology of such an anti-gravitational behaviour, which has not been fully explored yet, is probably rather interesting and may lead to additional interesting discoveries.\\

We are experiencing astonishing advances in experimental, observational and theoretical physics. We should be very excited at what the next years and decades will bring, and I personally cannot wait until the next discovery shakes the physics world again. Among the very many problems modern physicists must face, that of dark energy is possibly the most important, the most difficult, and perhaps the farthest from being solved. The solution to this problem will likely require contributions from a variety of sectors of experimental and theoretical physics, and those contributions will surely help us shed light on other related problems. It is entirely possible that we will need to abandon the current paradigms of quantum and gravitational physics.

I am honoured that I have had the chance to do my part in this fantastic journey, and wish to continue to do so in the future.

\cleardoublepage

\appendix
\chapter{Field Equations for Modified Gravitational Theories}
\label{App:field_eqs_mod_grav}
\section{General Modified Gravitational Action}
The variation of the action
\be
 A=\idx\,\-g\,f(R,\phi)\label{eq:act}
\ee
with respect to the metric yields
\begin{align}
\delta A&=\idx\,\left[f\delta\-g+\-g\delta f\right]\notag\\
&=\idx\,\left[f\delta\-g+\-gf_{,R}\left(R_\mn\,\delta g^\mn+g^\mn\,\delta R_\mn\right)\right]\,,
\end{align}
where $f_{,R}\equiv\partial f/\partial R$ and $\delta f=f_{,R}\,\delta R$ is the functional derivative of $f$ with respect to $R$. Using
\be
\delta\-g=-\frac{1}{2}\,\-g\,g_\mn\,\delta g^\mn\,,\label{eq:var-g}
\ee
we obtain
\be
\delta A=\idx\,\-g\left[\left( f_{,R}\,R_\mn-\frac{1}{2}\,fg_\mn\right)\delta g^\mn+f_{,R}\,g^\mn\,\delta R_\mn\right]\,.
\ee
The Ricci tensor is
\be
R_\mn\equiv R^\alpha_{\mu\alpha\nu}=\partial_\alpha\Gamma^\alpha_\mn-\partial_\nu\Gamma^\alpha_{\alpha\mu}+\Gamma^\lambda_{\lambda\sigma} \Gamma^\sigma_{\nu\mu}-\Gamma^\alpha_{\sigma\nu}\Gamma^\sigma_{\alpha\mu}\,,
\ee
and its variation is given by the Palatini identity
\be
\delta R_\mn = \D_\lambda(\delta\Gamma^\lambda_\mn)-\D_\nu(\delta\Gamma^\lambda_{\lambda\mu})\,.
\ee
Please note that although the affine connection is not a tensor, its variation is a tensor, and hence its covariant derivative makes sense (see e.g.~\cite{Landau_Lifshitz:2}). The metric compatibility condition $\D_\alpha g_\mn=0$ also yields
\be
g^\mn\,\delta R_\mn=\D_\lambda\left(g^\mn\,\delta\Gamma^\lambda_\mn - g^{\mu\lambda}\,\delta\Gamma^\nu_{\mn}\right)\,.
\ee
Now we need to calculate the variation of the affine connection, or Christoffel symbols, defined by
\be
\Gamma^\alpha_\mn\equiv\frac{1}{2}\,g^{\alpha\lambda}\left(\partial_\mu g_{\lambda\nu}+\partial_\nu g_{\lambda\mu}-\partial_\lambda g_\mn\right)\,.
\ee
In all calculations, we will assume that we are working in a torsion-free manifold, namely that the Christoffel symbols are symmetric under permutation of the lower indeces. Assuming that the variation ($\delta$) and partial derivative ($\partial$) operators commute, we obtain
\begin{align}
 \delta \Gamma^\alpha_\mn =\frac{1}{2}\,&\left[\delta g^{\alpha\lambda}\left(\partial_\mu g_{\lambda\nu}+\partial_\nu g_{\lambda\mu}-\partial_\lambda g_\mn\right)\right]+\notag\\
&+\frac{1}{2}\,\left[g^{\alpha\lambda}\left(\partial_\mu\,\delta g_{\lambda\nu} + \partial_\nu\,\delta g_{\lambda\mu} - \partial_\lambda\,\delta g_\mn\right)\right]\,.\label{eq:varchrist}
\end{align}
Substituting the expression for the covariant derivative of the metric tensor
\be
\D_\alpha \delta g_\mn = \partial_\alpha \delta g_\mn-\Gamma^\lambda_{\alpha\mu}\delta g_{\lambda\nu}-\Gamma^\lambda_{\alpha\nu}\delta g_{\lambda\mu}
\ee
and using \rif{eq:var-g}, equation \rif{eq:varchrist} becomes
\begin{align}&
\begin{aligned}\delta\Gamma^\alpha_\mn =\frac{1}{2}\,&\left[\delta g^{\alpha\lambda}\left(\partial_\mu g_{\lambda\nu}+\partial_\nu g_{\lambda\mu}-\partial_\lambda g_\mn\right)\right]+\notag\\
&+\frac{1}{2}\,g^{\alpha\lambda}\left[\D_\mu\delta g_{\nu\lambda}+\D_\nu\delta g_{\mu\lambda}-\D_\lambda \delta g_\mn+2\,\Gamma^\sigma_\mn \delta g_{\sigma\lambda}\right]\notag
\end{aligned}\\
&\qquad\,\,
\begin{aligned}
=\frac{1}{2}\,&\delta g^{\alpha\beta} g_{\beta\sigma}g^{\sigma\lambda}\left[\partial_\mu g_{\lambda\nu}+\partial_\nu g_{\lambda\mu}-\partial_\lambda g_\mn\right]+\notag\\
&+\frac{1}{2}\,g^{\alpha\lambda}\left[\D_\mu\delta g_{\nu\lambda}+\D_\nu\delta g_{\mu\lambda}-\D_\lambda \delta g_\mn\right]-\,g_{\sigma\beta}\delta g^{\beta\alpha}\Gamma^\sigma_\mn
\end{aligned}\\
&\qquad\,\,=\frac{1}{2}\,g^{\alpha\lambda}\left[\D_\mu\delta g_{\nu\lambda}+\D_\nu\delta g_{\mu\lambda}-\D_\lambda \delta g_\mn\right]\,.
\end{align}
Evidently, in the last step, the first term has canceled with the third. In similar fashion, one can prove that
\be
\delta\Gamma^\nu_\mn = \frac{1}{2}\,g^{\nu\lambda}\D_\mu\delta g_{\nu\lambda}\,,
\ee
so that
\begin{align}
g^\mn\delta R_\mn &= \frac{1}{2}\,\D_\lambda\left[g^\mn g^{\lambda\alpha}\left(\D_\mu\delta g_{\nu\alpha}+\D_\nu\delta g_{\mu\alpha}-\D_\alpha\delta g_\mn\right)-\right.\notag \\
&\left. \qquad\qquad\qquad -g^{\mu\lambda}g^{\nu\alpha}\D_\mu\delta g_{\nu\alpha}\right]\notag\\
&=\D^\mu\D^\nu\delta g_\mn-g^\mn\D^2\delta g_\mn\notag\\
&=g_\mn\D^2\delta g^\mn-\D_\mu\D_\nu\delta g^\mn\,,
\end{align}
where in obvious notation $\D^2\equiv\D_\mu\D^\mu$ is the covariant D'Alambertian operator. In order to be able to integrate by parts, it is useful to define the following quantities:
\be
\begin{gathered}
 M_\tau \equiv f_{,R}\,g_\mn\D_\tau\left(\delta g^\mn\right)-\delta g^\mn g_\mn\D_\tau\left(f_{,R}\right)\,,\\
N^\sigma \equiv f_{,R}\,\D_\mu(\delta g^{\mu\sigma})-\delta g^{\mu\sigma}\D_\mu\left(f_{,R}\right)\,.
\end{gathered}
\ee
One can see that
\begin{align}
\delta\idx\,\-g &\left[f_{,R}(R)g^\mn\delta R_\mn\right]=\notag\\
&=\idx\,\-g\,\delta g^\mn\left(g_\mn\D^2-\D_\mu\D_\nu\right)f_{,R}(R)\,+\notag\\
&\qquad +\idx\,\-g\,\left(\D^\tau M_\tau+\D_\sigma N^\sigma\right)\,.
\end{align}
The second term on the right-hand side, with the aid of the Gauss-Stokes theorem, can be expressed as a surface integral, and can be basically put to zero by requiring that the variations $\delta g_\mn$ vanish on the boundary\footnote{More formally, one can see that the term arising from the evaluation of the integral perfectly cancels out with the Gibbons-York-Hawking like boundary term \cite{Guarnizo:2010xr}.}.

Collecting all terms, we obtain
\be
\delta A=\idx\,\-g\delta g^\mn\left[f_{,R} R_\mn-\frac{1}{2}\,f g_\mn+\left(g_\mn\D^2-\D_\mu\D_\nu\right)f_{,R}\right]\,.
\ee
Adding a matter action $S_m$, and defining the energy-momentum tensor as
\be
T_\mn\equiv \frac{2}{\-g}\,\frac{\delta A_m}{\delta g^\mn}\,,
\ee
the complete field equation take the form
\be
f_{,R} R_\mn-\frac{1}{2}\,fg_\mn+\left(g_\mn\D^2-\D_\mu\D_\nu\right)f_{,R} = -\frac{1}{2}\,T_\mn\,.\label{eq:modfieldeq}
\ee
General relativity is recovered with the choice
\be
f=-\frac{\mpl^2}{16\pi}\,R\,,
\ee
which leads to the standard Einstein equations
\be
G_\mn\equiv R_\mn-\frac{1}{2}\,R\, g_\mn = \frac{8\pi}{\mpl^2}\,T_\mn\,.
\ee
\section{\titlefR Theories}
It is now straightforward to compute the field equations for more particular cases. With $f(R)$, the trick is trivial, since one only needs to put $f=f(R)$ in \rif{eq:modfieldeq}. Thus, the action
\be
A^{f(R)} = -\frac{\mpl^2}{16\pi}\idx\,\-g\,f(R) + A_M(\psi;g_\mn)
\ee
produces the field equations
\be
f_{,R}(R)R_\mn-\frac{1}{2}\,f(R)g_\mn + (g_\mn\D^2-\D_\mu\D_\nu)f_{,R}(R)=8\pi G\,T_\mn\,.
\ee
\section{Scalar-Tensor Theories}
In the case of scalar-tensor gravities, this is just slightly more complicated. We substitute $f=f(\phi)$ in \rif{eq:modfieldeq} and consider the complete action
\be
A^\text{ST} = \idx\,\-g\,\left[-\frac{\mpl^2\,f(\phi)}{16\pi}R+\frac{1}{2}\,(\partial\phi)^2-V(\phi)\right] + A_M(\psi;g_\mn)\,,
\ee
where
\be
(\partial\phi)^2 = g^\mn\,\partial_\mu\phi\,\partial_\nu\phi\,.
\ee
Variation with respect to metric yields the field equations
\be
f(\phi)G_\mn = 8\pi G\,\left(T_\mn^{(m)} + T_\mn^\phi\right)+(\D_\mu\D_\nu-g_\mn\D^2)f(\phi)\,,
\ee
where
\be
T_\mn^\phi=\partial_\mu\phi\,\partial_\nu\phi-\frac{1}{2}\,g_\mn(\partial\phi)^2+g_\mn\,V(\phi)
\ee
is the usual energy-momentum tensor for a minimally-coupled scalar field.

\cleardoublepage
\manualmark
\titolino{\bibname}
\cleardoublepage

\bibliography{./Biblio}%

\end{document}